\documentclass[onecolumn]{revtex4}
\usepackage{amsmath}
\usepackage{amsfonts,color}
\usepackage{amssymb,float}
\usepackage{mathtools}
\usepackage[utf8]{inputenc}
\usepackage{url}
\usepackage{graphicx}
\usepackage{refstyle}
\usepackage{wasysym}
\usepackage{accents}
\usepackage{graphicx}
\usepackage{color}
\usepackage[dvipsnames]{xcolor}
\usepackage{hyperref}

\hypersetup{
    colorlinks=true,
    linkcolor=blue,
    filecolor=magenta,      
    citecolor=red
}

\makeatletter
\long\def\dddddot#1{%
  {\mathop {#1}\limits ^{\vbox to-1.4\ex@ {\kern -\tw@ \ex@ \hbox {\normalfont .....}\vss }}}%
}
\long\def\multidots#1#2{%
  \count@=0
  {{\mathop {#2}\limits ^{\vbox to-1.4\ex@ {\kern -\tw@ \ex@ \hbox {\normalfont %
  \loop%
  \ifnum#1>\count@%
  .%
  \advance\count@ by1%
  \repeat%
  }\vss }}}}%
}
\makeatother

\newcommand{\udt}[3]{#1^{#2}_{\phantom{#2}#3}}
\newcommand{\udut}[4]{#1^{#2\phantom{#3}#4}_{\phantom{#2}#3\phantom{#4}}}

\newcommand{\dut}[3]{#1_{#2}^{\phantom{#2}#3}}
\newcommand{\dudt}[4]{#1_{#2\phantom{#3}#4}^{\phantom{#2}#3}}

\newcommand{\lc}[1]{\accentset{\circ}{#1}}

\newcommand{\dd}{{\rm d}}

\begin{document}

\title{Teleparallel scalar-tensor gravity through cosmological dynamical systems}

\author{S.A. Kadam}
\email{k.siddheshwar47@gmail.com}
\affiliation{Department of Mathematics, Birla Institute of Technology and Science-Pilani, Hyderabad Campus, Hyderabad-500078, India}

\author{B. Mishra}
\email{bivu@hyderabad.bits-pilani.ac.in}
\affiliation{Department of Mathematics, Birla Institute of Technology and Science-Pilani, Hyderabad Campus, Hyderabad-500078, India}

\author{Jackson Levi Said}
\email{jackson.said@um.edu.mt}
\affiliation{Institute of Space Sciences and Astronomy, University of Malta, Malta, MSD 2080}
\affiliation{Department of Physics, University of Malta, Malta}

\date{\today}

\begin{abstract}
Scalar-tensor theories offer the prospect of explaining the cosmological evolution of the Universe through an effective description of dark energy as a quantity with a non-trivial evolution. In this work, we investigate this feature of scalar-tensor theories in the teleparallel gravity context. Teleparallel gravity is a novel description of geometric gravity as a torsional- rather than curvature-based quantity which presents a new foundational base for gravity. Our investigation is centered on the impact of a nontrivial input from the kinetic term of the scalar field. We consider a number of model settings in the context of the dynamical system to reveal their evolutionary behavior. We determine the critical points of these systems and discuss their dynamics.
\end{abstract}

\maketitle

\section{Introduction}

The last several decades have seen cosmology radically altered with unprecedented observational evidence, first with the discovery of a Universe that is accelerating \cite{SupernovaSearchTeam:1998fmf,SupernovaCosmologyProject:1998vns} due to some form of dark energy and more recently with the increasingly convincing Hubble tension \cite{Abdalla:2022yfr,DiValentino:2021izs,Brout:2022vxf,Bernal:2016gxb,Benisty:2021cmq}. This tension has arisen between local measurements of the Hubble constant $H_0$ \cite{Riess:2021jrx,Wong:2019kwg} and predictions of this cosmological parameter from observations from the early Universe \cite{Planck:2018vyg,DES:2021wwk} which require the use of the concordance model or $\Lambda$CDM. To a lesser extent, the tension also appears in other cosmological parameters related to large scale structure measurements \cite{DiValentino:2020vhf,DiValentino:2020zio,DiValentino:2020vvd}, which has prompted various attempts in the community to resolve the possible issue using additional contributions from the matter sector, as well as renewed interest in physics beyond general relativity (GR) which are now becoming mainstream.

GR acts as the base gravitational model on which $\Lambda$CDM rests as a foundation. However, there exist many possible directions for modified gravity to work toward. Recently, there have been a plethora of possible observationally motivated theories in the literature \cite{Sotiriou:2008rp,Clifton:2011jh,CANTATA:2021ktz,Nojiri:2017ncd} which are yet to show promise of behaving better than $\Lambda$CDM when confronted with observations. By and large, these are mostly built as correction terms to the Einstein-Hilbert action of GR \cite{Faraoni:2008mf,Capozziello:2011et}. In these works, gravitational interactions continue to be described through the curvature associated with the Levi-Civita connection which is the sole source of curvature in GR \cite{misner1973gravitation}. However, this is not the only way to construct gravitational models. In recent years there has been growing interest in teleparallel gravity (TG) where torsion is considered as the base mode of interactions for the gravitational sector \cite{Bahamonde:2021gfp,Aldrovandi:2013wha,Cai:2015emx,Krssak:2018ywd}.

In TG, the Levi-Civita connection is replaced with the teleparallel connection \cite{Weitzenbock1923,Bahamonde:2021gfp} which expresses geometric deformations through torsion rather than curvature. In this regime, all measures of curvature vanish identically such as the Ricci scalar $R(\udt{\Gamma}{\sigma}{\mu\nu}) = 0$, where as its regular curvature form does not vanish $\lc{R}(\udt{\lc{\Gamma}}{\sigma}{\mu\nu}) \neq 0$ (over-circles represent quantities calculated with the Levi-Civita connection). By relating the connections together, a torsion scalar $T$ can be produced which is equal to the regular Ricci scalar up to a boundary term, meaning that it will produce field equations that are dynamically equivalent to GR, also called the \textit{Teleparallel equivalent of General Relativity} (TEGR). Thus, observations are indistinguishable between GR and TEGR. This boundary term is important because it embodies the fourth-order terms of the Ricci scalar, which are boundary terms in the Einstein-Hilbert action. Its only when generalizations such as $f(\lc{R})$ gravity are considered \cite{Capozziello:2011et} do these terms impact the order of the field equations. 

The division of the torsion scalar and boundary term means that a weaker Lovelock theorem is developed \cite{Lovelock:1971yv,Gonzalez:2015sha,Bahamonde:2019shr} where much more general theories that are second order can be produced such as $f(T)$ gravity \cite{Krssak:2018ywd}. $f(T)$ theories of gravity \cite{Ferraro:2006jd,Ferraro:2008ey,Bengochea:2008gz,Linder:2010py,Chen:2010va,Bahamonde:2019zea,Paliathanasis:2021nqa,Leon:2022oyy,Duchaniya:2022rqu} emerged using the same rationale as $f(\lc{R})$ gravity \cite{Sotiriou:2008rp,Faraoni:2008mf,Capozziello:2011et} where the TEGR Lagrangian is generalized to an arbitrary function of the torsion scalar. These theories are generically second order and have shown promise in meeting some observational challenges in the literature \cite{Cai:2015emx,Farrugia:2016qqe,Finch:2018gkh,Farrugia:2016xcw,Iorio:2012cm,Deng:2018ncg}. In this work, we explore the space of cosmological models that feature a scalar field. In curvature-based gravity, there has been extensive analyses of scalar couplings in this setting \cite{Copeland:2006wr,Bamba:2012cp} that have shown real promise in constructing viable models. An interesting collection of such models is that of Horndeski gravity \cite{Horndeski:1974wa} in which a single scalar field is allowed to couple arbitrarily to the Einstein-Hilbert action provided that it produces second order field equations \cite{Kobayashi:2019hrl,Deffayet:2009wt,Kobayashi:2011nu}. In the TG regime, a teleparallel analogue of Horndeski gravity has been proposed in the literature \cite{Bahamonde:2019shr} which has the added benefit that it allows for a much wider range of models that are consistent with recent measurements of the gravitational propagation speed \cite{Bahamonde:2019ipm}, as well as hosting a vast array of models that are consistent with solar system tests \cite{Bahamonde:2020cfv}. These models also show a rich array of possible gravitational wave polarizations \cite{Bahamonde:2021dqn} which is a growing topic for gravitational wave astronomy. More recently, the teleparallel analogue of Horndeski gravity has been used to construct well motivated models using well-tempered cosmological methods in either Minkowski \cite{Bernardo:2021bsg} or cosmological backgrounds \cite{Bernardo:2021izq}. Another interesting direction related to constructing models in this framework is that of using Noether symmetries to construct cosmology oriented models as was performed in Ref.~\cite{Dialektopoulos:2021ryi}.

In this study, we probe possible cosmological behaviours using a dynamical systems approach which can reveal the evolution of the individual models under consideration in the context of their potential to attain the known critical points of the Universe \cite{Bahamonde:2017ize}. By taking a flat homogeneous and isotropic background solution, dynamical systems can be used to determine the number and nature of the critical points of the system which express whether these positions in the cosmic evolution are stable or not. In the literature, scalar couplings with torsion have shown interesting results such as Ref.~\cite{Gonzalez-Espinoza:2020jss} where a nontrivial scalar field is allowed to contribute while the kinetic term remains canonical. In the present study, we extend these efforts to allow for several models with a nontrivial kinetic term. Our aim is to assess whether this type of generalization can produce a dynamical system consistent with our expectations for the Universe. To that end, we first briefly review the literature on TG and the structure of the teleparallel analogue of Horndeski gravity in Sec.~\ref{sec:review}. The cosmological system is then introduced in Sec.~\ref{sec:backgroundexpressions}, while the dynamical system analysis for the models is conducted in Secs.~\ref{sec:model-I},\ref{sec:model-II}. In these models we explore the impact of a power-law kinetic term coupled with the torsion scalar and another nonzero (for this background) scalar of the theory respectively. Finally, we summarize our results in Sec.~\ref{sec:conclusion} where we discuss how these results compare with the present literature.

\section{Scalar-Tensor Teleparallel Gravity}\label{sec:review}

The curvature associated with GR and other models based on the Levi-Civita connection \cite{Clifton:2011jh} is reformulated in TG through the teleparallel connection where torsion replaces curvature as the means by which gravity is expressed \cite{Bahamonde:2021gfp}. The origin of curvature in GR and related theories is not the metric, but rather the Levi-Civita connection $\udt{\lc{\Gamma}}{\sigma}{\mu\nu}$ (over-circles are used throughout to denote quantities determined using the Levi-Civita connection) which characterizes how any geometric deformation is to be characterized. On the other hand, TG characterizes gravitation as torsion through the teleparallel connection $\udt{\Gamma}{\sigma}{\mu\nu}$ which is curvature-less and satisfies metricity \cite{Hayashi:1979qx,Aldrovandi:2013wha}. While the regular Riemann tensor $\udt{\lc{R}}{\beta}{\mu\nu\alpha}$ does not vanish, its teleparallel analogue does, as do all quantitative measures of curvature, meaning that an entirely new approach to forming gravitational theory needs to be adopted (see reviews in Refs. \cite{Krssak:2018ywd,Cai:2015emx,Aldrovandi:2013wha}).

The most efficient path to forming teleparallel theories of gravity is through the tetrad $\udt{e}{A}{\mu}$ (and its inverses $\dut{E}{A}{\mu}$) which takes the place of the metric as the fundamental variable of theory through the relations
\begin{align}\label{metric_tetrad_rel}
    g_{\mu\nu}=\udt{e}{A}{\mu}\udt{e}{B}{\nu}\eta_{AB}\,,& &\eta_{AB} = \dut{E}{A}{\mu}\dut{E}{B}{\nu}g_{\mu\nu}\,,
\end{align}
where Latin indices represent coordinates on the tangent space while Greek indices continue to represent indices on the general manifold \cite{Cai:2015emx}. While they do appear in GR, the direct use of tetrads in GR is largely suppressed \cite{Chandrasekhar:1984siy}. Similar to the metric, the tetrads must satisfy orthogonality conditions which take of the form of
\begin{align}
    \udt{e}{A}{\mu}\dut{E}{B}{\mu}=\delta^A_B\,,&  &\udt{e}{A}{\mu}\dut{E}{A}{\nu}=\delta^{\nu}_{\mu}\,,
\end{align}
preserving internal consistency.

The teleparallel connection can be directly defined as \cite{Weitzenbock1923,Krssak:2015oua}
\begin{equation}
    \udt{\Gamma}{\sigma}{\nu\mu} := \dut{E}{A}{\sigma}\left(\partial_{\mu}\udt{e}{A}{\nu} + \udt{\omega}{A}{B\mu}\udt{e}{B}{\nu}\right)\,,
\end{equation}
where contributions of the tetrad are complemented by $\udt{\omega}{A}{B\mu}$ which is a flat spin connection and is responsible for incorporating local Lorentz invariance into teleparallel theories. This arises due to the explicit appearance of the Lorentz indices and thus the Lorentz frames. As tetrads appear in GR, so too do spin connections but one important distinction is that they are not flat in the GR case \cite{misner1973gravitation}. The tetrad-spin connection pair represent the gravitational and local degrees of freedom in the equations of motion of TG. Now, analogous to the way that the Levi-Civita connection builds up to the Riemann tensor, the teleparallel connection directly leads to the torsion tensor \cite{Hayashi:1979qx}
\begin{equation}
    \udt{T}{\sigma}{\mu\nu} :=2\udt{\Gamma}{\sigma}{[\nu\mu]}\,,
\end{equation}
where square brackets denote the anti-symmetry operator, and where $\udt{T}{\sigma}{\mu\nu}$ acts as the field strength of gravity \cite{Aldrovandi:2013wha}. This tensor is covariant under both diffeomorphisms and local Lorentz transformations. By an appropriate combination of contractions of torsion tensors, a torsion scalar can be written down such that \cite{Krssak:2018ywd,Cai:2015emx,Aldrovandi:2013wha,Bahamonde:2021gfp}
\begin{equation}
    T:=\frac{1}{4}\udt{T}{\alpha}{\mu\nu}\dut{T}{\alpha}{\mu\nu} + \frac{1}{2}\udt{T}{\alpha}{\mu\nu}\udt{T}{\nu\mu}{\alpha} - \udt{T}{\alpha}{\mu\alpha}\udt{T}{\beta\mu}{\beta}\,,
\end{equation}
which is the result of a requirement that $T$ be equivalent to the curvature scalar $\lc{R}$ (up to a boundary term). Similar to the way in which the curvature scalar is dependent only on the Levi-Civita connection, the torsion tensor is only dependent on the teleparallel connection.

Exchanging the Levi-Civita connection with the teleparallel connection means that measures of curvature identically vanish, such as $R\equiv 0$ (where we emphasize that $R = R(\udt{\Gamma}{\sigma}{\mu\nu})$ and $\lc{R}=\lc{R}(\udt{\lc{\Gamma}}{\sigma}{\mu\nu})$). In this context, we can write the following relation for the curvature and torsion scalars \cite{Bahamonde:2015zma,Farrugia:2016qqe}
\begin{equation}\label{LC_TG_conn}
    R=\lc{R} + T - B = 0\,.
\end{equation}
where $B = (2/e)\partial_{\rho}\left(e\udut{T}{\mu}{\mu}{\rho}\right)$ is a total divergence term, where $e=\det\left(\udt{e}{a}{\mu}\right)=\sqrt{-g}$ is the determinant of the tetrad. This guarantees that the GR and TEGR actions will generate identical field equations.

Taking a similar rationale as with many other extensions to GR, such as $f(\lc{R})$ gravity \cite{DeFelice:2010aj,Capozziello:2011et}, TEGR can be directly generalized to $f(T)$ by raising the TEGR Lagrangian to an arbitrary function \cite{Ferraro:2006jd,Ferraro:2008ey,Bengochea:2008gz,Linder:2010py,Chen:2010va,RezaeiAkbarieh:2018ijw}
\begin{equation}\label{f_T_ext_Lagran}
    \mathcal{S}_{\mathcal{F}(T)}^{} =  \frac{1}{2\kappa^2}\int \mathrm{d}^4 x\; e\left(-T + \mathcal{F}(T)\right) + \int \mathrm{d}^4 x\; e\mathcal{L}_{\text{m}}\,,
\end{equation}
where $\kappa^2=8\pi G$, and $\mathcal{L}_{\text{m}}$ is the matter Lagrangian in the Jordan frame. The tetrad and spin connection are independent variables in TG and thus produce independent field, which correspond to the ten metrical field equations and the six local Lorentz degrees of freedom. However, the tetrad variation $\dut{W}{a}{\mu} = \delta \mathcal{S}_{\mathcal{F}(T)}^{}/ \delta \udt{e}{A}{\mu}$ has an interesting property that when acted on by the anti-symmetric operator, it produces the spin connection field equations. Thus, all the equations of motion can be determined with this variation as \cite{Bahamonde:2021gfp}
\begin{equation}
    W_{(\mu\nu)} = \kappa^2 \Theta_{\mu\nu}\,, \quad \text{and} \quad W_{[\mu\nu]} = 0\,,
\end{equation}
which also holds for any other teleparallel gravity theory, and where $\dut{\Theta}{\rho}{\nu}$ is the regular energy-momentum tensor for matter. In this setting, the Weitzenb\"{o}ck gauge can defined as the tetrad choice in which the spin connection vanishes, i.e. where the anti-symmetric field equations are identically zero for the choice~(\ref{metric_tetrad_rel}) of tetrad.

In the present work, we explore the dynamical systems of a particular class of scalar-tensor models which uniquely appear in teleparallel gravity. To form the broader class of scalar-tensor extensions in TG, we first consider the irreducibles pieces of the torsion tensor \cite{Hayashi:1979qx,Bahamonde:2017wwk}
\begin{align}
    a_{\mu} & :=\frac{1}{6}\epsilon_{\mu\nu\lambda\rho}T^{\nu\lambda\rho}\,,\\
    v_{\mu} & :=\udt{T}{\lambda}{\lambda\mu}\,,\\
    t_{\lambda\mu\nu} & :=\frac{1}{2}\left(T_{\lambda\mu\nu}+T_{\mu\lambda\nu}\right)+\frac{1}{6}\left(g_{\nu\lambda}v_{\mu}+g_{\nu\mu}v_{\lambda}\right)-\frac{1}{3}g_{\lambda\mu}v_{\nu}\,,
\end{align}
which are respectively the axial, vector, and purely tensorial parts, and where $\epsilon_{\mu\nu\lambda\rho}$ is the totally antisymmetric Levi-Civita tensor in four dimensions. Taking appropriate contractions leads to the scalar invariants \cite{Bahamonde:2015zma}
\begin{align}
    T_{\text{ax}} & := a_{\mu}a^{\mu} = -\frac{1}{18}\left(T_{\lambda\mu\nu}T^{\lambda\mu\nu}-2T_{\lambda\mu\nu}T^{\mu\lambda\nu}\right)\,,\\
    T_{\text{vec}} & :=v_{\mu}v^{\mu}=\udt{T}{\lambda}{\lambda\mu}\dut{T}{\rho}{\rho\mu}\,,\\
    T{_{\text{ten}}} & :=t_{\lambda\mu\nu}t^{\lambda\mu\nu}=\frac{1}{2}\left(T_{\lambda\mu\nu}T^{\lambda\mu\nu}+T_{\lambda\mu\nu}T^{\mu\lambda\nu}\right)-\frac{1}{2}\udt{T}{\lambda}{\lambda\mu}\dut{T}{\rho}{\rho\mu}\,.
\end{align}
These three scalars form the most general parity preserving scalars that are quadratic in contractions of the torsion tensor, and even reproduce the torsion scalar $T:=\frac{3}{2}T_{\text{ax}}+\frac{2}{3}T_{\text{ten}}-\frac{2}{3}T{_{\text{vec}}}$. Recently, this has led to the proposal of a teleparallel analogue of Horndeski gravity \cite{Bahamonde:2019shr,Bahamonde:2019ipm,Bahamonde:2020cfv,Dialektopoulos:2021ryi,Bahamonde:2021dqn,Bernardo:2021izq,Bernardo:2021bsg}, also called Bahamonde-Dialektopoulos-Levi Said (BDLS) theory. As in curvature-baed gravity, this is grounded on the Lovelock theorem \cite{Lovelock:1971yv,Gonzalez:2015sha,Gonzalez:2019tky}, and leads to the linear torsion scalar contraction scalar invariants \cite{Bahamonde:2019shr}
\begin{equation}
    I_2 = v^{\mu} \phi_{;\mu}\,,\label{eq:lin_contrac_scalar}
\end{equation}
where $\phi$ is the scalar field, and while for the quadratic scenario, we find
\begin{align}
    J_{1} & =a^{\mu}a^{\nu}\phi_{;\mu}\phi_{;\nu}\,,\label{eq:quad_contrac_scal1}\\
    J_{3} & =v_{\sigma}t^{\sigma\mu\nu}\phi_{;\mu}\phi_{;\nu}\,,\\
    J_{5} & =t^{\sigma\mu\nu}\dudt{t}{\sigma}{\alpha}{\nu}\phi_{;\mu}\phi_{;\alpha}\,,\\
    J_{6} & =t^{\sigma\mu\nu}\dut{t}{\sigma}{\alpha\beta}\phi_{;\mu}\phi_{;\nu}\phi_{;\alpha}\phi_{;\beta}\,,\\
    J_{8} & =t^{\sigma\mu\nu}\dut{t}{\sigma\mu}{\alpha}\phi_{;\nu}\phi_{;\alpha}\,,\\
    J_{10} & =\udt{\epsilon}{\mu}{\nu\sigma\rho}a^{\nu}t^{\alpha\rho\sigma}\phi_{;\mu}\phi_{;\alpha}\,,\label{eq:quad_contrac_scal10}
\end{align}
where semicolons represent covariant derivatives with respect to the Levi-Civita connection. The Levi-Civita connection enters into the scalar field sector through the minimal coupling prescription of TG (See Ref.~\cite{Bahamonde:2019shr} for further details).

Naturally, the regular Horndeski terms from curvature-based gravity also appear in this framework \cite{Horndeski:1974wa}
\begin{align}
    \mathcal{L}_{2} & :=G_{2}(\phi,X)\,,\label{eq:LagrHorn1}\\
    \mathcal{L}_{3} & :=G_{3}(\phi,X)\mathring{\Box}\phi\,,\\
    \mathcal{L}_{4} & :=G_{4}(\phi,X)\left(-T+B\right)+G_{4,X}(\phi,X)\left(\left(\mathring{\Box}\phi\right)^{2}-\phi_{;\mu\nu}\phi^{;\mu\nu}\right)\,,\\
    \mathcal{L}_{5} & :=G_{5}(\phi,X)\mathring{G}_{\mu\nu}\phi^{;\mu\nu}-\frac{1}{6}G_{5,X}(\phi,X)\left(\left(\mathring{\Box}\phi\right)^{3}+2\dut{\phi}{;\mu}{\nu}\dut{\phi}{;\nu}{\alpha}\dut{\phi}{;\alpha}{\mu}-3\phi_{;\mu\nu}\phi^{;\mu\nu}\,\mathring{\Box}\phi\right)\,,\label{eq:LagrHorn5}
\end{align}
where the kinetic term is defined as $X:=-\frac{1}{2}\partial^{\mu}\phi\partial_{\mu}\phi$. BDLS theory simply adds the further Lagrangian component \cite{Bahamonde:2019shr}
\begin{equation}
    \mathcal{L}_{\text{{\rm Tele}}}:= G_{\text{{\rm Tele}}}\left(\phi,X,T,T_{\text{ax}},T_{\text{vec}},I_2,J_1,J_3,J_5,J_6,J_8,J_{10}\right)\,.
\end{equation}
This results in the BDLS action given by
\begin{equation}\label{action}
    \mathcal{S}_{\text{BDLS}} = \frac{1}{2\kappa^2}\int d^4 x\, e\mathcal{L}_{\text{{\rm Tele}}} + \frac{1}{2\kappa^2}\sum_{i=2}^{5} \int d^4 x\, e\mathcal{L}_i+ \int d^4x \, e\mathcal{L}_{\rm m}\,,
\end{equation}
with $\lc{G}_{\mu\nu}$ is the standard Einstein tensor. The curvature-based regular Horndeski theory is recovered for the limit where $G_{\text{{\rm Tele}}} = 0$. BDLS theory is invariant under local Lorentz transformations and diffeomorphisms due to it being based on the torsion tensor. One minor difference with regular Horndeski theory is that calculations are now based on the tetrad and spin connection components rather than the metric tensor, but this will not impact the values for the $\mathcal{L}_{2} - \mathcal{L}_{5}$ contributions. Another important point to highlight is that this version of the popular scalar-tensor theory provides a much more general framework on which to construct models. Indeed, this allows for a path to circumvent the strong constraints imposed in regular Horndeski gravity from gravitational wave observations~\cite{TheLIGOScientific:2017qsa}.

\section{Teleparallel Scalar-Tensor flat FLRW Cosmology} \label{sec:backgroundexpressions}

We consider a flat isotropic and homogeneous background cosmology through the Friedmann--Lema\^{i}tre--Robertson--Walker (FLRW) metric \cite{misner1973gravitation}
\begin{align}\label{eq:28}
    \dd s^2 = - \dd t^2 + a(t)^2(\dd x^2 + \dd y^2 + \dd z^2)\,,
\end{align}
Where $a(t)$ is the scale factor depends on cosmic time $t$. This can be described by the tetrad
\begin{align}\label{eq:29}
    \udt{e}{A}{\mu}=(1,a(t),a(t),a(t))\,,
\end{align}
which is consistent with the Weitzenb\"{o}ck gauge described in Sec.~\ref{sec:review}. We take the standard definition of the Hubble parameter $H= \frac{\dot{a}}{a}$, where dots refer to derivatives with respect to cosmic time. We also consider the equation of state (EoS) for matter $\omega_{m}=\frac{p_{\rm m}}{\rho_{\rm m}} = 0$ and radiation $\omega_{r}=\frac{p_{r}}{\rho_{\rm r}} = 1/3$, which will both contribute to our representation of cosmology. In this work, we consider the class of models in which
\begin{align}
    G_2 &= X - V(\phi)\,,\\
    G_3 &= 0 = G_5\,,\\
    G_4 &= 1/2\kappa^2\,,
\end{align}
where we take a generalization of a canonical scalar field together with a TEGR term. This then lets use probe different forms of the $G_{\rm Tele}$ term in the action~(\ref{action}). As discussed, our aim is to probe the nature of power-law couplings with the kinetic term. To that end, we consider two models that embody nonvanishing terms for an FLRW background cosmology, which are
\begin{align}
    G_{{\rm Tele}_1} &= X^{\alpha} T\,,\\
    G_{{\rm Tele}_2} &= X^{\alpha} I_2\,,
\end{align}
where the other terms effectively do not contribute to the Friedmann equations \cite{Bahamonde:2019shr}, and where
\begin{align}
    T &= 6H^2\,,\\
    I_2 &= 3H\dot{\phi}\,.\label{eq:I_2_scalar}
\end{align}
Thus, we can write the effective Friedmann equations as
\begin{align}
    \dfrac{3}{\kappa^{2}}H^{2} &= \rho_{\rm m} + \rho_{\rm r} + X + V + 6H\dot{\phi} G_{{\rm Tele},I_2} + 12H^2 G_{{\rm Tele},T} + 2X G_{{\rm Tele},X} - G_{{\rm Tele}}\,,\label{eq:30}\\
    -\dfrac{2}{\kappa^{2}}\dot{H} &= \rho_{\rm m} + \frac{4}{3}\rho_{\rm r} + 2X + 3H\dot{\phi}G_{{\rm Tele},I_2} + 2XG_{{\rm Tele},X} - \frac{d}{dt}\left(4HG_{{\rm Tele},T} + \dot{\phi} G_{{\rm Tele},I_2}\right) \,,\label{eq:31}
\end{align}
and where the scalar field equation is given by
\begin{equation}
    \frac{1}{a^3}\frac{d}{dt}\left[a^3 \dot{\phi} \left( 1+ G_{{\rm Tele},X}\right)\right] = -V'(\phi) - 9H^2 G_{{\rm Tele},I_2} + G_{{\rm Tele},\phi} - 3\frac{d}{dt}\left(HG_{{\rm Tele},I_2}\right)\,.\label{eq:kg_eq}
\end{equation}
This effective fluid observes the energy-conservation equation
\begin{align}\label{eq:40}
    \dot{\rho}_{DE}+3H(\rho_{\rm DE}+p_{\rm DE}) = 0\,,
\end{align}
but this description breaks down at perturbative level. Moreover, in what follows we will use the density parameters
\begin{align}\label{eq:41}
    \Omega_{m}=\frac{\kappa^2\rho_m}{3H^2}, \quad \Omega_{\rm DE}=\frac{\kappa^2\rho_{\rm DE}}{3H^2}, \quad \Omega_{r}=\frac{\kappa^2\rho_r}{3H^2}\,.
\end{align}
which satisfies the conservation relation
\begin{align}\label{eq:42}
    \Omega_m + \Omega_{\rm DE} + \Omega_r = 1\,.
\end{align}

\section{Model 1 -- Kinetic Term Coupled with Torsion Scalar}\label{sec:model-I}

The action for Model 1 is given by a coupling between a power-law-like term and the torsion scalar represented by
\begin{equation}\label{eq:action_model_1}
    S = \int d^4x e [X-V(\phi)-\frac{T}{2\kappa^{2}}+X^{\alpha}T]+S_{m}+S{r}\,,
\end{equation}
where $V(\phi)$ is the scalar potential, $S_m$ represents the action of for matter and $S_r$ describes the action for the radiation component. We shall perform the dynamical system analysis for the general case $\alpha$, followed by two examples with $\alpha=1$, $\alpha=2$. Taking background FLRW cosmology~(\ref{eq:29}) and the action above, we can obtain the Friedmann equations in Eqs.~(\ref{eq:44}--\ref{eq:45}) whereas the Klein-Gordon equation can be obtained in Eq.~(\ref{eq:46}), altogether giving
\begin{align}
    \frac{T}{2\kappa^{2}}-V(\phi)-X-X^{\alpha}T-2\alpha X^{\alpha}T &= \rho_{\rm m}+\rho_{\rm r}\,,\label{eq:44}\\
    -V(\phi)+\frac{T}{2\kappa^{2}}+ X-X^{\alpha}T+\frac{2\dot{H}}{\kappa^{2}}-4X^{\alpha}\dot{H}-8\alpha X^{\alpha}H\frac{\ddot{\phi}}{\dot{\phi}} &= -p_{r}\,,\label{eq:45}\\
    V^{'}(\phi)+3H\dot{\phi}+\frac{6X^{\alpha}T\alpha H}{\dot{\phi}}+\frac{4\alpha X^{\alpha}T \dot{H}}{\dot{\phi} H}+\ddot{\phi}[1-\alpha X^{\alpha-1}T+\alpha^{2}X^{\alpha-2}T] &= 0\,.\label{eq:46}
\end{align}
Now using the background expressions defined in Sec-\ref{sec:backgroundexpressions}, we can obtain the expression for energy density and pressure of effective dark energy as
\begin{align}
    \rho_{\rm DE} &= X^{\alpha}T+2\alpha X^{\alpha}T+V(\phi)+X\,,\label{eq:47}\\
    p_{\rm DE} &= -V(\phi)+X-X^{\alpha}T-4X^{\alpha}\dot{H}-\frac{8\alpha X^{\alpha} H \ddot{\phi}} {\dot{\phi}}\,.\label{eq:48}
\end{align}
To study the phases of cosmic evolution, the autonomous dynamical system for the above set of cosmological expressions can be defined using following dimensionless variables.
\begin{align}\label{eq:49}
    x=\dfrac{\kappa\dot{\phi}}{\sqrt{6}H}\,,\quad y=\frac{\kappa\sqrt{V}}{\sqrt{3}H}\,,\quad u=2 X^\alpha \kappa^2\,,\quad \rho= \frac{\kappa\sqrt{\rho_{\rm r}}}{\sqrt{3}H}\,,\quad \lambda=\frac{-V^{'}(\phi)}{\kappa V(\phi)}\,,\quad \Gamma=\dfrac{V(\phi)V^{''}(\phi)}{V^{'}(\phi)^{2}}\,,
\end{align}
where the constraint equation for the dimensionless variables can be obtained as,
\begin{align}\label{eq:50}
    x^{2}+y^{2}+(1+2\alpha)u+\Omega_{m}+\rho^{2}=1\,.
\end{align}
Now the autonomous dynamical system can be defined by differentiating the dimensionless variables with respect to $N=\ln a$ as,
\begin{align}
    \dfrac{dx}{dN} &= \frac{x \left(-x^2 \left(\rho ^2+3 (\alpha  (2 \alpha +5)+1) u-3 y^2-3\right)+\sqrt{6} \lambda  x y^2 (2 \alpha  u+u-1)-3 x^4\right)}{2 (u-1) x^2-2 \alpha  u (2 \alpha  (u+1)+u-1)}\nonumber\\
    & - \frac{\alpha  u x \left(2 \alpha  \left(\rho ^2+3\right)+\rho ^2+(6 \alpha +3) u-3 (2 \alpha +1) y^2-3\right)}{2 (u-1) x^2-2 \alpha  u (2 \alpha  (u+1)+u-1)}\,,\label{eq:dx_dN_model_1}\\
    \dfrac{dy}{dN} &= \frac{-y \left(x^2 \left(\rho ^2+\left(6 \alpha ^2+9 \alpha -3\right) u-3 y^2+3\right)-2 \sqrt{6} \alpha  \lambda  u x y^2+3 x^4\right)}{2 (u-1) x^2-2 \alpha  u (2 \alpha  (u+1)+u-1)} \nonumber\\
    & + \frac{\alpha  u (-y) \left((6 \alpha +3) u-(2 \alpha -1) \left(-\rho ^2+3 y^2-3\right)\right)}{2 (u-1) x^2-2 \alpha  u (2 \alpha  (u+1)+u-1)}-y\sqrt{\frac{3}{2}}  \lambda  x \,,\\
    \dfrac{du}{dN} &= \frac{\alpha  u \left(2 \alpha  u \left(\rho ^2+3 x^2-3 y^2\right)+(u-1) x \left(6 x-\sqrt{6} \lambda y^2\right)\right)}{\alpha u (2 \alpha (u+1)+u-1)-(u-1) x^2}\,,\\
    \frac{d\rho}{dN} &= \frac{\rho \left(-x^2 \left(\rho ^2+6 \alpha ^2 u+9 \alpha  u+u-3 y^2-1\right)+2 \sqrt{6} \alpha \lambda  u x y^2-3 x^4\right)}{2 (u-1) x^2-2 \alpha  u (2 \alpha  (u+1)+u-1)} \nonumber\\
    & + \frac{\alpha \rho u \left(2 \alpha  u+u+(2 \alpha -1) \left(-\rho ^2+3 y^2+1\right)\right)}{2 (u-1) x^2-2 \alpha  u (2 \alpha  (u+1)+u-1)}\,,\\
    \frac{d\lambda}{dN} &= \sqrt{6}(\Gamma-1)\lambda^{2}x\,.\label{eq:dl_dN_model_1}
\end{align}

Unless the parameters $\Gamma$ is known, the dynamical systems presented in this work are not an autonomous systems. We will now on wards focus on the exponential potential $V(\phi) =V_{0}e^{-\lambda \kappa \phi}$  with $\lambda$ is a dimensionless constant. This particular form of potential function leads to, $\Gamma = 1$ and can give rise to an accelerated expansion of the universe. We obtain the critical points (or fixed points) for autonomous dynamical system presented in Eqs.~(\ref{eq:dx_dN_model_1}--\ref{eq:dl_dN_model_1}) by imposing conditions $\frac{dx}{dN}=0$, $\frac{dy}{dN}=0$, $\frac{du}{dN}=0$, $\frac{d\rho}{dN}=0$. The critical points are titled with capital letters and presented in corresponding tables. To study the cosmological implications, the value of the deceleration parameter and value for the total EoS $\omega_{tot}$ are also presented in the tables. From the Table~\ref{TABLE-I} observation, it can be concluded that parameter $\alpha$ contributes in the co-ordinates of critical points $D$ and $E$ and represents the de Sitter solution for the dynamical system presented in Eqs.~(\ref{eq:dx_dN_model_1}--\ref{eq:dl_dN_model_1}). While analyzing the critical points for the case general $\alpha$, there is a chance to get more number of critical points than for a particular value of $\alpha$. The critical points $L$, $M$, $F$, $G$ and $A$ represent same cosmological implication. These critical points show deceleration parameter value $q=\frac{1}{2}$ and $\omega_{tot}=0$, hence explaining the cold dark matter-dominated era. Similarly the critical points $B$, $C$, and $N$ represent the same phase of evolution with value of $\omega_{tot} =1$, hence  cannot describe current accelerated phase of evolution and behave as stiff matter. The critical points $J$ and $K$ are defined at $\lambda=2$ and show value of $\omega_{tot}=\frac{1}{3}$ hence  describe the radiation dominated era. Since the critical points $H$ and $I$ represent deceleration parameter value, $q=-1+\frac{\lambda^2}{2}$, these critical points can describe current acceleration of the universe for any real value of $\lambda$ and are compatible with current observational data.\\

The stability of critical points can be studied by obtaining eigenvalues of linear perturbation matrix at critical points. Depending upon the signature of eigenvalues one can classify the stability properties as, if all the eigenvalues possesses positive signature it is an \textit{unstable node}; if all the eigenvalues possesses negative signature then it is a \textit{stable node}; if one of the eigenvalue possesses positive signature and other possesses negative sign in this case it is \textit{saddle point} and if the determinant of linear perturbation matrix is negative and the real parts of all the eigenvalues possesses negative signature then it is a \textit{stable spiral}. The eigenvalues and stability conditions for dynamical system in Eqs.~(\ref{eq:dx_dN_model_1}--\ref{eq:dl_dN_model_1}) are presented in Table~\ref{TABLE-II}. The existence of positive and negative eigenvalues for permutation matrix at the critical points $A$, $B$, $C$, $J$, $K$ and $N$ describe saddle point behaviour at these critical points hence these critical points are unstable. However value of deceleration parameter at these critical points clarify that these critical points can not explain accelerated expansion phase of the universe. Critical points $F$ and $G$ ensure stability in the range of parameter $\alpha >0$ and $\left(-2 \sqrt{\frac{6}{7}}\leq \lambda <-\sqrt{3}\lor \sqrt{3}<\lambda \leq 2 \sqrt{\frac{6}{7}}\right)$ these critical points explain the standard matter dominated era. The critical points $D$ and $E$ shows stable behaviour and explain the de Sitter solution.

\begin{table}[H]
\caption{Critical Points for Dynamical System Corresponding to Model-I, for General $\alpha$. } 
\centering 
\begin{tabular}{|c|c|c|c|c|c|c|} 
\hline\hline 
Name of Critical Point & $x_{c}$ & $y_{c}$ & $u_{c}$ & $\rho_{c}$ & Deceleration Parameter ($q$) & $\omega_{tot}$\\ [0.5ex] 
\hline\hline 
$A,$ \begin{tabular}{@{}c@{}}$2 \alpha ^2 \tau^2$\\ $+\alpha  \tau^2+2 \alpha ^2 \tau-\alpha  \tau\neq 0 $\end{tabular}& 0 & $0$ & $\tau$ & $0$ & $\frac{1}{2}$ & $0$\\
\hline
$B$ & $1$ & $0$ & $0$ & $0$ & $2$ & $1$\\
\hline
$C$ & $-1$ & $0$ & $0$ & $0$ & $2$ & $1$\\
\hline
$D$, in this case $\lambda=0$ & $\zeta, (\alpha +1)\zeta^2 \neq \left(\alpha -1\right) \alpha$ & $\frac{\sqrt{(\alpha +1) \zeta ^2+\alpha }}{\sqrt{\alpha }}$ & $-\frac{\zeta ^2}{\alpha }$ & $0$ & $-1$ & $-1$\\
\hline
$E$, in this case $\lambda=0$ & $\zeta, (\alpha +1)\zeta^2 \neq \left(\alpha -1\right) \alpha$ &- $\frac{\sqrt{(\alpha +1) \zeta ^2+\alpha }}{\sqrt{\alpha }}$ & $-\frac{\zeta ^2}{\alpha }$ & $0$ & $-1$ & $-1$\\
\hline
$F$ & $\frac{\sqrt{\frac{3}{2}}}{\lambda }$ & $\sqrt{\frac{3}{2}} \sqrt{\frac{1}{\lambda ^2}}$ & $0$ & $0$ & $\frac{1}{2}$ & $0$\\
\hline
$G$ & $\frac{\sqrt{\frac{3}{2}}}{\lambda }$ & $-\sqrt{\frac{3}{2}} \sqrt{\frac{1}{\lambda ^2}}$ & $0$ & $0$ & $\frac{1}{2}$ & $0$\\
\hline
$H$ & $\frac{\lambda }{\sqrt{6}}$ & $\sqrt{1-\frac{\lambda ^2}{6}}$ & $0$ & $0$& $\frac{1}{2} \left(\lambda ^2-2\right)$ & $-1+\frac{\lambda^2}{3}$\\
\hline
$I$ & $\frac{\lambda }{\sqrt{6}}$ & $-\sqrt{1-\frac{\lambda ^2}{6}}$ & $0$ & $0$& $\frac{1}{2} \left(\lambda ^2-2\right)$ & $-1+\frac{\lambda^2}{3}$\\
\hline
$J$, in this case $\lambda=2$ & $\sqrt{\frac{2}{3}}$ & $\sqrt{\frac{1}{3}}$ & $0$ & $0$ & $1$ & $\frac{1}{3}$\\
\hline
$K$, in this case $\lambda=2$ & $\sqrt{\frac{2}{3}}$ & $-\sqrt{\frac{1}{3}}$ & $0$ & $0$ & $1$ & $\frac{1}{3}$\\
\hline
$L$ & $\frac{\sqrt{\frac{3}{2}}}{\lambda }$ & $\frac{\sqrt{\frac{3}{2}}}{\lambda }$ & $\chi, \chi -1\neq 0$ & $0$ & $\frac{1}{2}$ & $0$\\
\hline
$M$ & $\frac{\sqrt{\frac{3}{2}}}{\lambda }$ & $-\frac{\sqrt{\frac{3}{2}}}{\lambda }$ & $\chi, \chi -1\neq 0$ & $0$ & $\frac{1}{2}$ & $0$\\
\hline
$N,\alpha =0$ & $\delta$, $\delta\neq 0$ & $0$ & $1-\delta^2$ & $0$ & $2$ & $1$\\
[1ex] 
\hline 
\end{tabular}
\label{TABLE-I}
\end{table}

\begin{table}[H]
\caption{Eigenvalues and Stability of Eigenvalue at Corresponding Critical Point.} 
\centering 
\begin{tabular}{|c|c|c|} 
\hline 
Name of Critical Point & Corresponding Eigenvalues & Stability \\ [0.5ex] 
\hline 
$A$ & $\left\{\frac{3}{2},\frac{3}{2},-\frac{1}{2},0\right\}$ & Unstable\\
\hline
$B$ & $\left\{3,1,-6 \alpha ,\frac{1}{2} \left(6-\sqrt{6} \lambda \right)\right\}$ & Unstable\\
\hline
$C$ & $\left\{3,1,-6 \alpha ,\frac{1}{2} \left(6+\sqrt{6} \lambda \right)\right\}$ & Unstable\\
\hline
$D$ & $\{0,-3,-3,-2\}$ & Stable \\
\hline
$E$ &$\{0,-3,-3,-2\}$ & Stable \\
\hline
$F$ & $\left\{-\frac{1}{2},-3 \alpha ,\frac{3 \left(-\lambda ^2-\sqrt{24 \lambda ^2-7 \lambda ^4}\right)}{4 \lambda ^2},\frac{3 \left(\sqrt{24 \lambda ^2-7 \lambda ^4}-\lambda ^2\right)}{4 \lambda ^2}\right\}$ & \begin{tabular}{@{}c@{}} Stable for  $\alpha >0$\\ $\land \left(-2 \sqrt{\frac{6}{7}}\leq \lambda <-\sqrt{3}\lor \sqrt{3}<\lambda \leq 2 \sqrt{\frac{6}{7}}\right)$ \end{tabular}\\
\hline
$G$ & $\left\{-\frac{1}{2},-3 \alpha ,\frac{3 \left(-\lambda ^2-\sqrt{24 \lambda ^2-7 \lambda ^4}\right)}{4 \lambda ^2},\frac{3 \left(\sqrt{24 \lambda ^2-7 \lambda ^4}-\lambda ^2\right)}{4 \lambda ^2}\right\}$ & \begin{tabular}{@{}c@{}} Stable for  $\alpha >0$\\ $\land \left(-2 \sqrt{\frac{6}{7}}\leq \lambda <-\sqrt{3}\lor \sqrt{3}<\lambda \leq 2 \sqrt{\frac{6}{7}}\right)$ \end{tabular}\\
\hline
$H$ & $\left\{-\alpha  \lambda ^2,\frac{1}{2} \left(\lambda ^2-6\right),\frac{1}{2} \left(\lambda ^2-4\right),\lambda ^2-3\right\}$ & Stable for $\alpha >0\land \left(-\sqrt{3}<\lambda <0\lor 0<\lambda <\sqrt{3}\right)$\\
\hline
$I$ & $\left\{-\alpha  \lambda ^2,\frac{1}{2} \left(\lambda ^2-6\right),\frac{1}{2} \left(\lambda ^2-4\right),\lambda ^2-3\right\}$ & Stable for $\alpha >0\land \left(-\sqrt{3}<\lambda <0\lor 0<\lambda <\sqrt{3}\right)$\\
\hline
$J$ & $\{-1,1,0,-4 \alpha \}$ & Unstable \\
\hline
$K$ & $\{-1,1,0,-4 \alpha \}$ & Unstable \\
\hline
$L$ & \begin{tabular}{@{}c@{}}$\bigg\{0,-\frac{1}{2},\frac{3}{4} \left(\frac{\sqrt{-\lambda ^2 (\chi -1) \left(7 \lambda ^2 (\chi -1)+24\right)}}{\lambda ^2 (\chi -1)}-1\right),$ \\ $-\frac{3 \sqrt{-\lambda ^2 (\chi -1) \left(7 \lambda ^2 (\chi -1)+24\right)}}{4 \lambda ^2 (\chi -1)}-\frac{3}{4} \bigg\}$ \end{tabular} & Stable for $\lambda \in \mathbb{R}\land \lambda \neq 0\land \frac{7 \lambda ^2-24}{7 \lambda ^2}\leq \chi <\frac{\lambda ^2-3}{\lambda ^2}$ \\
\hline
$M$ & \begin{tabular}{@{}c@{}}$\bigg\{0,-\frac{1}{2},\frac{3}{4} \left(\frac{\sqrt{-\lambda ^2 (\chi -1) \left(7 \lambda ^2 (\chi -1)+24\right)}}{\lambda ^2 (\chi -1)}-1\right),$ \\ $-\frac{3 \sqrt{-\lambda ^2 (\chi -1) \left(7 \lambda ^2 (\chi -1)+24\right)}}{4 \lambda ^2 (\chi -1)}-\frac{3}{4} \bigg\}$ \end{tabular} & Stable for $\lambda \in \mathbb{R}\land \lambda \neq 0\land \frac{7 \lambda ^2-24}{7 \lambda ^2}\leq \chi <\frac{\lambda ^2-3}{\lambda ^2}$ \\
\hline
$N$ &  $\left\{0,1,3,\frac{1}{2} \left(6-\sqrt{6} \delta  \lambda \right)\right\}$ &  Unstable\\
[1ex] 
\hline 
\end{tabular}
\label{TABLE-II}
\end{table}
Note : During this study, for non-hyperbolic critical points, we have dimension of set of eigenvalues is one which is equal to the number of vanishing eigenvalues, hence set of eigenvalues is normally hyperbolic and critical point corresponding to set is stable but can not be a global attractor \cite{AAKoley1999}.\\

For any real value of $\lambda$, the critical points $H$ and $I$ represent dark energy dominated era  and this point shows stable behaviour for the parameters obey the value $\alpha >0$ and $\left(-\sqrt{3}<\lambda <0\lor 0<\lambda <\sqrt{3}\right)$. During the study of stability conditions, for critical points $F$, $G$, $H$ and $I$, we get condition on $\alpha$, $\alpha>0$. This implies stability of critical points can be assessed for positive value of $\alpha$ for Model-I. All the eigenvalues at critical points $L$ and $M$ are less than zero for $\lambda \in \mathbb{R}\land \lambda \neq 0\land \frac{7 \lambda ^2-24}{7 \lambda ^2}\leq \chi <\frac{\lambda ^2-3}{\lambda ^2}$, hence show stable behaviour at these parametric values.\\

\begin{figure}[H]
    \centering
    \includegraphics[width=74mm]{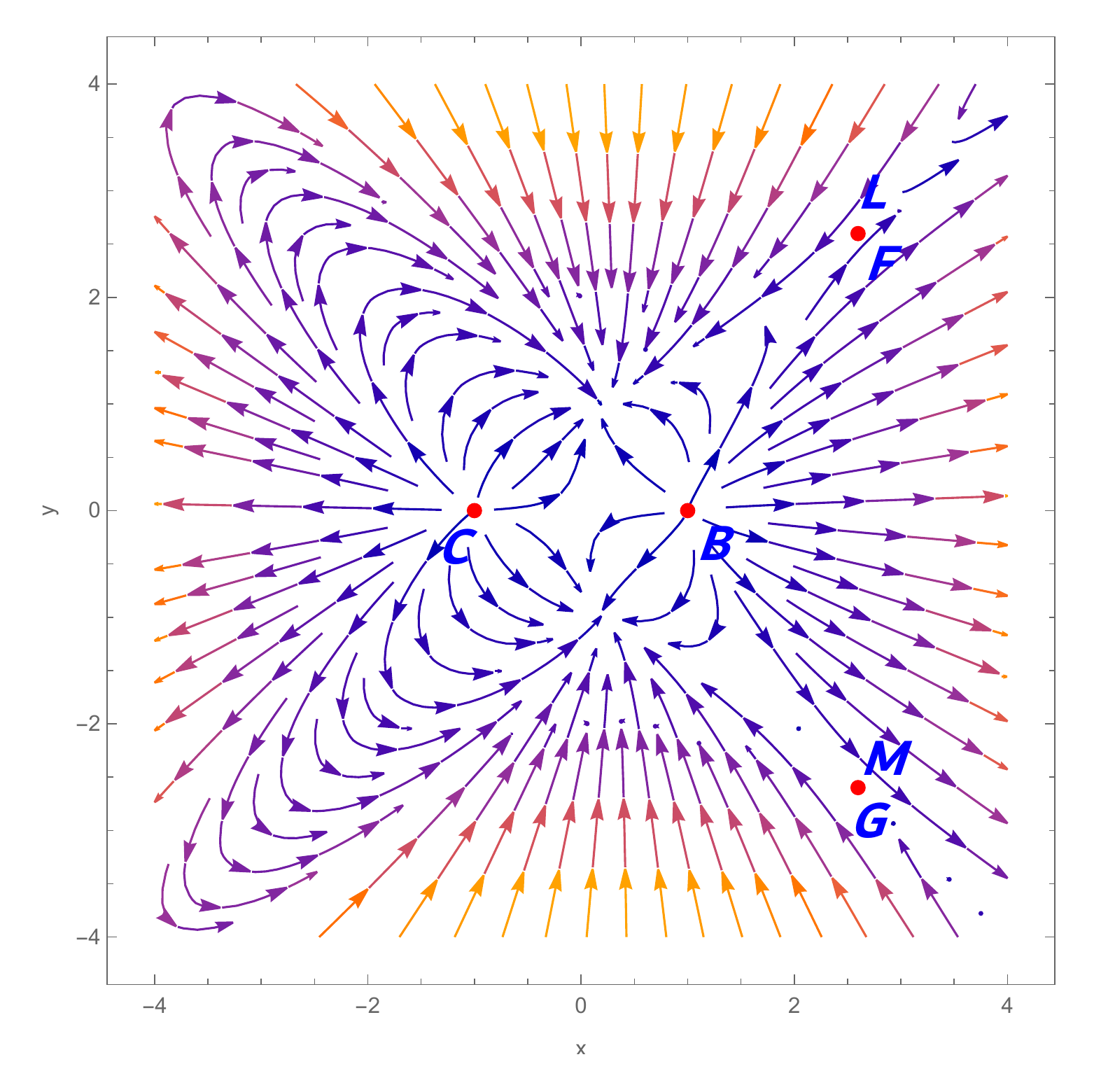}
    \includegraphics[width=74mm]{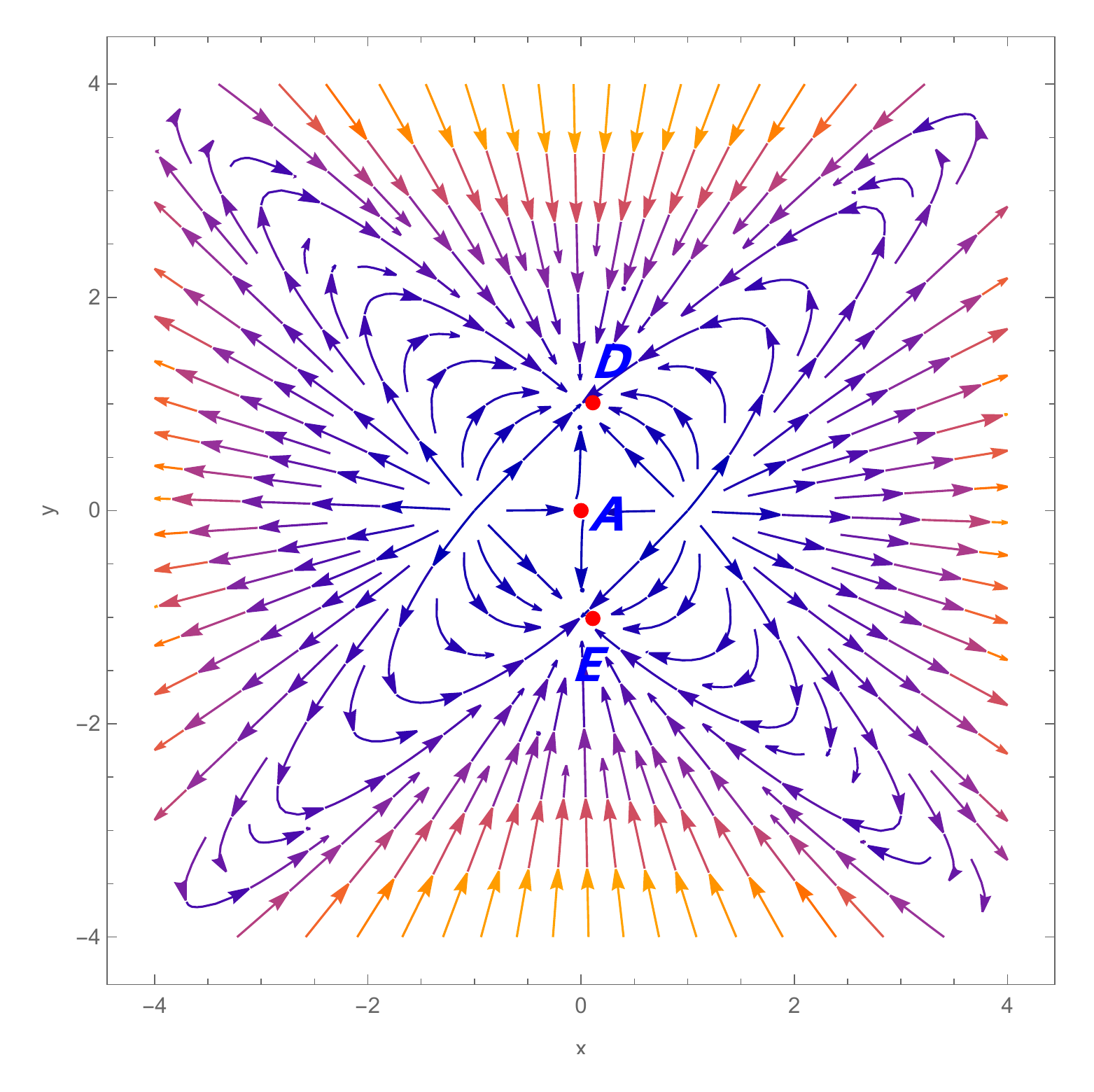}
    \includegraphics[width=74mm]{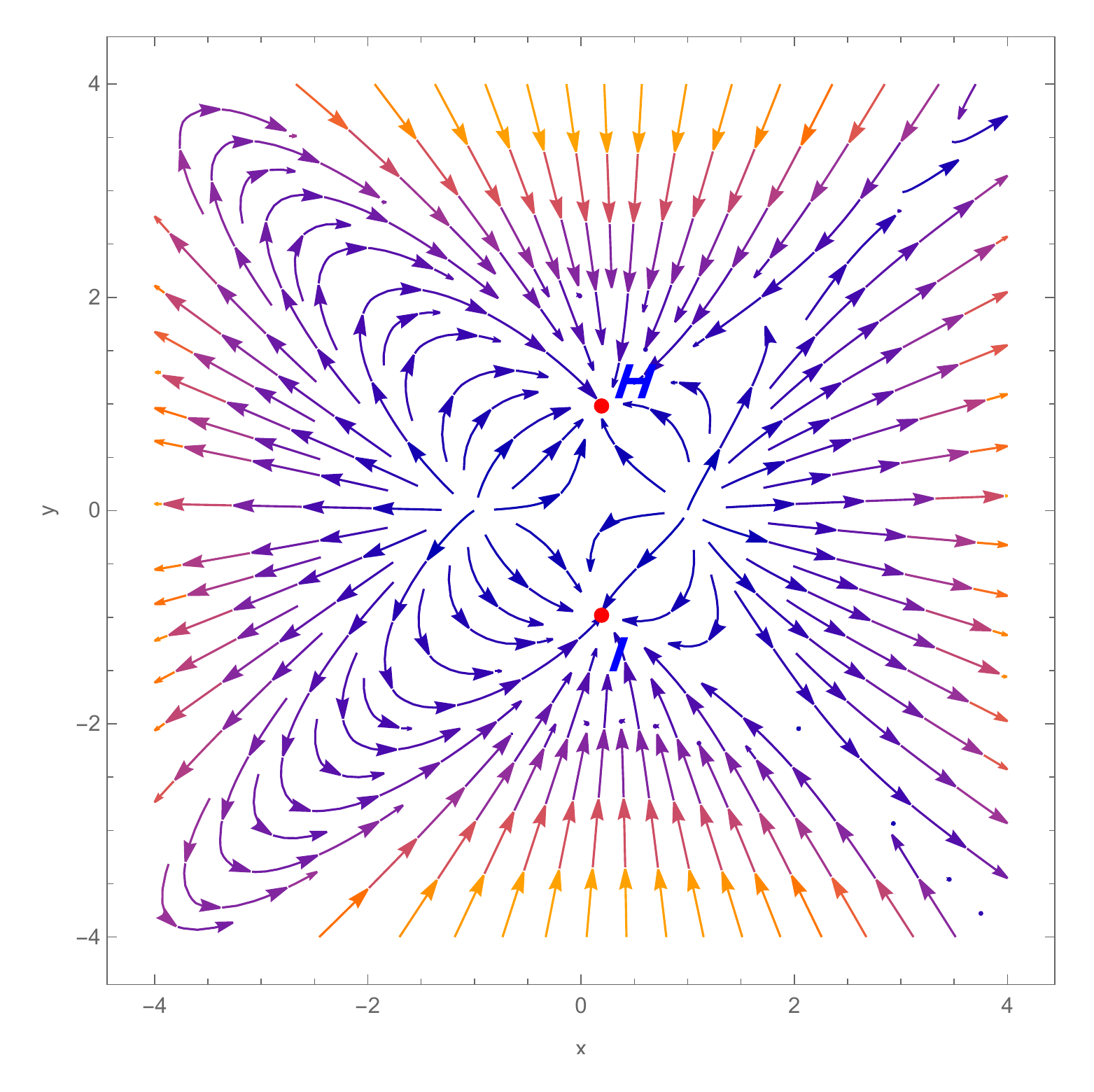}
    \includegraphics[width=74mm]{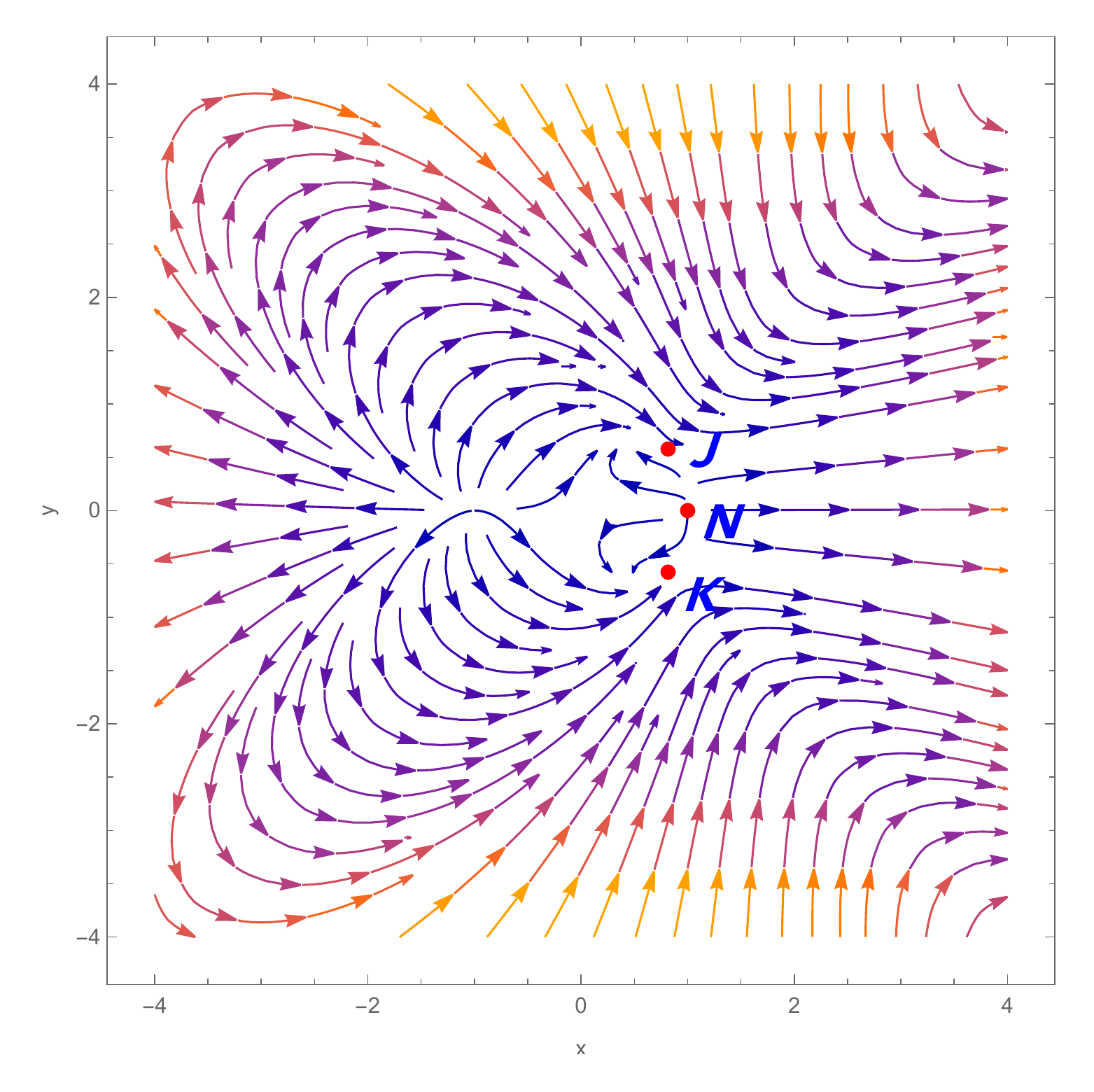}
    \caption{Phase portrait for above dynamical system, the upper left plot is for $u=0$, $\rho=0$, $\lambda=\sqrt{\frac{2}{9}}$. The upper right plot having parameter values $u=0$, $\rho=0$, $\zeta =\frac{1}{9}$, $\tau=1$, $\alpha=1.1$, lower left phase portrait is for $u=0$, $\rho=0$, $\lambda=\sqrt{\frac{2}{9}}$, lower right phase portrait is for $u=0$, $\rho=0$, $\delta=1$.} \label{Fig1}
\end{figure}

We have analyse the phase space for all the critical points by fixing some parameters to a appropriate value. The phase space plots for dynamical system described in Eqs.~(\ref{eq:dx_dN_model_1}--\ref{eq:dl_dN_model_1}) are presented in Fig.\ref{Fig1}. From phase space diagram for Model-I, we can conclude that the phase space trajectories for critical points $L$, $F$, $M$, $G$, $A$, $J$, $N$ and $K$ are moving away from the critical point hence confirm saddle point behaviour. Critical points $F$ and $G$ show stability  for $-2 \sqrt{\frac{6}{7}}\leq \lambda <-\sqrt{3}$ or $\sqrt{3}<\lambda \leq 2 \sqrt{\frac{6}{7}}$ but we choose $\lambda=\sqrt{\frac{2}{9}}$ hence these are showing saddle points nature in phase diagram. For the critical points $L$ and $M$ the stability conditions depend on $\chi$ which represents $u$ co-ordinate and the phase plots are analysed in xy-axis plane hence critical points $L$ and $M$ may show saddle point behaviour. From the phase diagram, it can observe that at critical points $H$, $I$, $D$ and $E$ trajectories are attracted towards the critical point hence describe attracting behaviour of these critical points, also these critical points can explain dark energy dominated universe. Although eigenvalues at $B$ and $C$ contains both positive and negative signature, from the phase portrait critical points $B$ and $C$ represent an unstable node leading to the positive eigenvalues only (due to consideration of $u=0$, $\rho=$ may the negative eigenvalues are not contributing in the phase space plot). We shall present below two examples of this model for $\alpha=1$ and $\alpha=2$.

\subsection{Case A: \texorpdfstring{$\alpha=1$}{}}\label{sec:model_1_case_1}

In this case we have consider $\alpha=1$, in the action equation Eq.~(\ref{eq:action_model_1}). The evolution equations can be obtained by limiting Eqs.~(\ref{eq:44}-\ref{eq:46}) from general $\alpha$ to $\alpha=1$. To study cosmic evolution through dynamical system analysis approach, the set of dimensionless variables associated with the above set of cosmological equations can be defined as follow \cite{Gonzalez-Espinoza:2020jss}
\begin{equation}\label{eq:56}
    x=\dfrac{\kappa\dot{\phi}}{\sqrt{6}H}\,,\quad y=\frac{\kappa\sqrt{V}}{\sqrt{3}H}\,, u=3\dot{\phi}^2 \kappa^2\,,\quad \rho= \frac{\kappa\sqrt{\rho_{\rm r}}}{\sqrt{3}H}\,,\quad \lambda=\frac{-V^{'}(\phi)}{\kappa V(\phi)}\,,\quad \Gamma=\dfrac{V(\phi)V^{''}(\phi)}{V^{'}(\phi)^{2}}\,.
\end{equation}
In this case the dimensionless variables are selected such that they can linked with each other in following constraint equation form, note that in this expression u is considered as it is (without any scalar multiplier).
\begin{align}\label{eq:57}
    x^{2}+y^{2}+u+\Omega_{m}+\rho^{2}=1
\end{align}
Using these dimensionless variables the cosmological expressions in this case can be written in terms of autonomous dynamical system as follow, 
\begin{align}
    \dfrac{dx}{dN} &= \frac{3 x y^2 \left(\sqrt{6} \lambda  (u-1) x+3 u+3 x^2\right)-3 x \left(u^2+u \left(\rho ^2+8 x^2+1\right)+x^2 \left(\rho ^2+3 x^2-3\right)\right)}{2 (u-3) x^2-2 u (u+1)}\,,\label{eq:dx_dN_model_1_alpha_1}\\
    \dfrac{dy}{dN} &= -y \left(\frac{-3 u^2-u \left(\rho ^2+12 x^2+3\right)+u y^2 \left(2 \sqrt{6} \lambda x+3\right)-3 x^2 \left(\rho ^2+3 x^2-3 y^2+3\right)}{2 u (u+1)-2 (u-3) x^2}+\sqrt{\frac{3}{2}} \lambda  x\right)\,,\\
    \dfrac{du}{dN} &= \frac{2 u \left(\rho ^2 u+(6 u-9) x^2\right)-u y^2 \left(\sqrt{6} \lambda  (u-3) x+6 u\right)}{u (u+1)-(u-3) x^2}\,,\\
    \frac{d\rho}{dN} &=\frac{\rho \left(-u^2+u \left(\rho ^2+16 x^2-y^2 \left(2 \sqrt{6} \lambda  x+3\right)-1\right)+3 x^2 \left(\rho ^2+3 x^2-3 y^2-1\right)\right)}{2 u (u+1)-2 (u-3) x^2}\,,\\
    \dfrac{d\lambda}{dN} &=-\sqrt{6}(\Gamma-1)\lambda^{2}x\label{eq:dl_dN_model_1_alpha_1}\,.
\end{align}

\begin{table}[H]
\caption{ Critical Points for Dynamical System Corresponding to Model-I, $\alpha=1$.} 
\centering 
\begin{tabular}{|c|c|c|c|c|c|c|} 
\hline\hline 
Name of Critical Point & $x_{c}$ & $y_{c}$ & $u_{c}$ & $\rho_{c}$ & Deceleration Parameter ($q$) & $\omega_{tot}$\\ [0.5ex] 
\hline\hline 
$A$ & $\tau, \tau ^2+\tau \neq 0$ & $0$ & $0$ & $0$ & $\frac{1}{2}$ & $0$\\
\hline
$B$ & $1$ & $0$ & $0$ & $0$ & $2$ & $1$\\
\hline
$C$ & $-1$ & $0$ & $0$ & $0$ & $2$ & $1$\\
\hline
$D$, in this case  $\lambda=0$ & $\zeta, \zeta \neq 0$ & $\sqrt{2 \zeta ^2+1}$ & $-3 \zeta ^2$ & $0$ & $-1$ & $-1$\\
\hline
$E$, in this case  $\lambda=0$ & $\zeta, \zeta \neq 0$ & $-\sqrt{2 \zeta ^2+1}$ & $-3 \zeta ^2$ & $0$ & $-1$ & $-1$\\
\hline
$F$ & $\frac{\sqrt{\frac{3}{2}}}{\lambda }$ & $\sqrt{\frac{3}{2}} \sqrt{\frac{1}{\lambda ^2}}$ & $0$ & $0$ & $\frac{1}{2}$ & $0$\\
\hline
$G$ & $\frac{\sqrt{\frac{3}{2}}}{\lambda }$ & $-\sqrt{\frac{3}{2}} \sqrt{\frac{1}{\lambda ^2}}$ & $0$ & $0$ & $\frac{1}{2}$ & $0$\\
\hline
$H$ & $\frac{\lambda }{\sqrt{6}}$ & $\sqrt{1-\frac{\lambda ^2}{6}}$ & $0$ & $0$& $\frac{1}{2} \left(\lambda ^2-2\right)$ & $-1+\frac{\lambda^2}{3}$\\
\hline
$I$ & $\frac{\lambda }{\sqrt{6}}$ & $-\sqrt{1-\frac{\lambda ^2}{6}}$ & $0$ & $0$& $\frac{1}{2} \left(\lambda ^2-2\right)$ & $-1+\frac{\lambda^2}{3}$\\
\hline
$J$, in this case  $\lambda=2$ & $\sqrt{\frac{2}{3}}$ & $\frac{1}{\sqrt{3}}$ & $0$ & $0$ & $1$ & $\frac{1}{3}$\\
\hline
$K$, in this case  $\lambda=2$ & $\sqrt{\frac{2}{3}}$ & $-\frac{1}{\sqrt{3}}$ & $0$ & $0$ & $1$ & $\frac{1}{3}$\\
[1ex] 
\hline 
\end{tabular}
\label{TABLE-III}
\end{table}

From Table~\ref{TABLE-III} observations, we can conclude that, at critical points $A$, $F$, $G$ the value of deceleration parameter is $\frac{1}{2}$ with $\omega_{tot}=0$. These critical points not represent the accelerating universe, but the cold dark matter-dominated era. The critical points $B$ and $C$ behave as stiff matter showing $\omega_{tot}=1$. The critical points $J$ and $K$ represent the radiation-dominated solutions. The critical points $D$, $E$, $H$ and $I$ show the value of the deceleration parameter negative, these critical points can represent the accelerating behavior of the universe. The critical points $D$ and $E$ are the de Sitter solutions with the value of $\omega_{tot}=-1$ and can be obtained only at particular value of $\lambda=0$.  At the critical points $H$ and $I$ deceleration parameter shows negative value for $-\sqrt{2}$ $<$ $\lambda$ $<$ $\sqrt{2}$, these points explains dark energy dominated universe. To analyse the stability behavior of all these critical points the eigenvalues and stability conditions are presented in Table~\ref{TABLE-IV} \\    

From Table~\ref{TABLE-IV}, we can conclude that, the critical points $D$, $E$ shows stable behaviour and these critical points can attract the universe at late time. At the critical points $H$ and $I$ eigenvalues show stability at $-\sqrt{3}<\lambda <0$ or $0<\lambda <\sqrt{3}$ these points explain dark energy domination at late time. The critical points $A$ to $C$ are saddle points and hence unstable for all values of $\lambda$. However these points cannot describe the current accelerated expansion of the universe. The radiation dominated representation belong to the critical points $J$ and $K$, these points are also saddle points for any value of $\lambda$ and hence unstable in nature. Although critical points $F$ and $G$ represents cold dark matter dominated universe these critical point obey stability at $-2 \sqrt{\frac{6}{7}}\leq \lambda <-\sqrt{3}$ or $\sqrt{3}<\lambda \leq 2 \sqrt{\frac{6}{7}}$.\\

\begin{table}[H]
\caption{ Eigenvalues and Stability of Corresponding Critical Point.} 
\centering 
\begin{tabular}{|c|c|c|} 
\hline 
Name of Critical Point & Corresponding Eigenvalues & Stability \\ [0.5ex] 
\hline 
$A$ & $\left\{\frac{3}{2},\frac{3}{2},-\frac{1}{2},0\right\}$ & Unstable\\
\hline
$B$ & $\left\{-6,3,1,\frac{1}{2} \left(6-\sqrt{6} \lambda \right)\right\}$ & Unstable\\
\hline
$C$ & $\left\{-6,3,1,\frac{1}{2} \left(6+\sqrt{6} \lambda \right)\right\}$ & Unstable\\
\hline
$D$ & $\{0,-3,-3,-2\}$ & Stable\\
\hline
$E$ &$\{0,-3,-3,-2\}$  & Stable\\
\hline
$F$ & $\left\{-3,-\frac{1}{2},\frac{3 \left(-\lambda ^4-\sqrt{24 \lambda ^6-7 \lambda ^8}\right)}{4 \lambda ^4},\frac{3 \left(\sqrt{24 \lambda ^6-7 \lambda ^8}-\lambda ^4\right)}{4 \lambda ^4}\right\}$ & Stable for $-2 \sqrt{\frac{6}{7}}\leq \lambda <-\sqrt{3}\lor \sqrt{3}<\lambda \leq 2 \sqrt{\frac{6}{7}}$\\
\hline
$G$ & $\left\{-3,-\frac{1}{2},\frac{3 \left(-\lambda ^4-\sqrt{24 \lambda ^6-7 \lambda ^8}\right)}{4 \lambda ^4},\frac{3 \left(\sqrt{24 \lambda ^6-7 \lambda ^8}-\lambda ^4\right)}{4 \lambda ^4}\right\}$ & Stable for $-2 \sqrt{\frac{6}{7}}\leq \lambda <-\sqrt{3}\lor \sqrt{3}<\lambda \leq 2 \sqrt{\frac{6}{7}}$\\
\hline
$H$ & $\left\{-\lambda ^2,\frac{1}{2} \left(\lambda ^2-6\right),\frac{1}{2} \left(\lambda ^2-4\right),\lambda ^2-3\right\}$ & Stable for $-\sqrt{3}<\lambda <0\lor 0<\lambda <\sqrt{3}$\\
\hline
$I$ & $\left\{-\lambda ^2,\frac{1}{2} \left(\lambda ^2-6\right),\frac{1}{2} \left(\lambda ^2-4\right),\lambda ^2-3\right\}$ & Stable for $-\sqrt{3}<\lambda <0\lor 0<\lambda <\sqrt{3}$\\
\hline
$J$ & $\{-4,-1,1,0\} $& Unstable\\
\hline
$K$ &  $\{-4,-1,1,0\}$& Unstable \\
[1ex] 
\hline 
\end{tabular}
\label{TABLE-IV}
\end{table}

\begin{figure}[H]
    \centering
    \includegraphics[width=75mm]{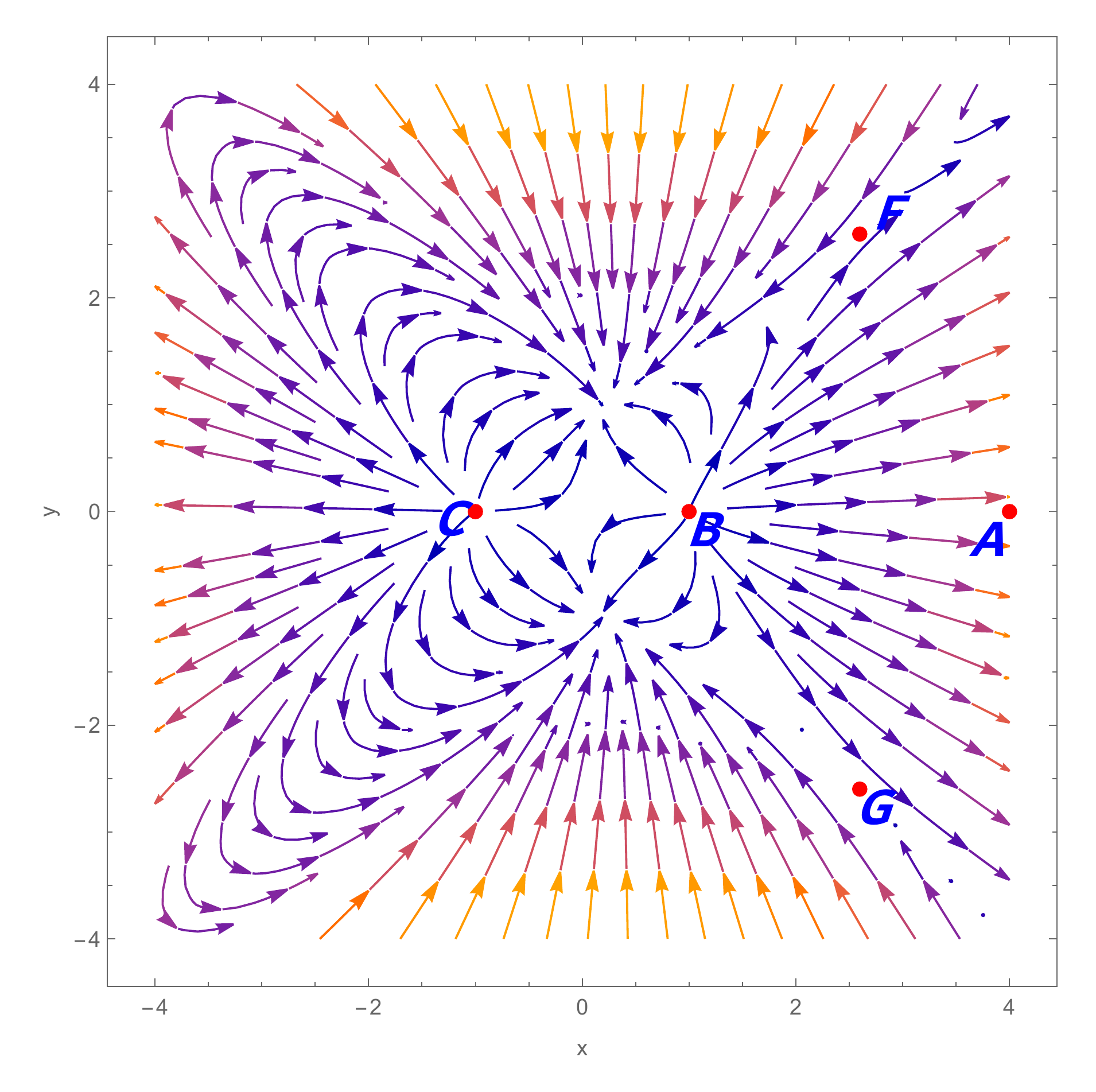}
    \includegraphics[width=75mm]{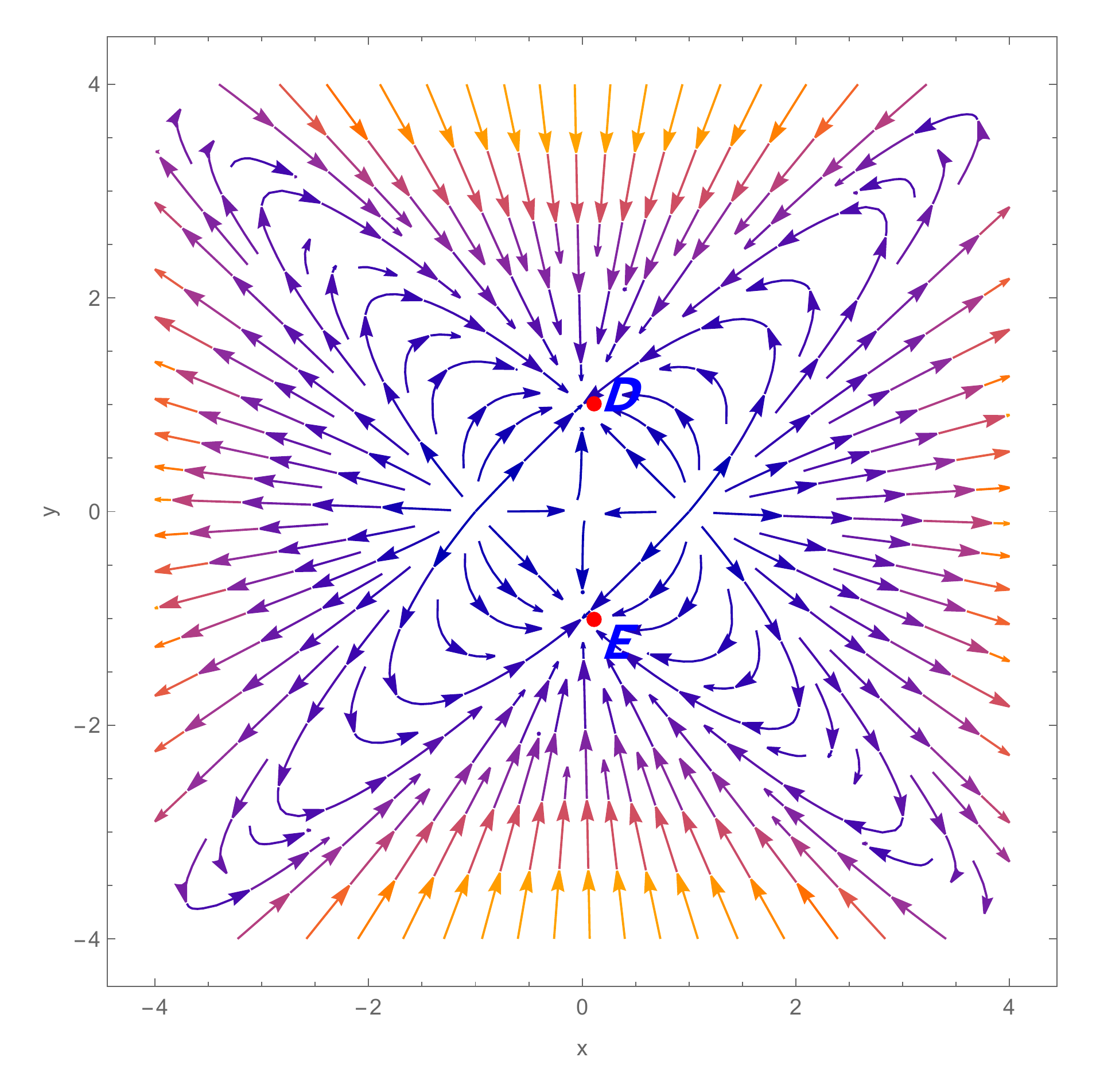}
    \includegraphics[width=75mm]{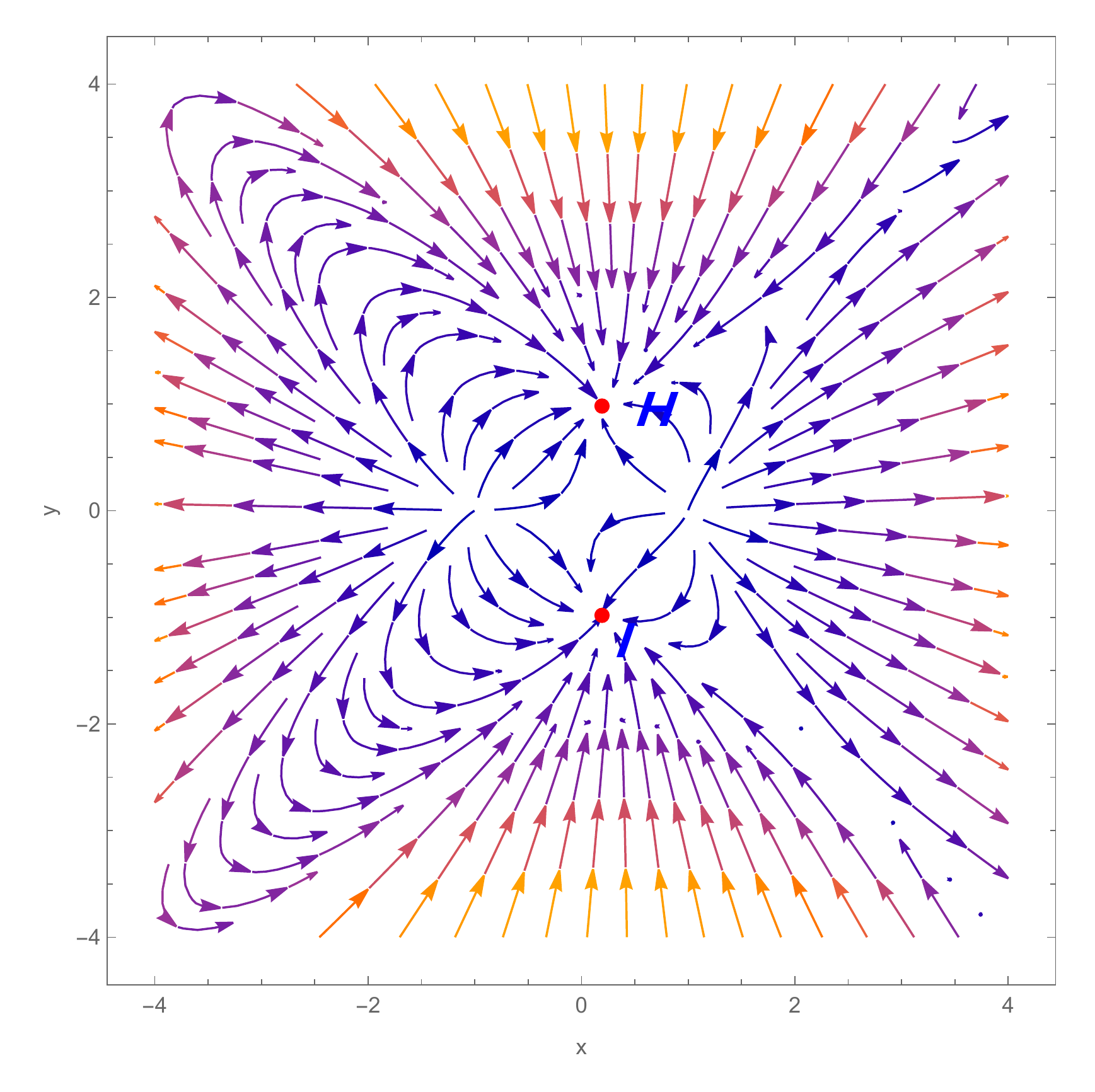}
    \includegraphics[width=75mm]{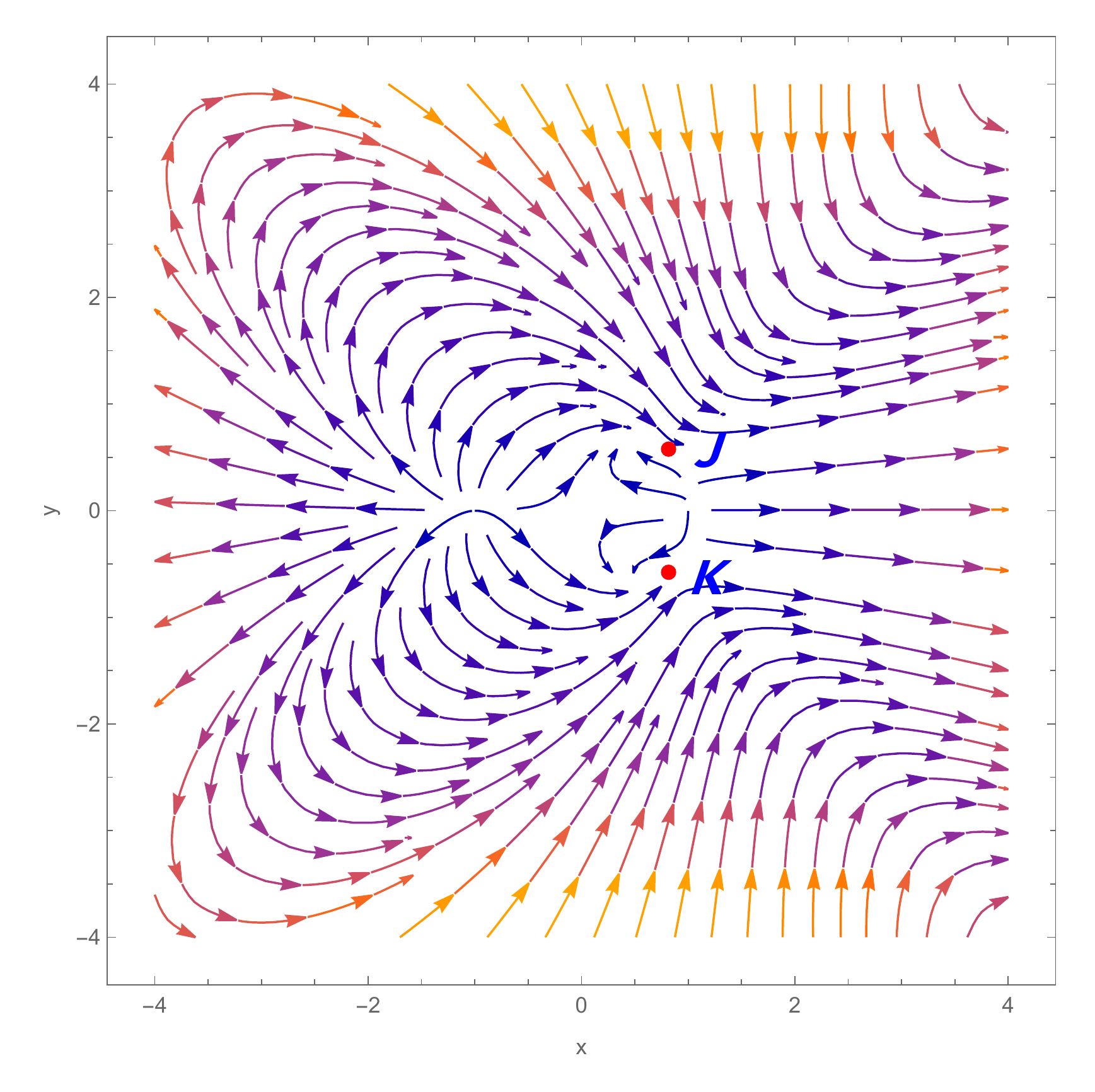}
    \caption{Phase portrait for dynamical system presented in Eqs(\ref{eq:dx_dN_model_1_alpha_1}-\ref{eq:dl_dN_model_1_alpha_1}), the upper left plot is for the parametric value $u=0$, $\rho=0$, $\tau=1$ and $\lambda=\sqrt{\frac{2}{9}}$. The upper right plot is for the parameteric values $u=0$, $\rho=0$ and $\zeta =\frac{1}{9}$. The lower left phase portrait is for $u=0$, $\rho=0$ and $\lambda=\sqrt{\frac{2}{9}}$ and lower right phase portrait is for $u=0$ and $\rho=0$.} \label{Fig2}
\end{figure}

From the Fig.~\ref{Fig2} observation, it can be conclude that critical points $H$, $I$ are attractors and can be analysed by setting $\lambda=\sqrt{\frac{2}{9}}$ which belongs to the stability range of $\lambda$ (that is $-\sqrt{3}<\lambda <0\lor 0<\lambda <\sqrt{3}$ ). Since $F$ and $G$ show stability  for $-2 \sqrt{\frac{6}{7}}\leq \lambda <-\sqrt{3}$ or $\sqrt{3}<\lambda \leq 2 \sqrt{\frac{6}{7}}$ and here, we have analyse phase plots at $\lambda=\sqrt{\frac{2}{9}}$ which does not belong to stability range of $\lambda$, the critical points $F$ and $G$ show saddle point behaviour. The critical points $A$, $J$, $K$ having eigenvalues with both positive and negative sign and hence these are saddle points. The particular parametric value $\lambda=2$ helps us to analyse the stability at $J$ and $K$ also from phase space analysis it can be observe that phase space trajectories are moving away at these critical points and hence show the saddle point behaviour. The de Sitter solution is represented by critical points $D$ and $E$, these critical points are exists only for parametric value $\lambda=0$ and from the phase space analysis it can conclude that these points represent attracting solution. The critical points $A$ to $K$ can obtained in particular case $\alpha=1$ similar to general $\alpha$, but the critical points  $L$, $M$ and $N$ discussed in general $\alpha$ case are not contributing in the case $\alpha=1$.

\subsection{Case B: \texorpdfstring{$\alpha=2$}{}}\label{sec:model_1_case_2}

In this case we have analyse cosmological implications by using dynamical system approach for particular value $\alpha=2$ in action Eq.~(\ref{eq:action_model_1}). In this case, the set of dimensionless variables to obtain autonomous dynamical system can be defined as follow,
\begin{equation}\label{eq:63}
    x=\dfrac{\kappa\dot{\phi}}{\sqrt{6}H}\,,\quad y=\frac{\kappa\sqrt{V}}{\sqrt{3}H}\,,\quad u=\frac{5}{2}\kappa^2\dot{\phi}^{4}\,,\quad \rho= \frac{\kappa\sqrt{\rho_{\rm r}}}{\sqrt{3}H}\,,\quad \lambda=\frac{-V^{'}(\phi)}{\kappa V(\phi)}\,,\quad \Gamma=\dfrac{V(\phi)V^{''}(\phi)}{V^{'}(\phi)^{2}}\,.
\end{equation}

One can observe that the dimensionless variables defined in the study of these scalar tensor models are not same in \cite{Gonzalez-Espinoza:2020jss}, but these types of variables are usually used to obtain viable critical points in cosmology.. The dimensionless variables defined in Eq.~(\ref{eq:63}) also satisfy  the constraint equation Eq.~(\ref{eq:57}) and the dynamical system in this case can be defined as follow,
\begin{align}
    \frac{dx}{dN} &= \frac{x \left(-5 \rho ^2 \left(2 u+x^2\right)+5 y^2 \left(\sqrt{6} \lambda  (u-1) x+6 u+3 x^2\right)+3 (5-19 u) x^2-6 u (u+3)-15 x^4\right)}{2 (u-5) x^2-4 u (u+3)}\,,\label{eq:dx_dN_model_1_alpha_2}\\
    \frac{dy}{dN} &= -y \left(\frac{-6 u^2-3 u \left(2 \rho ^2+13 x^2+6\right)+2 u y^2 \left(2 \sqrt{6} \lambda  x+9\right)-5 x^2 \left(\rho ^2+3 x^2-3 y^2+3\right)}{4 u (u+3)-2 (u-5) x^2}+\sqrt{\frac{3}{2}} \lambda  x\right)\,,\\
    \frac{du}{dN} &= \frac{4 u \left(2 \rho ^2 u+3 (3 u-5) x^2\right)-2 u y^2 \left(\sqrt{6} \lambda  (u-5) x+12 u\right)}{2 u (u+3)-(u-5) x^2}\,,\\
    \frac{d\rho}{dN} &= \frac{\rho  \left(-2 u^2+u \left(6 \rho ^2+43 x^2-2 y^2 \left(2 \sqrt{6} \lambda  x+9\right)-6\right)+5 x^2 \left(\rho ^2+3 x^2-3 y^2-1\right)\right)}{4 u (u+3)-2 (u-5) x^2}\,,\\
    \frac{d\lambda}{dN} &= -\sqrt{6}(\Gamma-1)\lambda^{2}x \,.\label{eq:dl_dN_model_1_alpha_2}
\end{align}

To study cosmological implications, the study of dynamical system at critical points obtained from cosmological evolution equations is very important. Critical points for autonomous dynamical system presented in Eqs.~(\ref{eq:dx_dN_model_1_alpha_2}-\ref{eq:dl_dN_model_1_alpha_2}) are presented in Table~\ref{TABLE-V}. From the table observations it can conclude that, although critical points have different co-ordinates than the critical points in the Table~\ref{TABLE-III} (for $\alpha=1$) case, but the cosmological implications are almost similar in nature. When we observe critical points in Table~\ref{TABLE-III} and Table~\ref{TABLE-V}, we can easily see that the deceleration parameter ($q$) and $\omega_{tot}$ for critical points with the same name are same. From Table~\ref{TABLE-V} it can be clearly observe that critical points $D$, $E$, $H$ and $I$ can show deceleration parameter value in negative range and hence these critical points can deal with the dark energy dominated era. For critical points $H$ and $I$, we get accelerating behaviour for $-\sqrt{2}$ $<$ $\lambda$ $<$ $\sqrt{2}$ and critical points $D$ and $E$ are defined only for parametric value $\lambda=0$ and represent de Sitter solution for the system. The other critical points do not gives negative value for deceleration parameter and hence defines non-accelerating phase of evolution. The critical points $A$, $F$ and $G$ represents cold dark matter dominated era with $\omega_{tot}=0$. In this case (for $\alpha=2$) also we are getting critical points $B$ and $C$ representing stiff matter. The critical points $J$ and $K$ are defined for $\lambda=2$ and deliver value for $\omega_{tot}=\frac{1}{3}$, hence represent radiation-dominated era.

\begin{table}[H]
\caption{Critical Points for Dynamical System Corresponding to Model-I, $\alpha=2$.} 
\centering 
\begin{tabular}{|c|c|c|c|c|c|c|} 
\hline\hline 
Name of Critical Point & $x_{c}$ & $y_{c}$ & $u_{c}$ & $\rho_{c}$ & Deceleration Parameter ($q$) & $\omega_{tot}$\\ [0.5ex] 
\hline\hline 
$A$ & $\tau.\tau ^2+3\tau \neq 0$ & $0$ & $0$ & $0$ & $\frac{1}{2}$ & $0$\\
\hline
$B$ & $1$ & $0$ & $0$ & $0$ & $2$ & $1$\\
\hline
$C$ & $-1$ & $0$ & $0$ & $0$ & $2$ & $1$\\
\hline
$D$, in this case $\lambda=0$ & $\zeta, 3 \zeta ^3-2 \zeta \neq 0$ & $\sqrt{\frac{3 \zeta ^2}{2}+1}$ & $-\frac{1}{2} \left(5 \zeta ^2\right)$ & $0$ & $-1$ & $-1$\\
\hline
$E$, in this case $\lambda=0$ &  $\zeta, 3 \zeta ^3-2 \zeta \neq 0$ & $-\sqrt{\frac{3 \zeta ^2}{2}+1}$ & $-\frac{1}{2} \left(5 \zeta ^2\right)$ & $0$ & $-1$ & $-1$\\
\hline
$F$ & $\frac{\sqrt{\frac{3}{2}}}{\lambda }$ & $\sqrt{\frac{3}{2}} \sqrt{\frac{1}{\lambda ^2}}$ & $0$ & $0$ & $\frac{1}{2}$ & $0$\\
\hline
$G$ & $\frac{\sqrt{\frac{3}{2}}}{\lambda }$ & $-\sqrt{\frac{3}{2}} \sqrt{\frac{1}{\lambda ^2}}$ & $0$ & $0$ & $\frac{1}{2}$ & $0$\\
\hline
$H$ & $\frac{\lambda }{\sqrt{6}}$ & $\sqrt{1-\frac{\lambda ^2}{6}}$ & $0$ & $0$& $\frac{1}{2} \left(\lambda ^2-2\right)$ & $-1+\frac{\lambda^2}{3}$\\
\hline
$I$ & $\frac{\lambda }{\sqrt{6}}$ & $-\sqrt{1-\frac{\lambda ^2}{6}}$ & $0$ & $0$& $\frac{1}{2} \left(\lambda ^2-2\right)$ & $-1+\frac{\lambda^2}{3}$\\
\hline
$J$, in this case $\lambda=2$ & $\sqrt{\frac{2}{3}}$ & $\sqrt{\frac{1}{3}}$ & $0$ & $0$ & $1$ & $\frac{1}{3}$\\
\hline
$K$, in this case $\lambda=2$ & $\sqrt{\frac{2}{3}}$ & $-\sqrt{\frac{1}{3}}$ & $0$ & $0$ & $1$ & $\frac{1}{3}$\\
[1ex] 
\hline 
\end{tabular}
\label{TABLE-V}
\end{table}

The stability conditions for critical points corresponding to dynamical system in Eqs.~(\ref{eq:dx_dN_model_1_alpha_2}-\ref{eq:dl_dN_model_1_alpha_2}) are presented in Table~\ref{TABLE-VI}. The signature of eigenvalues confirms the stability of corresponding critical point.  From the table observations, we can conclude that critical points $A$, $B$ and $C$ are unstable for all values of $\lambda$ and are saddle points. At the critical points $H$ and $I$ eigenvalues show stability at $-\sqrt{3}<\lambda <0$ or $0<\lambda <\sqrt{3}$ and these points explain dark energy domination at late time. Critical points $D$, $E$ show stable behaviour and in the further analysis it is noticed that they can attract the universe at late time. The critical points $J$ and $K$ represent radiation dominated and from signature of eigenvalues these points are saddle points for any value of $\lambda$, hence unstable in nature. Critical points $F$ and $G$ represents cold dark matter dominated universe and these critical point obey stability at $-2 \sqrt{\frac{6}{7}}\leq \lambda <-\sqrt{3}$ or $\sqrt{3}<\lambda \leq 2 \sqrt{\frac{6}{7}}$. From Table~\ref{TABLE-IV} and Table~\ref{TABLE-VI} it can observe that, the  stability conditions for the case ($\alpha=1$) and ($\alpha=2$) are showing similar nature to explain the evolution of the universe.\\

\begin{table}[H]
\caption{Eigenvalues and Stability of Eigenvalue at Corresponding Critical Point.} 
\centering 
\begin{tabular}{|c|c|c|} 
\hline 
Name of Critical Point & Corresponding Eigenvalues & Stability \\ [0.5ex] 
\hline 
$A$ & $\left\{\frac{3}{2},\frac{3}{2},-\frac{1}{2},0\right\}$ & Unstable\\
\hline
$B$ & $\left\{-12,3,1,\frac{1}{2} \left(6-\sqrt{6} \lambda \right)\right\}$ & Unstable\\
\hline
$C$ & $\left\{-12,3,1,\frac{1}{2} \left(\sqrt{6} \lambda +6\right)\right\}$ & Unstable\\
\hline
$D$ & $\{0,-3,-3,-2\}$ & Stable \\
\hline
$E$ &$\{0,-3,-3,-2\}$ & Stable \\
\hline
$F$ & $\left\{-6,-\frac{1}{2},\frac{3 \left(-\lambda ^2-\sqrt{24 \lambda ^2-7 \lambda ^4}\right)}{4 \lambda ^2},\frac{3 \left(\sqrt{24 \lambda ^2-7 \lambda ^4}-\lambda ^2\right)}{4 \lambda ^2}\right\}$ & Stable for $-2 \sqrt{\frac{6}{7}}\leq \lambda <-\sqrt{3}\lor \sqrt{3}<\lambda \leq 2 \sqrt{\frac{6}{7}}$\\
\hline
$G$ & $\left\{-6,-\frac{1}{2},\frac{3 \left(-\lambda ^2-\sqrt{24 \lambda ^2-7 \lambda ^4}\right)}{4 \lambda ^2},\frac{3 \left(\sqrt{24 \lambda ^2-7 \lambda ^4}-\lambda ^2\right)}{4 \lambda ^2}\right\}$ & Stable for $-2 \sqrt{\frac{6}{7}}\leq \lambda <-\sqrt{3}\lor \sqrt{3}<\lambda \leq 2 \sqrt{\frac{6}{7}}$\\
\hline
$H$ & $\left\{-2 \lambda ^2,\frac{1}{2} \left(\lambda ^2-6\right),\frac{1}{2} \left(\lambda ^2-4\right),\lambda ^2-3\right\}$ & Stable for $-\sqrt{3}<\lambda <0\lor 0<\lambda <\sqrt{3}$\\
\hline
$I$ & $\left\{-2 \lambda ^2,\frac{1}{2} \left(\lambda ^2-6\right),\frac{1}{2} \left(\lambda ^2-4\right),\lambda ^2-3\right\}$ & Stable for $-\sqrt{3}<\lambda <0\lor 0<\lambda <\sqrt{3}$\\
\hline
$J$ & \{-8,-1,1,0\}& Unstable \\
\hline
$K$ & \{-8,-1,1,0\}& Unstable \\
[1ex] 
\hline 
\end{tabular}
\label{TABLE-VI}
\end{table}

\begin{figure}[H]
    \centering
    \includegraphics[width=75mm]{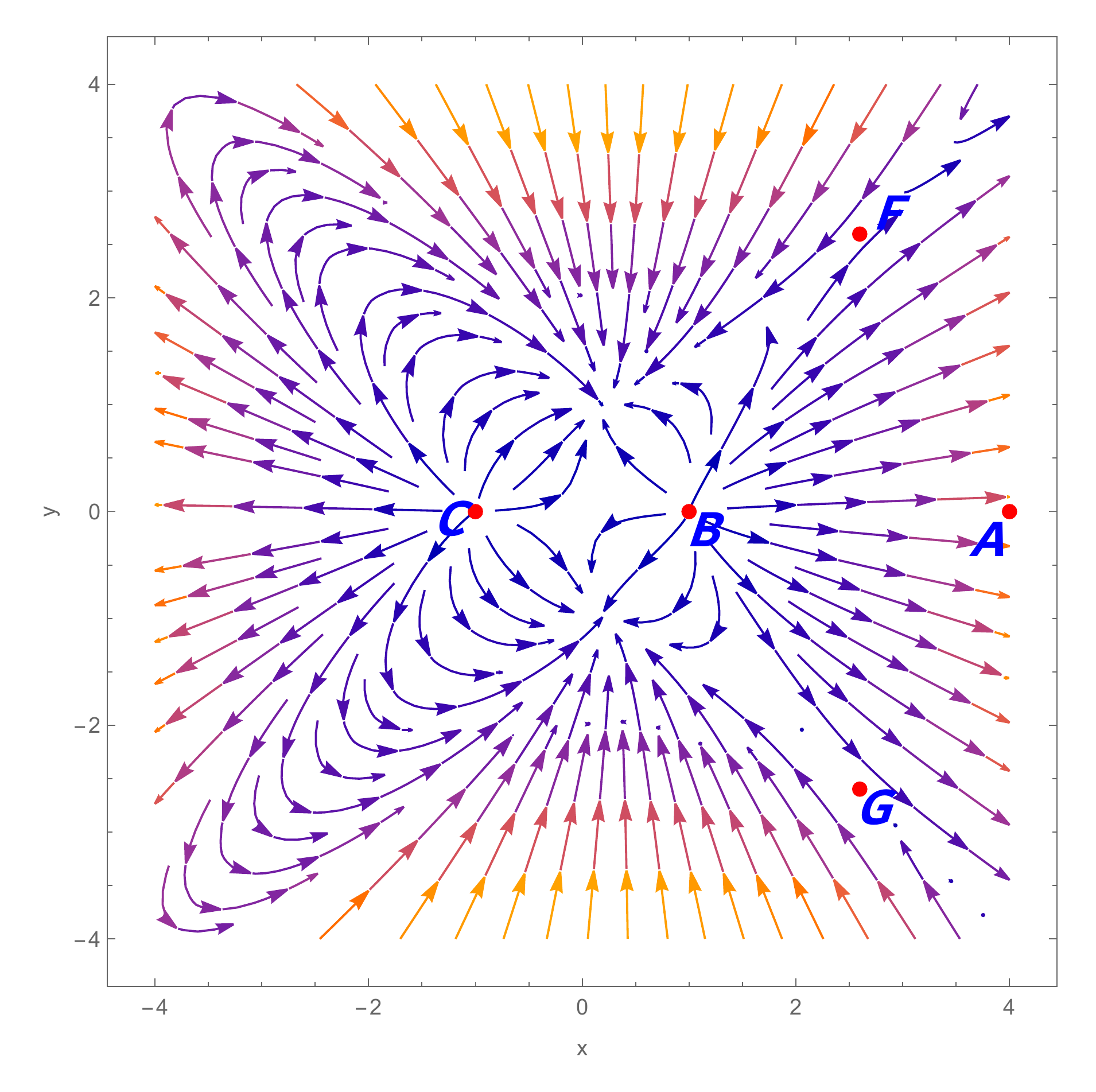}
    \includegraphics[width=75mm]{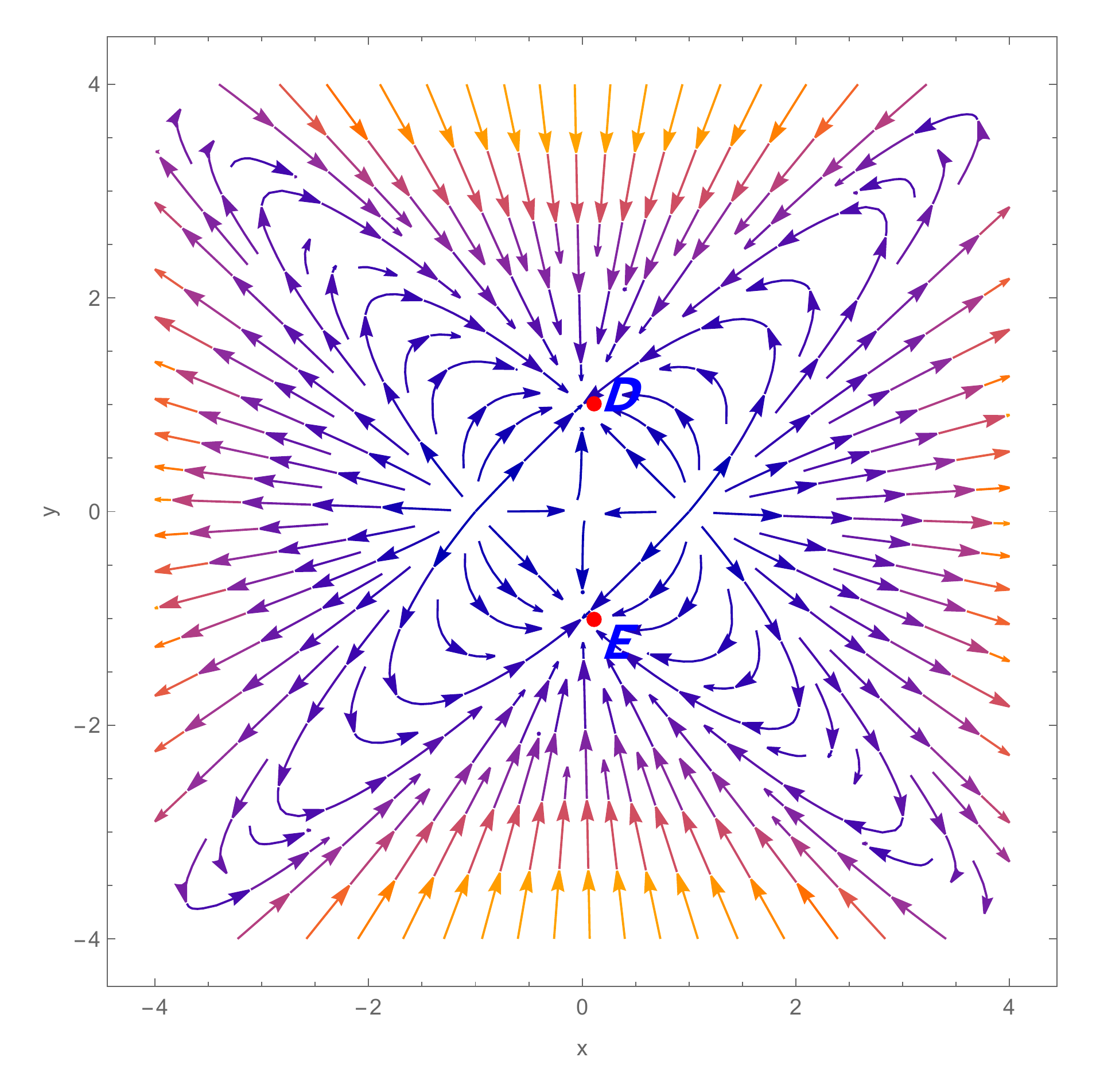}
    \caption{Phase portrait for dynamical system in Eqs.~(\ref{eq:dx_dN_model_1_alpha_2}-\ref{eq:dl_dN_model_1_alpha_2}), the left plot is for $u=0$, $\rho=0$, $\tau=1$, $\lambda=\sqrt{\frac{2}{9}}$, the right plot having parametric values $u=0$, $\rho=0$, $\zeta =\frac{1}{9}$.} 
    \label{Fig3}
\end{figure}

\begin{figure}[H]
    \centering
    \includegraphics[width=75mm]{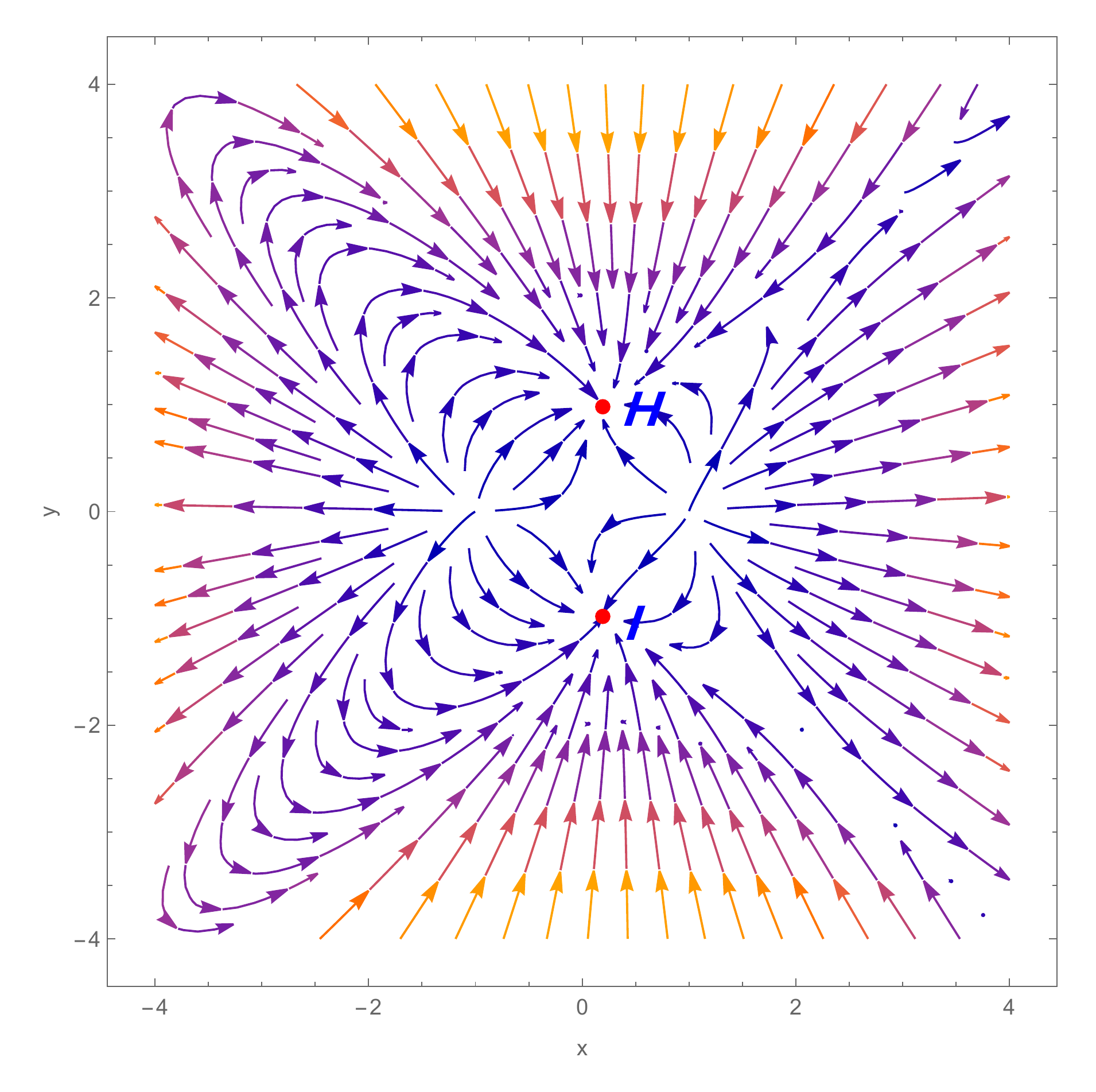}
    \includegraphics[width=75mm]{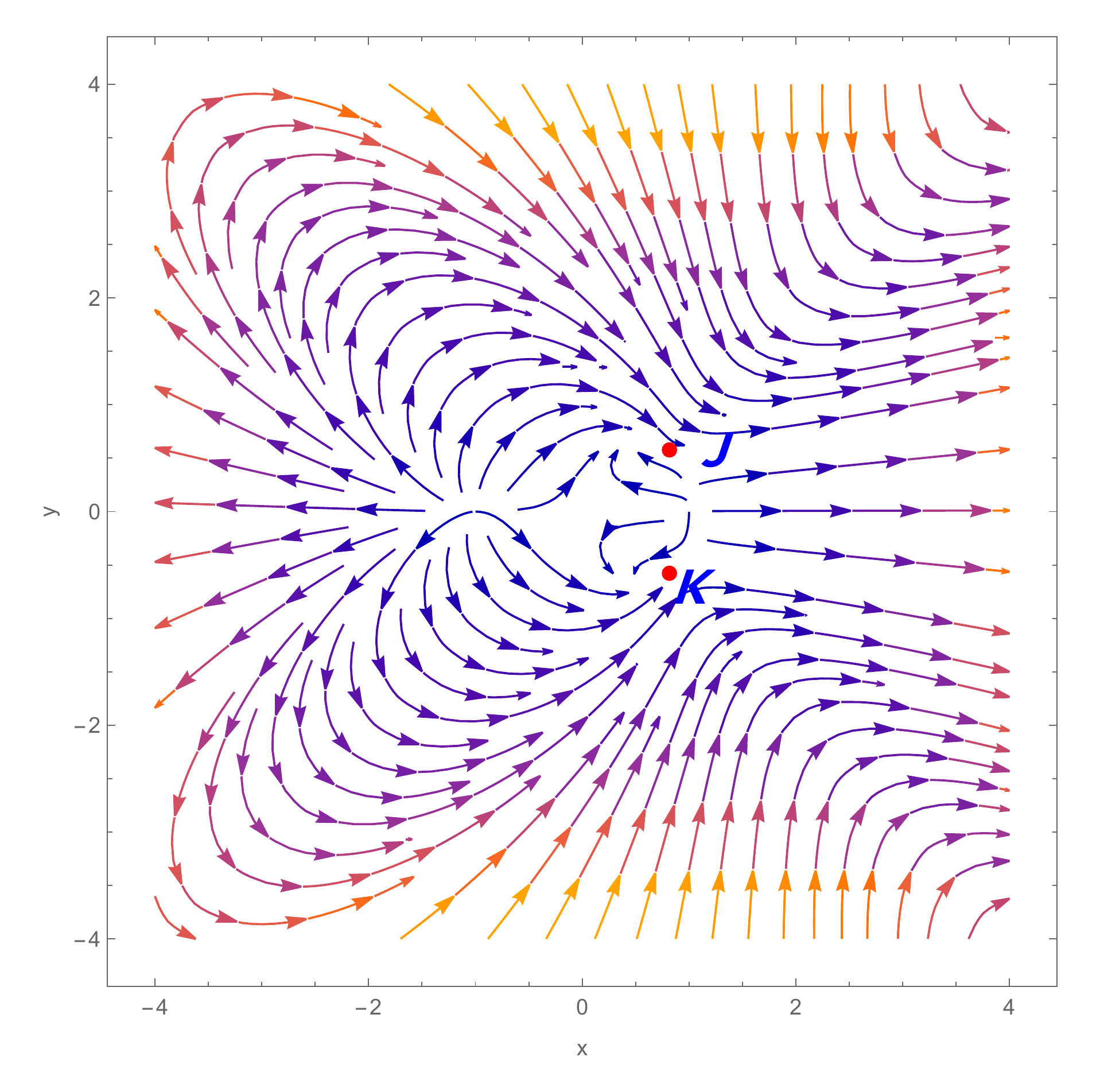}
    \caption{Phase portrait for dynamical system in Eqs.~(\ref{eq:dx_dN_model_1_alpha_2}-\ref{eq:dl_dN_model_1_alpha_2}), left phase portrait is for $u=0$, $\rho=0$, $\lambda=\sqrt{\frac{2}{9}}$ and the right phase portrait is for $u=0$, $\rho=0$.} \label{Fig4}
\end{figure}

The phase space diagram is presented in Fig-\ref{Fig3} and Fig-\ref{Fig4} for the parametric values $\lambda=\sqrt{\frac{2}{9}}$ which belongs to the stability range for $\lambda$ for critical points $H$ and $I$ and other parameters $\tau=1$, $\zeta=\frac{1}{9}$ are chosen such that the phase space diagram explain the stability conditions for the corresponding critical points. Critical points $D$ and $E$, represent de Sitter solution  the phase space analysis confirms the attracting nature of these critical points. From the Fig.\ref{Fig2} observations, it can be conclude that due to different co-ordinates the critical point $A$ is presented in Fig-{\ref{Fig3}} moves in positive X-axis than in Fig.\ref{Fig2}, but it does not impact on it's stability nature. Critical points $F$ and $G$ show stability  for $-2 \sqrt{\frac{6}{7}}\leq \lambda <-\sqrt{3}$ or $\sqrt{3}<\lambda \leq 2 \sqrt{\frac{6}{7}}$ but we choose $\lambda=\sqrt{\frac{2}{9}}$ hence these are a saddle points. The phase space diagram allows us to conclude that critical points $J$ and $K$ are saddle points.

\section{Model 2 -- Kinetic Term Coupled with \texorpdfstring{$I_2$}{}}\label{sec:model-II}

The model second action equation consist of a coupling between the $X^\alpha$ term with $I_2$, the action for Model 2 is described as below with
\begin{equation}\label{eq:action_model_2}
    S = \int d^4x e [X-V(\phi)-\frac{T}{2 \kappa^{2}}+X^{\alpha}I_{2}]+S_{m}+S{r}\,,
\end{equation}
In the action equation, we have  $I_2=3H\dot{\phi}$ and other notations are same as in the first model. The expression for sum of energy density for matter and radiation and negative of pressure for radiation can be obtained on varying of action equation for Model 2 with respect to the tetrad field presented in Eqs.~(\ref{eq:70}--\ref{eq:71}) respectively. The motion equation in this case can be obtained by taking variation of action equation  with respect to $\phi$, presented in Eq.~(\ref{eq:72}).
\begin{align}
    \frac{T}{2\kappa^{2}}-V(\phi)-X-X^{\alpha}I_2-2\alpha X^{\alpha}I_2 &= \rho_{\rm m}+\rho_{\rm r} \,,\label{eq:70}\\
    -V(\phi)+\frac{T}{2\kappa^{2}}+ X+\frac{2\dot{H}}{\kappa^{2}}-X^{\alpha}\ddot{\phi}-2\alpha X^{\alpha}\ddot{\phi} &= -p_{r} \,,\label{eq:71}\\
    V^{'}(\phi)+3H\dot{\phi}+\frac{(3+6\alpha)}{2}X^\alpha T + (3+6\alpha)X^\alpha\dot{H} +\ddot{\phi}[1+\frac{\alpha X^{\alpha}6H}{\dot{\phi}}+\frac{\alpha^{2} X^{\alpha}12H}{\dot{\phi}}] &= 0 \,.\label{eq:72}
\end{align}
Using background expressions discussed in Sec.~\ref{sec:backgroundexpressions}, we can obtain expression for energy density and pressure for effective dark energy are presented in Eq.~(\ref{eq:73}) and Eq.~(\ref{eq:74}) respectively.
\begin{align}
    \rho_{\rm DE} &= V(\phi)+X+X^{\alpha}I_2+2\alpha X^{\alpha}I_2 \,,\label{eq:73}\\
    p_{\rm DE} &= -V(\phi)+X-(1+2\alpha)X^{\alpha}\ddot{\phi} \,,\label{eq:74}
\end{align}
To define autonomous dynamical system, the dimensionless variables defined in this case are as follow
\begin{equation}\label{eq:model_2_dynamic_var}
    x=\dfrac{\kappa\dot{\phi}}{\sqrt{6}H},\quad y=\frac{\kappa\sqrt{V}}{\sqrt{3}H}, \quad
    u=\frac{\kappa^2 X^\alpha \dot{\phi} }{H},\quad
    \rho= \frac{\kappa\sqrt{\rho_{\rm r}}}{\sqrt{3}H}, \quad \lambda=\frac{-V^{'}(\phi)}{\kappa V(\phi)}, \quad \Gamma=\dfrac{V(\phi)V^{''}(\phi)}{V^{'}(\phi)^{2}} \,.
\end{equation}
These dimensionless variables also satisfy constraint equation Eq.~(\ref{eq:50}), and dynamical system can be obtained as presented below
\begin{align}
    \dfrac{dx}{dN} &= \frac{x \left((2 \alpha +1) u \left(2 \alpha  \rho ^2+6 \alpha +\rho ^2+6 \alpha  x^2+9 x^2-y^2 \left(6 \alpha +\sqrt{6} \lambda  x+3\right)-3\right)\right)}{(2 \alpha +1) u (2 \alpha  (u+2)+u)+4 x^2} \nonumber \\
    & +\frac{x \left(3 (2 \alpha  u+u)^2+2 x \left(3 x^3+x \left(\rho ^2-3 y^2-3\right)+\sqrt{6} \lambda  y^2\right)\right)}{(2 \alpha +1) u (2 \alpha  (u+2)+u)+4 x^2}\,,\label{eq:model_2_dx_dN}\\
    \dfrac{dy}{dN} &= -\sqrt{\frac{3}{2}} \lambda  x y+\frac{y \left(3 (2 \alpha  u+u)^2+2 x^2 \left(\rho ^2+3 x^2-3 y^2+3\right)\right)}{(2 \alpha +1) u (2 \alpha  (u+2)+u)+4 x^2}\nonumber \\
    & +\frac{(2 \alpha +1) u y \left(6 (\alpha +1) x^2-\sqrt{6} \lambda  x y^2+2 \alpha  \left(\rho ^2-3 y^2+3\right)\right)}{(2 \alpha +1) u (2 \alpha  (u+2)+u)+4 x^2}\,, \\
    \dfrac{du}{dN} &= \frac{u \left(3 (2 \alpha  u+u)^2+(2 \alpha +1) u \left(4 \alpha  \rho ^2+\rho ^2+12 \alpha  x^2+9 x^2-y^2 \left(12 \alpha +\sqrt{6} \lambda  x+3\right)-3\right)\right)}{(2 \alpha +1) u (2 \alpha  (u+2)+u)+4 x^2}\nonumber \\ 
    & + \frac{2 u x \left(3 x^3+x \left(\rho ^2-3 \left(4 \alpha +y^2+1\right)\right)+\sqrt{6} (2 \alpha +1) \lambda  y^2\right)}{(2 \alpha +1) u (2 \alpha  (u+2)+u)+4 x^2}\,, \\
    \frac{d\rho}{dN} &= \frac{\rho  \left((2 \alpha  u+u)^2+2 x^2 \left(\rho ^2+3 x^2-3 y^2-1\right)\right)}{(2 \alpha +1) u (2 \alpha  (u+2)+u)+4 x^2}\nonumber\\ 
    & +\frac{\rho  u \left((2 \alpha +1) \left(6 (\alpha +1) x^2-\sqrt{6} \lambda  x y^2+2 \alpha  \left(\rho ^2-3 y^2-1\right)\right)\right)}{(2 \alpha +1) u (2 \alpha  (u+2)+u)+4 x^2}\,, \\
    \frac{d\lambda}{dN} &= \sqrt{6}(\Gamma-1)\lambda^{2}x\,.\label{eq:model_2_dl_dN}
\end{align}

The critical points for the dynamical system described in Eqs.~(\ref{eq:model_2_dx_dN}-\ref{eq:model_2_dl_dN}) are presented in Table~\ref{TABLE-VII}. From table observations it can conclude that, critical points $J$ to $O$ represent similar cosmological implications in terms of deceleration parameter $q=1$ and $\omega_{tot}=\frac{1}{3}$ describe radiation dominated phase of the universe. Amongst these critical points the critical points $N$ and $O$ are defined for $\alpha=\frac{1}{2}$. The critical points $F$, $G$ and $P$ also describe the same cosmological implication and represent cold dark matter dominated era with $\omega_{tot}=0$. $\alpha$ play role in the co-ordinate representation of critical points $D$ and $E$, these critical points describe de Sitter solution and defined for $\lambda=0$. The critical points $H$ and $I$ represent value for deceleration parameter $q=-1+\frac{\lambda^2}{2}$, these critical points can explain dark energy dominated universe. Critical points $A$, $B$ and $C$ deliver same value for $q$ and $\omega_{tot}$ hence these critical points also represent similar phase of universe evolution and behaves as stiff matter.\\

The stability conditions of the critical points are discussed in Table~\ref{TABLE-VIII}. The critical points $J$ to $O$ are presenting radiation dominated universe, eigenvalues for the linear perturbation matrix at these critical points having at least one eigenvalue with positive signature hence showing unstable behaviour. The critical points $F$, $G$ show eigenvalues in negative range for  $\alpha >0\land \left(-2 \sqrt{\frac{6}{7}}\leq \lambda <-\sqrt{3}\lor \sqrt{3}<\lambda \leq 2 \sqrt{\frac{6}{7}}\right)$ and critical point $P$ is stable at $\left(\lambda <0\land \sigma <\frac{\sqrt{\frac{3}{2}}}{\lambda }\right)\lor \left(\lambda >0\land \sigma >\frac{\sqrt{\frac{3}{2}}}{\lambda }\right)$, hence show stability within this range, also these points express $\omega_{tot}=0$, hence addressing cold dark matter phase of the universe evolution. The critical points $B$ and $C$ represent stiff matter era and possessing at least one eigenvalue with positive signature hence are unstable in nature, where critical point $A$ show stable behaviour in the parametric range same as critical point $P$. The critical points $D$, $E$ , $H$ and $I$ represent dark energy dominated era, where $D$ and $E$ are non-hyperbolic critical points but are stable in nature and critical points $H$ and $I$ show their stability in the parametric range $\alpha >0\land \left(-\sqrt{3}<\lambda <0\lor 0<\lambda <\sqrt{3}\right)$. The stability conditions of critical points $F$, $G$, $H$ and $I$ show $\alpha>0$ condition which implies stability of critical points representing cold dark matter and dark energy dominated era can be obtained for positive value of $\alpha$.

\begin{table}[H]
\caption{Critical Points for Dynamical System Corresponding to Model-II, for General $\alpha$.} 
\centering 
\begin{tabular}{|c|c|c|c|c|c|c|} 
\hline\hline 
Name of Critical Point & $x_{c}$ & $y_{c}$ & $u_{c}$ & $\rho_{c}$ & Deceleration Parameter $(q)$ & $\omega_{tot}$\\ [0.5ex] 
\hline\hline 
$A$,\begin{tabular}{@{}c@{}} $3 \mu  \sigma ^2-\mu +$\\ $2 \sigma ^4-6 \sigma ^2\neq 0$ \end{tabular}& $\sigma$ & $0$ & $1-\sigma ^2$ & $0$ & $2$ & $1$\\
\hline
$B$ & $1$ & $0$ & $0$ & $0$ & $2$ & $1$\\
\hline
$C$ & $-1$ & $0$ & $0$ & $0$ & $2$ & $1$\\
\hline
$D$, For $\lambda=0$, $ 2 \alpha +1\neq 0$ & $\varphi, -2 \alpha  \varphi +\varphi ^3+\varphi \neq 0$ & $\sqrt{\varphi ^2+1}$ & $-\frac{2 \varphi ^2}{2 \alpha +1}$ & $0$ & $-1$ & $-1$\\[1.5ex] 
\hline
$E$, For $\lambda=0$, $ 2 \alpha +1\neq 0$ & $\varphi, -2 \alpha  \varphi +\varphi ^3+\varphi \neq 0$ & $-\sqrt{\varphi ^2+1}$ & $-\frac{2 \varphi ^2}{2 \alpha +1}$ & $0$ & $-1$ & $-1$\\
\hline
$F$ & $\frac{\sqrt{\frac{3}{2}}}{\lambda }$ & $\sqrt{\frac{3}{2}} \sqrt{\frac{1}{\lambda ^2}}$ & $0$ & $0$ & $\frac{1}{2}$ & $0$\\
\hline
$G$ & $\frac{\sqrt{\frac{3}{2}}}{\lambda }$ & $-\sqrt{\frac{3}{2}} \sqrt{\frac{1}{\lambda ^2}}$ & $0$ & $0$ & $\frac{1}{2}$ & $0$\\
\hline
$H$ & $\frac{\lambda }{\sqrt{6}}$ & $\sqrt{1-\frac{\lambda ^2}{6}}$ & $0$ & $0$& $\frac{1}{2} \left(\lambda ^2-2\right)$ & $-1+\frac{\lambda^2}{3}$\\
\hline
$I$ & $\frac{\lambda }{\sqrt{6}}$ & $-\sqrt{1-\frac{\lambda ^2}{6}}$ & $0$ & $0$& $\frac{1}{2} \left(\lambda ^2-2\right)$ & $-1+\frac{\lambda^2}{3}$\\
\hline
$J$, in this case $\lambda=2$ & $\sqrt{\frac{2}{3}}$ & $\sqrt{\frac{1}{3}}$ & $0$ & $0$ & $1$ & $\frac{1}{3}$\\
\hline
$K$, in this case $\lambda=2$ & $\sqrt{\frac{2}{3}}$ & $-\sqrt{\frac{1}{3}}$ & $0$ & $0$ & $1$ & $\frac{1}{3}$\\
\hline
$L$ & $\eta, \eta ^3+\eta \neq 0$ & $0$ & $-2 \eta ^2$ & $\sqrt{\eta ^2+1}$ & $1$ & $\frac{1}{3}$\\
\hline
$M$&$\eta, \eta ^3+\eta \neq 0$ & $0$ & $-2 \eta ^2$ & $-\sqrt{\eta ^2+1}$ & $1$ & $\frac{1}{3}$\\
\hline
$N$, in this case $\alpha=\frac{1}{2}$ & $0$ & $0$ & $\epsilon, \epsilon ^2+\epsilon \neq 0$ & $\sqrt{1-2 \epsilon }$ & $1$ & $\frac{1}{3}$\\
\hline
$O$, in this case $\alpha=\frac{1}{2}$ & $0$ & $0$ & $\epsilon, \epsilon ^2+\epsilon \neq 0$ & $-\sqrt{1-2 \epsilon }$ & $1$ & $\frac{1}{3}$\\
\hline
$P$& $\sigma$ & $0$ & $-2 \sigma ^2$ & $0$ & $\frac{1}{2}$ & $0$\\
[1ex] 
\hline 
\end{tabular}
\label{TABLE-VII}
\end{table}

\begin{table}[H]
\caption{Eigenvalues and Stability of Eigenvalue at Corresponding Critical Point.} 
\centering 
\begin{tabular}{|p{2cm}|p{8cm}|p{8cm}|} 
\hline 
Name of Critical Point & Corresponding Eigenvalues & Stability \\ [0.5ex] 
\hline 
$A$ &  $\left\{0,-\frac{3}{2},-\frac{1}{2},\frac{1}{2} \left(3-\sqrt{6} \lambda  \sigma \right)\right\}$ & Stable for $\left(\lambda <0\land \sigma <\frac{\sqrt{\frac{3}{2}}}{\lambda }\right)\lor \left(\lambda >0\land \sigma >\frac{\sqrt{\frac{3}{2}}}{\lambda }\right)$\\
\hline
$B$ & $\left\{3,1,-6 \alpha ,\frac{1}{2} \left(6-\sqrt{6} \lambda \right)\right\}$ & Unstable\\
\hline
$C$ & $\left\{3,1,-6 \alpha ,\frac{1}{2} \left(6+\sqrt{6} \lambda \right)\right\}$ & Unstable\\
\hline
$D$ & $\{0,-3,-3,-2\}$ & Stable \\
\hline
$E$ &$\{0,-3,-3,-2\}$ & Stable \\
\hline
$F$ & $\left\{-\frac{1}{2},-3 \alpha ,\frac{3 \left(-\lambda ^2-\sqrt{24 \lambda ^2-7 \lambda ^4}\right)}{4 \lambda ^2},\frac{3 \left(\sqrt{24 \lambda ^2-7 \lambda ^4}-\lambda ^2\right)}{4 \lambda ^2}\right\}$ & Stable for  \begin{tabular}{@{}c@{}} $\alpha >0\land$ \\ $\left(-2 \sqrt{\frac{6}{7}}\leq \lambda <-\sqrt{3}\lor \sqrt{3}<\lambda \leq 2 \sqrt{\frac{6}{7}}\right)$\end{tabular}\\
\hline
$G$ & $\left\{-\frac{1}{2},-3 \alpha ,\frac{3 \left(-\lambda ^2-\sqrt{24 \lambda ^2-7 \lambda ^4}\right)}{4 \lambda ^2},\frac{3 \left(\sqrt{24 \lambda ^2-7 \lambda ^4}-\lambda ^2\right)}{4 \lambda ^2}\right\}$ & Stable for \begin{tabular}{@{}c@{}} $\alpha >0\land$\\ $\left(-2 \sqrt{\frac{6}{7}}\leq \lambda <-\sqrt{3}\lor \sqrt{3}<\lambda \leq 2 \sqrt{\frac{6}{7}}\right)$\end{tabular}\\
\hline
$H$ & $\left\{-\alpha  \lambda ^2,\frac{1}{2} \left(\lambda ^2-6\right),\frac{1}{2} \left(\lambda ^2-4\right),\lambda ^2-3\right\}$ & Stable for $\alpha >0\land \left(-\sqrt{3}<\lambda <0\lor 0<\lambda <\sqrt{3}\right)$\\
\hline
$I$ & $\left\{-\alpha  \lambda ^2,\frac{1}{2} \left(\lambda ^2-6\right),\frac{1}{2} \left(\lambda ^2-4\right),\lambda ^2-3\right\}$ & Stable for $\alpha >0\land \left(-\sqrt{3}<\lambda <0\lor 0<\lambda <\sqrt{3}\right)$\\
\hline
$J$ & $\{-1,1,0,-4 \alpha \}$ & Unstable \\
\hline
$K$ & $\{-1,1,0,-4 \alpha \}$ & Unstable \\
\hline
$L$ & $\left\{0,1,\frac{1}{2} \left(4-\sqrt{6} \lambda  \eta \right),-\frac{\eta ^3+\eta }{\eta  \left(\eta ^2+1\right)}\right\}$ & Unstable \\
\hline
$M$ & $\left\{0,1,\frac{1}{2} \left(4-\sqrt{6} \lambda  \eta \right),-\frac{\eta ^3+\tau }{\eta  \left(\eta ^2+1\right)}\right\}$ & Unstable \\
\hline
$N$ & $\left\{\frac{1-2 \epsilon }{\epsilon +1},\frac{3 \epsilon }{\epsilon +1},1,2\right\}$ & Unstable \\
\hline
$O$ & $\left\{\frac{1-2 \epsilon }{\epsilon +1},\frac{3 \epsilon }{\epsilon +1},1,2\right\}$ & Unstable \\
\hline
$P$ & $\left\{0,-\frac{3}{2},-\frac{1}{2},\frac{1}{2} \left(3-\sqrt{6} \lambda  \sigma \right)\right\}$ & Stable for $\left(\lambda <0\land \sigma <\frac{\sqrt{\frac{3}{2}}}{\lambda }\right)\lor \left(\lambda >0\land \sigma >\frac{\sqrt{\frac{3}{2}}}{\lambda }\right)$ \\
[1ex] 
\hline 
\end{tabular}
\label{TABLE-VIII}
\end{table}

\begin{figure}[H]
    \centering
    \includegraphics[width=75mm]{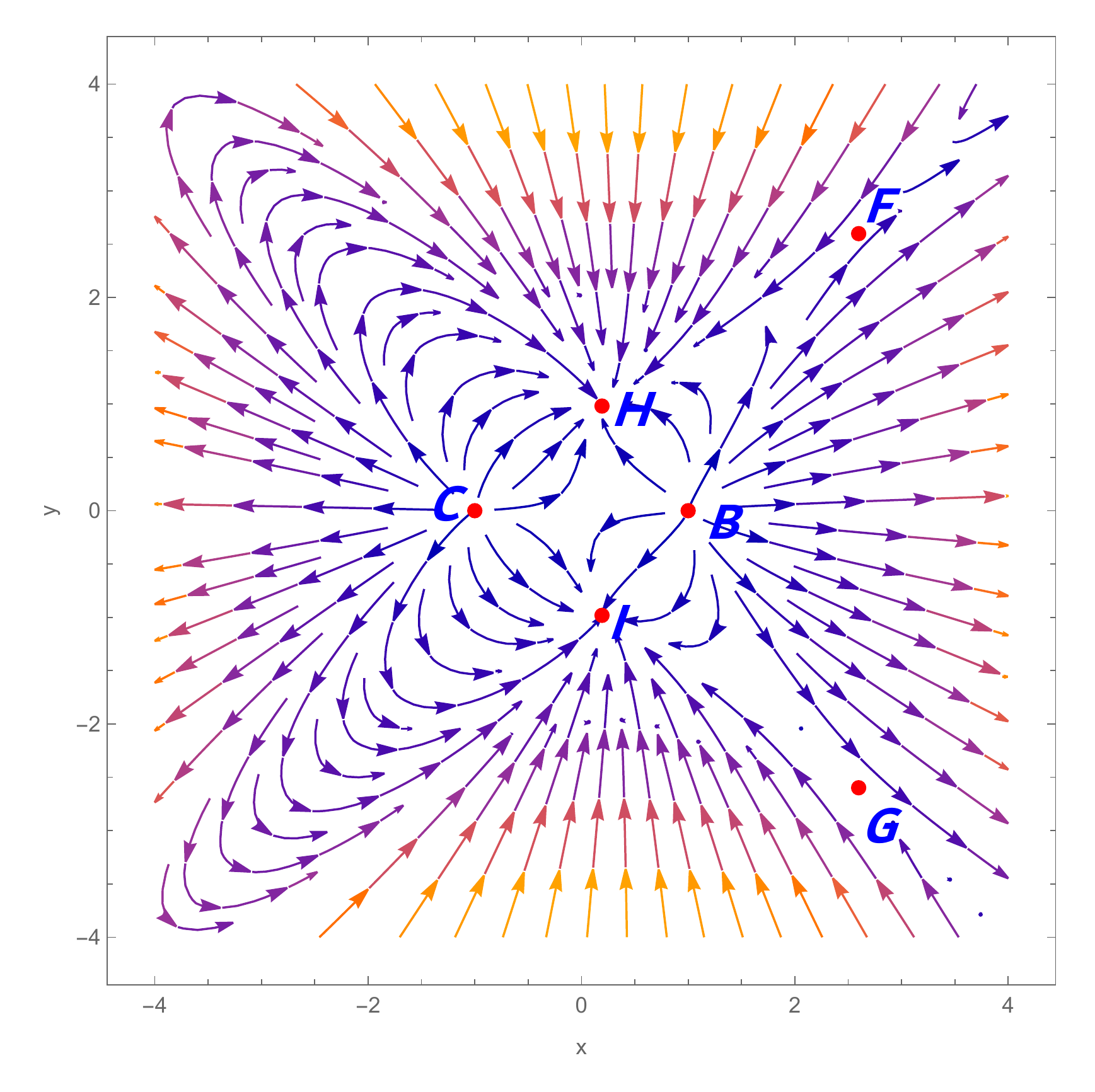}
    \includegraphics[width=75mm]{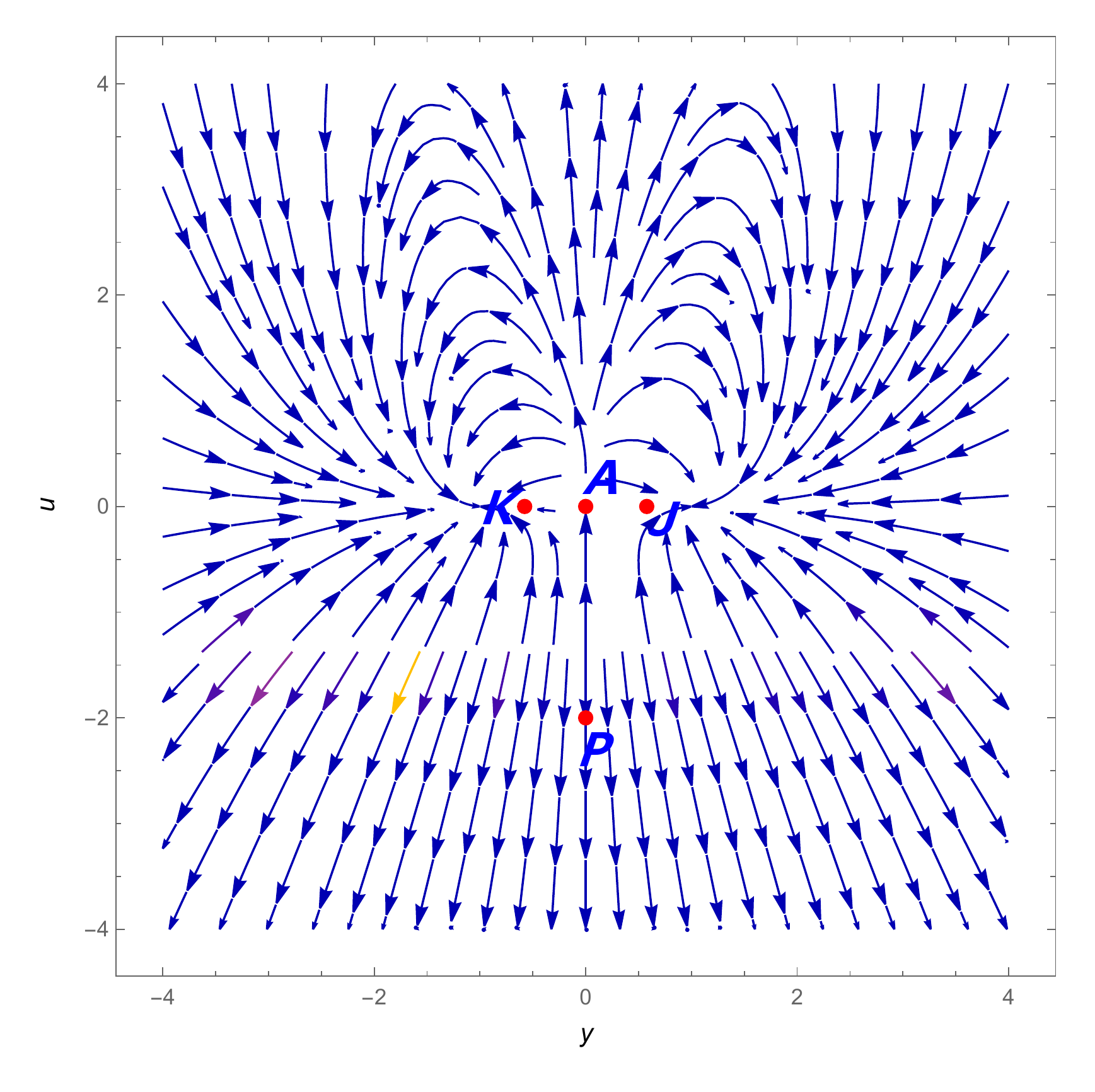}
    \includegraphics[width=75mm]{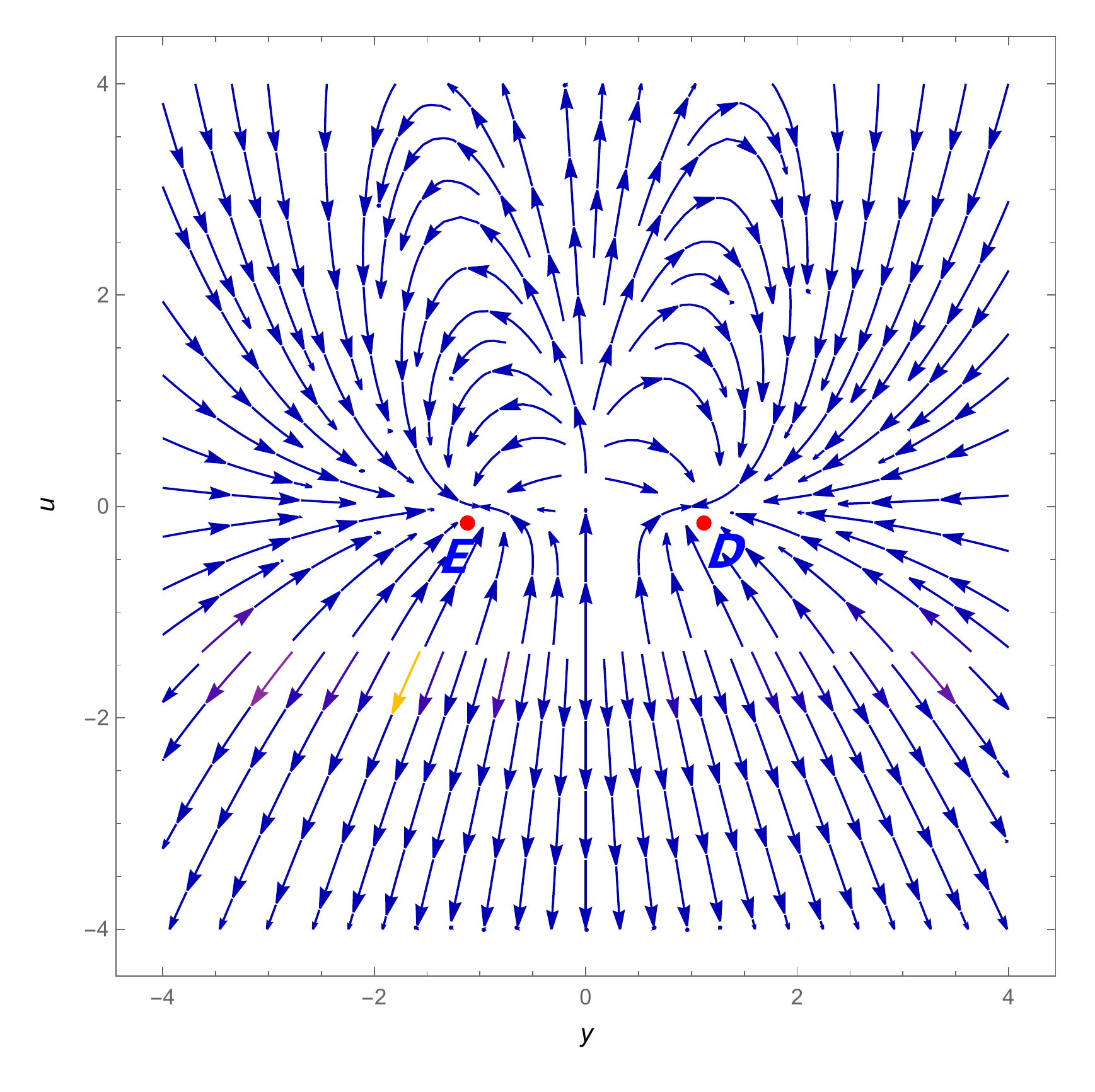}
    \includegraphics[width=75mm]{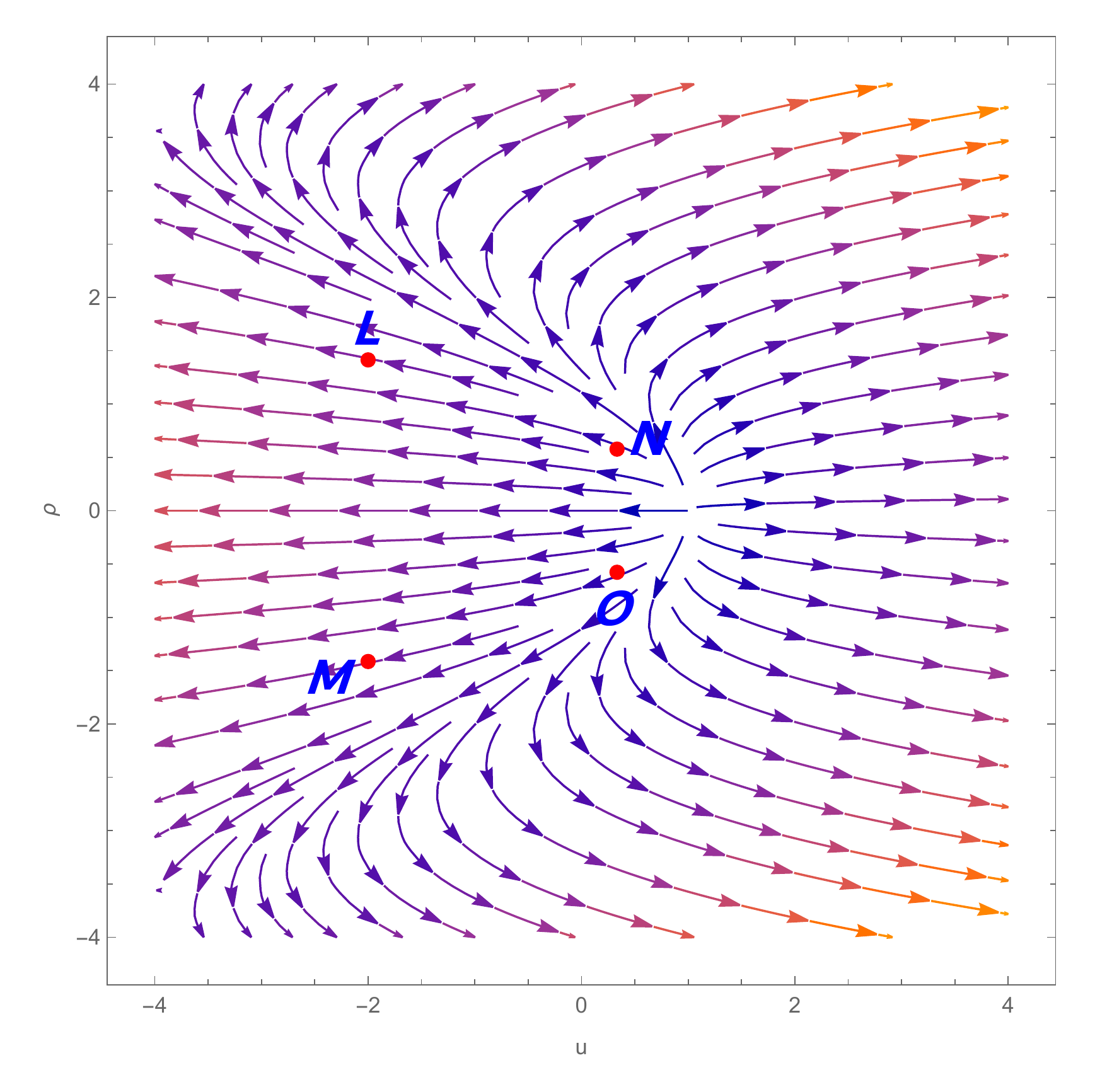}
    \caption{Phase portrait for above dynamical system Eqs.~(\ref{eq:model_2_dx_dN}-\ref{eq:model_2_dl_dN}), the upper left plot is for $u=0$, $\rho=0$, $\lambda=\sqrt{\frac{2}{9}}$ and the upper right plot having parameter values $x=0$, $\rho=0$,, $\sigma=1$, $\alpha=1.1$. The lower left phase portrait is for $x=0$ , $\rho=0$, $\alpha=1.1$, $\varphi=\frac{1}{2}$, the lower right phase portrait is for $u=0$, $\rho=0$, $\eta=1$, $\epsilon=\frac{1}{3}$.} \label{Fig5}
\end{figure}

The critical points are plotted in the phase diagram Fig.\ref{Fig5}. These phase plots are plotted for the dynamical system presented in Eqs.~(\ref{eq:model_2_dx_dN}-\ref{eq:model_2_dl_dN}). The lower right plot shows that the phase space trajectories are moving away from critical points $L$, $M$, $N$ and $O$ hence these points represent instability with saddle point behaviour. The critical points $N$ and $O$ are defined for $\alpha=\frac{1}{2}$. The critical points $D$ and $E$ are de Sitter solutions and phase diagram clarify that these points behaves as attracting solutions. The upper left phase plot describe phase space trajectories behaviour of critical points $H$, $I$, $B$, $C$, $F$ and $G$. We can observe that the critical points $B$ and $C$ which is unstable saddle point but showing unstable node behaviour which is leading to the positive eigenvalues at critical points $B$ and $C$. Also point $H$ and $I$ showing attracting point behaviour, these critical points represent dark energy dominated era with stability as described in Table~\ref{TABLE-VIII}. Critical points $F$ and $G$ represent cold dark matter dominated era, we have plotted plots for $\lambda=\sqrt{\frac{2}{9}}$ which is not in the stability range of $F$ and $G$. The upper right diagram represent phase space trajectories at critical points $K$, $J$ and $P$, since the phase space trajectories are moving away from these critical points, these critical points are showing saddle point behaviour hence unstable. Since the parametric value $\sigma=1$ do not follow stability range of critical point $A$ hence getting saddle point behaviour.

\subsection{Case A: \texorpdfstring{$\alpha=1$}{}}

We have consider $\alpha=1$ in the action equation Eq.~(\ref{eq:action_model_2}) to discuss role of $\alpha$ in getting number of critical points and their participation in describing different phases of Universe evolution. In this case, the evolution equations can be obtained by limiting the general $\alpha$ to $\alpha=1$ from Eqs.~(\ref{eq:70}--\ref{eq:72}) the set of dimensionless variables can be defined as follow
\begin{equation}\label{eq:65}
    x=\dfrac{\kappa\dot{\phi}}{\sqrt{6}H}\,,\quad y=\frac{\kappa\sqrt{V}}{\sqrt{3}H}\,,\quad u=\frac{3\kappa^2\dot{\phi}^3}{2H}\,,\quad \rho= \frac{\kappa\sqrt{\rho_{\rm r}}}{\sqrt{3}H}\,, \quad \lambda=\frac{-V^{'}(\phi)}{\kappa V(\phi)}\,,\quad \Gamma=\dfrac{V(\phi)V^{''}(\phi)}{V^{'}(\phi)^{2}}\,.
\end{equation}
These dimensionless variables satisfy constraint equation Eq.~(\ref{eq:57}), and the dynamical system in this case can be defined as follow,
\begin{align}
    \dfrac{dx}{dN} &= \frac{x \left(3 u^2+3 \rho ^2 u-y^2 \left(\sqrt{6} \lambda  (u-2) x+9 u+6 x^2\right)+15 u x^2+3 u+2 \rho ^2 x^2+6 x^4-6 x^2\right)}{u (u+4)+4 x^2}\,,\label{eq:model_2_csae_1_dx_dN}\\
    \dfrac{dy}{dN} &= -y \left(\frac{-3 u^2-2 u \left(\rho ^2+6 x^2+3\right)+u y^2 \left(\sqrt{6} \lambda  x+6\right)-2 x^2 \left(\rho ^2+3 x^2-3 y^2+3\right)}{u (u+4)+4 x^2}+\sqrt{\frac{3}{2}} \lambda  x\right)\,,\\
    \dfrac{du}{dN} &= \frac{u \left(3 u^2+5 \rho ^2 u-y^2 \left(\sqrt{6} \lambda  (u-6) x+15 u+6 x^2\right)+21 u x^2-3 u+2 \rho ^2 x^2+6 x^4-30 x^2\right)}{u (u+4)+4 x^2}\,,\\
    \frac{d\rho}{dN} &= \frac{\rho  \left(u^2+2 u \left(\rho ^2+6 x^2-1\right)-u y^2 \left(\sqrt{6} \lambda  x+6\right)+2 x^2 \left(\rho ^2+3 x^2-3 y^2-1\right)\right)}{u (u+4)+4 x^2}\,,\\
    \dfrac{d\lambda}{dN} &= -\sqrt{6}(\Gamma-1)\lambda^{2}x\,.\label{eq:model_2_csae_1_dl_dN}
\end{align}

The critical points with value of deceleration parameter and $\omega_{tot}$ are presented in the Table~\ref{TABLE-IX}. From the table observations we can conclude that we get less number of critical points than the general case. The critical points $J$ and $K$ show deceleration parameter value $q=1$, hence describe radiation dominated era. The critical points $F$ and $G$ are showing similar cosmological implication in terms of value of deceleration parameter and value of $\omega_{tot}$, representing cold dark matter dominated era. Amongst all the critical points, an accelerated phase of evolution can be described by the critical points $D$, $E$, $H$ and $I$. The critical points $D$ and $E$ describe de Sitter solution with $q=-1$ and critical points $H$ and $I$ deliver deceleration parameter value $q=-1+\frac{\lambda^2}{2}$ hence represent the dark energy dominated era. The critical points $B$ and $C$ gives $q=2$, these points can describe stiff matter. The critical point $A$ showing deceleration parameter value in positive range, hence can not describe the accelerated phase of the evolution.

\begin{table}[H]
\caption{ Critical points for the dynamical system corresponding to updated calculations for Model-I} 
\centering 
\begin{tabular}{|c|c|c|c|c|c|c|} 
\hline\hline 
Name of Critical Point & $x_{c}$ & $y_{c}$ & $u_{c}$ & $\rho_{c}$ & Deceleration Parameter ($q$) & $\omega_{tot}$\\ [0.5ex] 
\hline\hline 
$A$ & $0$ & $0$ & $1$ & $0$ & $\frac{4}{5}$ & $\frac{1}{5}$\\
\hline
$B$ & $1$ & $0$ & $0$ & $0$ & $2$ & $1$\\
\hline
$C$ & $-1$ & $0$ & $0$ & $0$ & $2$ & $1$\\
\hline
$D$, in this case $\lambda=0$  & $\sigma , \sigma ^3- \sigma \neq 0$ & $\sqrt{\sigma ^2+1}$ & $-2 \sigma ^2$ & $0$ & $-1$ & $-1$\\
\hline
$E$, in this case $\lambda=0$ & $\sigma , \sigma ^3- \sigma \neq 0$ & $-\sqrt{\sigma ^2+1}$ & $-2 \sigma ^2$ & $0$ & $-1$ & $-1$\\
\hline
$F$ & $\frac{\sqrt{\frac{3}{2}}}{\lambda }$ & $\sqrt{\frac{3}{2}} \sqrt{\frac{1}{\lambda ^2}}$ & $0$ & $0$ & $\frac{1}{2}$ & $0$\\
\hline
$G$ & $\frac{\sqrt{\frac{3}{2}}}{\lambda }$ & $-\sqrt{\frac{3}{2}} \sqrt{\frac{1}{\lambda ^2}}$ & $0$ & $0$ & $\frac{1}{2}$ & $0$\\
\hline
$H$ & $\frac{\lambda }{\sqrt{6}}$ & $\sqrt{1-\frac{\lambda ^2}{6}}$ & $0$ & $0$& $\frac{1}{2} \left(\lambda ^2-2\right)$ & $-1+\frac{\lambda^2}{3}$\\
\hline
$I$ & $\frac{\lambda }{\sqrt{6}}$ & $-\sqrt{1-\frac{\lambda ^2}{6}}$ & $0$ & $0$& $\frac{1}{2} \left(\lambda ^2-2\right)$ & $-1+\frac{\lambda^2}{3}$\\
\hline
$J$, in this case $\lambda=2$ & $\sqrt{\frac{2}{3}}$ & $\frac{1}{\sqrt{3}}$ & $0$ & $0$ & $1$ & $\frac{1}{3}$\\
\hline
$K$, in this case $\lambda=2$ & $\sqrt{\frac{2}{3}}$ & $-\frac{1}{\sqrt{3}}$ & $0$ & $0$ & $1$ & $\frac{1}{3}$\\
[1ex] 
\hline 
\end{tabular}
\label{TABLE-IX}
\end{table}

To analyse the stability conditions eigenvalues for all critical points are presented in Table~\ref{TABLE-X}. The critical points $F$, $G$ show stability for the parametric values $-2 \sqrt{\frac{6}{7}}\leq \lambda <-\sqrt{3}\lor \sqrt{3}<\lambda \leq 2 \sqrt{\frac{6}{7}}$ and describe cold dark matter dominated era. The critical points $H$ and $I$ are shows stable behaviour for the parametric values $-\sqrt{3}<\lambda <0\lor 0<\lambda <\sqrt{3}$ these critical points represent dark energy dominated era for any real values of $\lambda$. The critical points $J$ and $K$ are defined at $\lambda=2$ and presence of eigenvalues with both positive and negative signature leads to saddle point hence unstable. The critical points $B$ and $C$ is also showing saddle point behaviour and unstable in nature. In this case, we get critical point $A$ with deceleration parameter value $q=\frac{4}{5}$ with existence of positive eigenvalues at linear perturbation matrix hence unstable in nature.

\begin{table}[H]
\caption{ eigenvalues and Stability of Eigenvalue at Corresponding Critical Point.} 
\centering 
\begin{tabular}{|c|c|c|} 
\hline 
Name of Critical Point & Corresponding Eigenvalues & Stability \\ [0.5ex] 
\hline 
$A$ & $\left\{\frac{9}{5},\frac{6}{5},\frac{3}{5},-\frac{1}{5}\right\}$ & Unstable\\
\hline
$B$ & $\left\{-12,3,1,\frac{1}{2} \left(6-\sqrt{6} \lambda \right)\right\}$ & Unstable\\
\hline
$C$ & $\left\{-12,3,1,\frac{1}{2} \left(\sqrt{6} \lambda +6\right)\right\}$ & Unstable\\
\hline
$D$ & $\{0,-3,-3,-2\}$ & Stable\\
\hline
$E$ &$\{0,-3,-3,-2\}$ & Stable \\
\hline
$F$ & $\left\{-3,-\frac{1}{2},\frac{3 \left(-\lambda ^2-\sqrt{24 \lambda ^2-7 \lambda ^4}\right)}{4 \lambda ^2},\frac{3 \left(\sqrt{24 \lambda ^2-7 \lambda ^4}-\lambda ^2\right)}{4 \lambda ^2}\right\}$ & Stable for $-2 \sqrt{\frac{6}{7}}\leq \lambda <-\sqrt{3}\lor \sqrt{3}<\lambda \leq 2 \sqrt{\frac{6}{7}}$\\
\hline
$G$ & $\left\{-3,-\frac{1}{2},\frac{3 \left(-\lambda ^2-\sqrt{24 \lambda ^2-7 \lambda ^4}\right)}{4 \lambda ^2},\frac{3 \left(\sqrt{24 \lambda ^2-7 \lambda ^4}-\lambda ^2\right)}{4 \lambda ^2}\right\}$ & Stable for $-2 \sqrt{\frac{6}{7}}\leq \lambda <-\sqrt{3}\lor \sqrt{3}<\lambda \leq 2 \sqrt{\frac{6}{7}}$\\
\hline
$H$ & $\left\{-\lambda ^2,\frac{1}{2} \left(\lambda ^2-6\right),\frac{1}{2} \left(\lambda ^2-4\right),\lambda ^2-3\right\}$ & Stable for $-\sqrt{3}<\lambda <0\lor 0<\lambda <\sqrt{3}$\\
\hline
$I$ & $\left\{-\lambda ^2,\frac{1}{2} \left(\lambda ^2-6\right),\frac{1}{2} \left(\lambda ^2-4\right),\lambda ^2-3\right\}$ & Stable for $-\sqrt{3}<\lambda <0\lor 0<\lambda <\sqrt{3}$\\
\hline
$J$ & $\{-4,-1,1,0\}$&Unstable \\
\hline
$K$ & $\{-4,-1,1,0\}$&Unstable\\
[1ex] 
\hline 
\end{tabular}
\label{TABLE-X}
\end{table}

\begin{figure}[H]
    \centering
    \includegraphics[width=75mm]{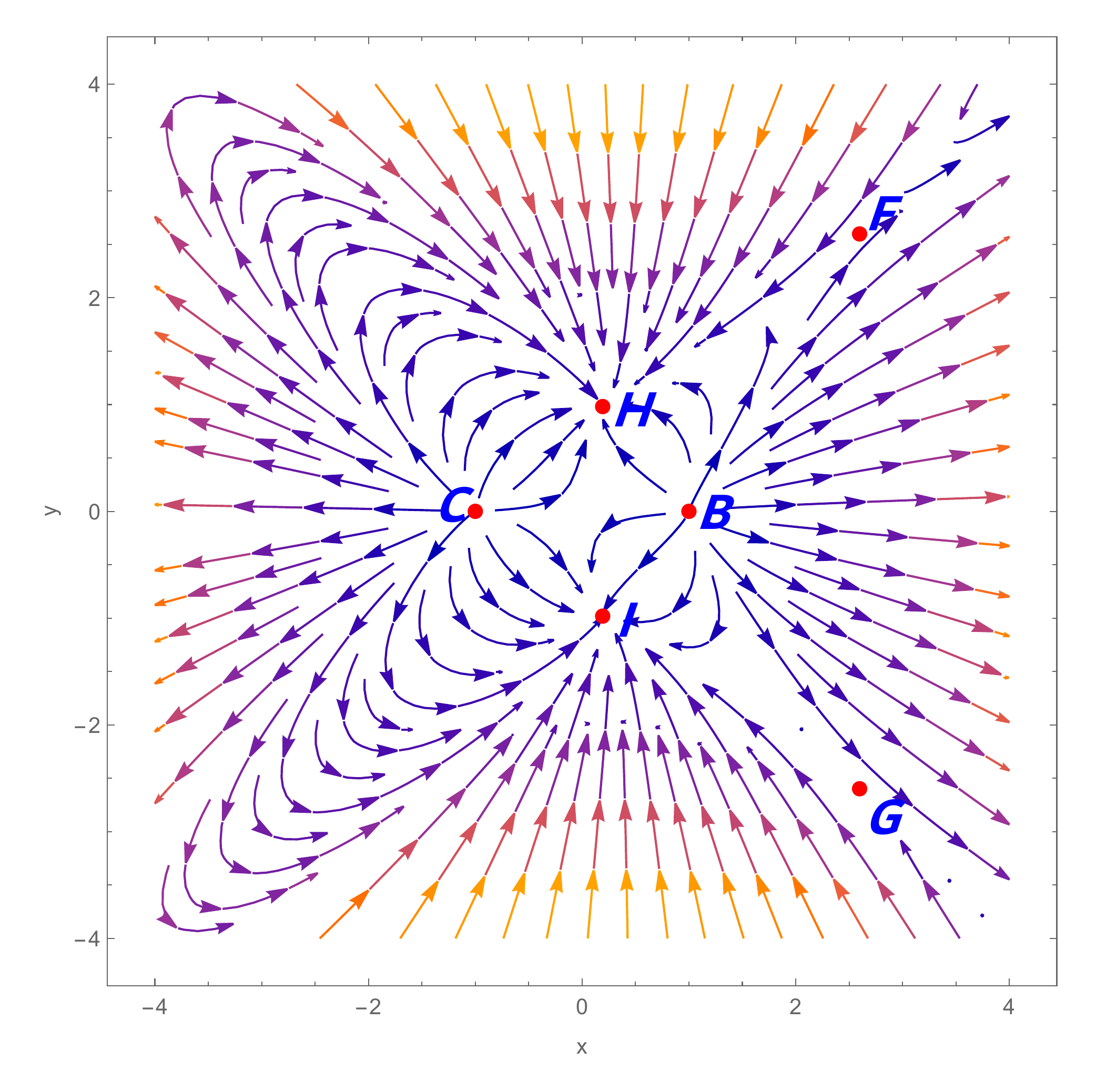}
    \includegraphics[width=75mm]{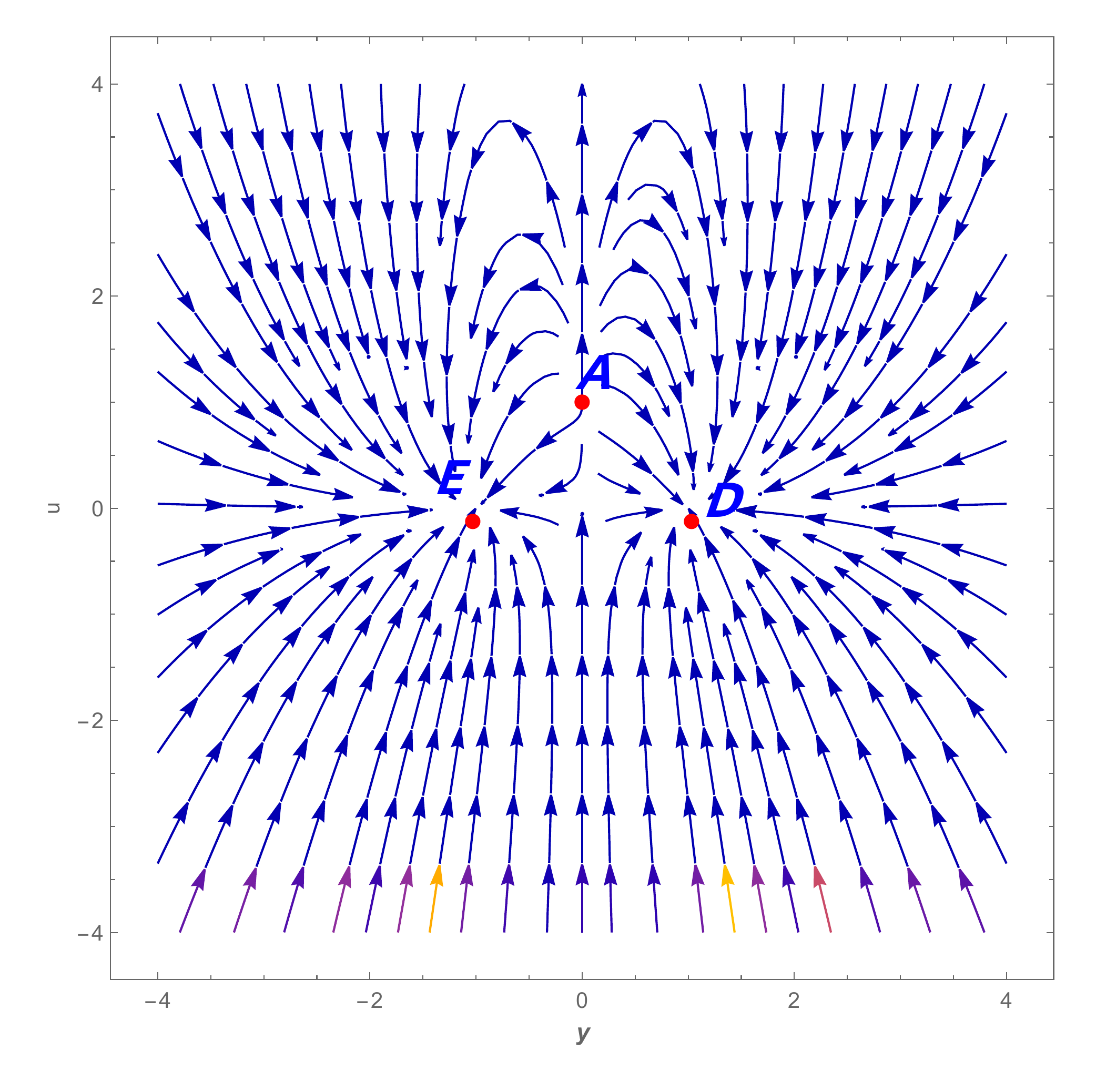}
    \caption{left Phase portrait is plotted for the parametric values $u=0$, $\rho=0$ and $\lambda=\sqrt{\frac{2}{9}}$ and right side plot parametric values are  $x=0$, $\rho=0$, $\sigma=\frac{1}{4}$ for right side figure.} \label{Fig6}
\end{figure}

\begin{figure}[H]
    \centering
    \includegraphics[width=75mm]{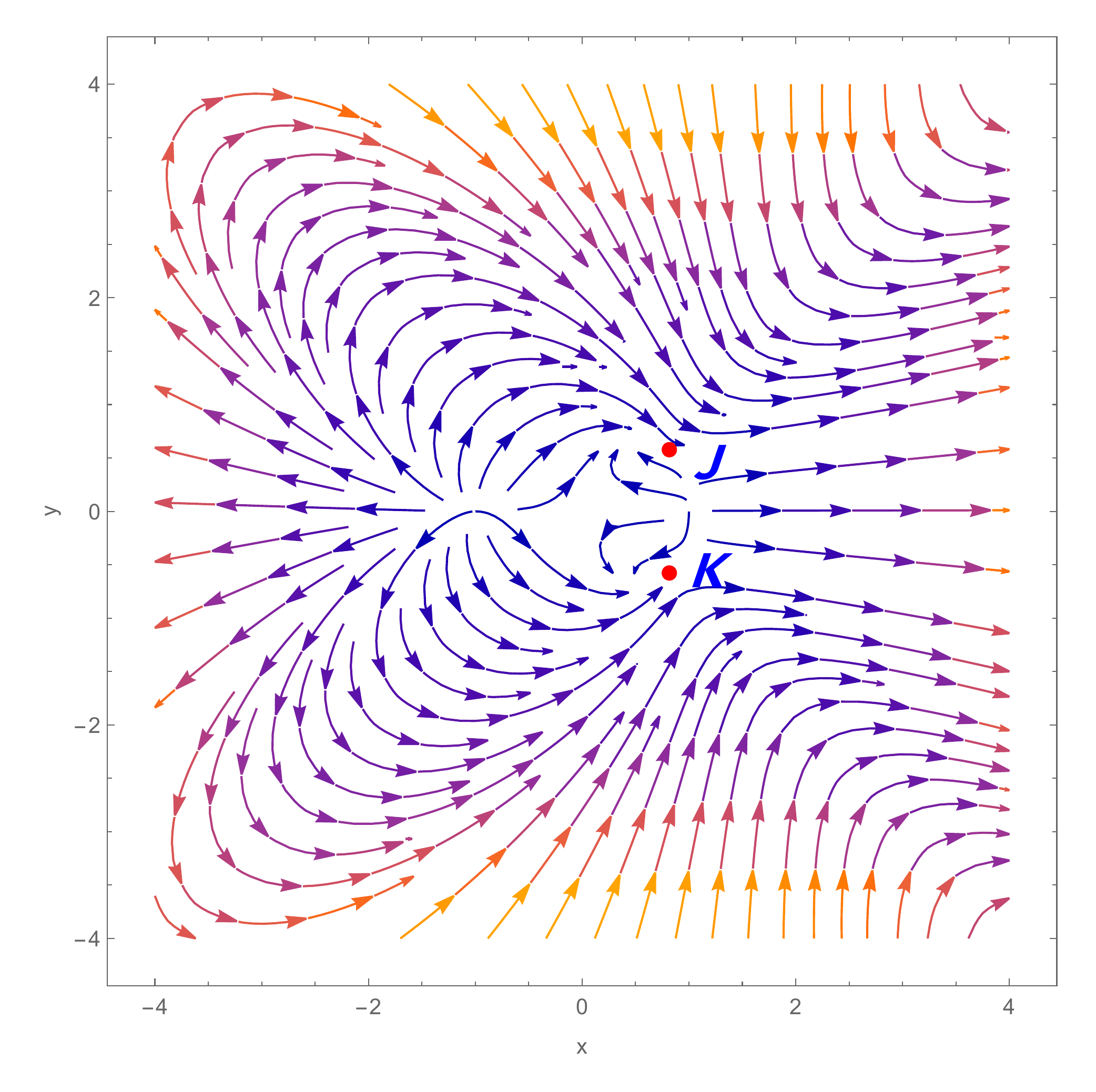}
    \caption{The phase portrait is plotted for the parametric values $\lambda=2$, $u=0$, $\rho=0$.}  \label{Fig7}
\end{figure}

The phase space plots for dynamical system presented in Eq.~(\ref{eq:model_2_csae_1_dx_dN}--\ref{eq:model_2_csae_1_dl_dN}) are described in Figs.(\ref{Fig6}) and (\ref{Fig7}). The phase space analysis concludes that critical points $H$ and $I$ shows attracting nature, the accelerating expansion of the universe can be described at these critical points. The phase trajectories at critical points $B$ and $C$ are moving away from the critical point hence showing unstable node behaviour leading to the positive eigenvalues. The phase space plot for critical points $F$ and $G$ describe saddle point behaviour and represent cold dark matter dominated era. The phase plots at critical points $D$ and $E$ show attracting behaviour and these critical points describe de Sitter solution. The critical point $A$ is a saddle point and can be confirmed by observing phase trajectories at $A$. The plot in Fig-\ref{Fig7} also describe that phase space trajectories are moving away from critical points hence showing unstable. behaviour.

\subsection{Case B: \texorpdfstring{$\alpha=2$}{}}

In this case we have discussed dynamical system analysis for Model-II, $\alpha=2$. The evolution expressions can be obtained by limiting Eqs.~(\ref{eq:70}--\ref{eq:72}) from general $\alpha$ to $\alpha=2$. The dimensionless variables to obtain autonomous dynamical system can be defined as follow
\begin{equation}\label{eq:87}
    x=\dfrac{\kappa\dot{\phi}}{\sqrt{6}H}\,,\quad y=\frac{\kappa\sqrt{V}}{\sqrt{3}H}\,,\quad u=\dfrac{5\kappa^{2}\dot{\phi}^{5}}{4H}\,,\quad \rho= \frac{\kappa\sqrt{\rho_{\rm r}}}{\sqrt{3}H}\,,\quad \lambda=\frac{-V^{'}(\phi)}{\kappa V(\phi)}\,,\quad \Gamma=\dfrac{V(\phi)V^{''}(\phi)}{V^{'}(\phi)^{2}}\,.
\end{equation}
These dimensionless variables satisfy constraint equation presented in Eq.~(\ref{eq:57}), the dynamical system in this case is as follow
\begin{align}
    \frac{dx}{dN} &= \frac{x \left(3 u^2+5 \rho ^2 u-y^2 \left(\sqrt{6} \lambda  (u-2) x+15 u+6 x^2\right)+21 u x^2+9 u+2 \rho ^2 x^2+6 x^4-6 x^2\right)}{u (u+8)+4 x^2}\,,\label{eq:model_2_case_2_dx_dN}\\
    \frac{dy}{dN} &= -y \left(\frac{-3 u^2-2 u \left(2 \rho ^2+9 x^2+6\right)+u y^2 \left(\sqrt{6} \lambda  x+12\right)-2 x^2 \left(\rho ^2+3 x^2-3 y^2+3\right)}{u (u+8)+4 x^2}+\sqrt{\frac{3}{2}} \lambda  x\right)\,,\\
    \frac{du}{dN} &= \frac{u \left(3 u^2+9 \rho ^2 u-y^2 \left(\sqrt{6} \lambda  (u-10) x+27 u+6 x^2\right)+33 u x^2-3 u+2 \rho ^2 x^2+6 x^4-54 x^2\right)}{u (u+8)+4 x^2}\,,\\
    \frac{d\rho}{dN} &= \frac{\rho  \left(u^2+2 u \left(2 \rho ^2+9 x^2-2\right)-u y^2 \left(\sqrt{6} \lambda  x+12\right)+2 x^2 \left(\rho ^2+3 x^2-3 y^2-1\right)\right)}{u (u+8)+4 x^2}\,,\\
    \frac{d\lambda}{dN} &= -\sqrt{6}(\Gamma-1)\lambda^{2}x\,.\label{eq:model_2_case_2_dl_dN}
\end{align}

The critical points with value of deceleration parameter and $\omega_{tot}$ for dynamical system in Eqs.~(\ref{eq:model_2_case_2_dx_dN}--\ref{eq:model_2_case_2_dl_dN}) are presented in Table~\ref{TABLE-XI}. From table observations in Model-II for critical point $A$ we get different positive deceleration parameter value for different values of $\alpha$. For $\alpha=2$ we are getting $q=\frac{2}{3}$ and $\omega_{tot}=\frac{1}{9}$. The critical points $D$, $E$ represent the de Sitter solution to the system and defined for $\lambda=0$. While critical points $H$ and $H$ deliver deceleration parameter value $q=-1+\frac{\lambda^2}{2}$ which may describe dark energy dominated era of the universe. Critical points $F$ and $G$ represent cold dark matter dominated era with $\omega_{tot}=0$. The critical points $J$ and $K$ are defined for $\lambda=2$ and describe radiation dominated era of the universe evolution. The critical points $B$ and $C$ behaves as stiff matter with $\omega_{tot}=1$.

\begin{table}[H]
\caption{ Critical points for the dynamical system corresponding to updated calculations for Model-I} 
\centering 
\begin{tabular}{|c|c|c|c|c|c|c|} 
\hline\hline 
Name of Critical Point & $x_{c}$ & $y_{c}$ & $u_{c}$ & $\rho_{c}$ & Deceleration Parameter ($q$) & $\omega_{tot}$\\ [0.5ex] 
\hline\hline 
$A$ & $0$ & $0$ & $1$ & $0$ & $\frac{2}{3}$ & $\frac{1}{9}$\\
\hline
$B$ & $1$ & $0$ & $0$ & $0$ & $2$ & $1$\\
\hline
$C$ & $-1$ & $0$ & $0$ & $0$ & $2$ & $1$\\
\hline
$D$, in this case $\lambda=0$ & $\sigma , \sigma ^3-3 \sigma \neq 0$ & $\sqrt{\sigma ^2+1}$ & $-2 \sigma ^2$ & $0$ & $-1$ & $-1$\\
\hline
$E$, in this case $\lambda=0$ & $\sigma , \sigma ^3-3 \sigma \neq 0$ & $-\sqrt{\sigma ^2+1}$ & $-2 \sigma ^2$ & $0$ & $-1$ & $-1$\\
\hline
$F$ & $\frac{\sqrt{\frac{3}{2}}}{\lambda }$ & $\sqrt{\frac{3}{2}} \sqrt{\frac{1}{\lambda ^2}}$ & $0$ & $0$ & $\frac{1}{2}$ & $0$\\
\hline
$G$ & $\frac{\sqrt{\frac{3}{2}}}{\lambda }$ & $-\sqrt{\frac{3}{2}} \sqrt{\frac{1}{\lambda ^2}}$ & $0$ & $0$ & $\frac{1}{2}$ & $0$\\
\hline
$H$ & $\frac{\lambda }{\sqrt{6}}$ & $\sqrt{1-\frac{\lambda ^2}{6}}$ & $0$ & $0$& $\frac{1}{2} \left(\lambda ^2-2\right)$ & $-1+\frac{\lambda^2}{3}$\\
\hline
$I$ & $\frac{\lambda }{\sqrt{6}}$ & $-\sqrt{1-\frac{\lambda ^2}{6}}$ & $0$ & $0$& $\frac{1}{2} \left(\lambda ^2-2\right)$ & $-1+\frac{\lambda^2}{3}$\\
\hline
$J$, in this case $\lambda=2$ &  $\sqrt{\frac{2}{3}}$ & $-\frac{1}{\sqrt{3}}$ & $0$ & $0$ & $1$ & $\frac{1}{3}$\\
\hline
$K$, in this case $\lambda=2$ & $\sqrt{\frac{2}{3}}$ & $\frac{1}{\sqrt{3}}$ & $0$ & $0$ & $1$ & $\frac{1}{3}$\\
[1ex] 
\hline 
\end{tabular}
\label{TABLE-XI}
\end{table}

\begin{table}[H]
\caption{ eigenvalues and Stability of Eigenvalue at Corresponding Critical Point.} 
\centering 
\begin{tabular}{|c|c|c|} 
\hline 
Name of Critical Point & Corresponding Eigenvalues & Stability \\ [0.5ex] 
\hline 
$A$ & $\left\{\frac{5}{3},\frac{4}{3},-\frac{1}{3},\frac{1}{3}\right\}$ & Unstable\\
\hline
$B$ & $\left\{-12,3,1,\frac{1}{2} \left(6-\sqrt{6} \lambda \right)\right\}$ & Unstable\\
\hline
$C$ & $\left\{-12,3,1,\frac{1}{2} \left(\sqrt{6} \lambda +6\right)\right\}$ & Unstable\\
\hline
$D$ & $\{0,-3,-3,-2\}$ & Stable \\
\hline
$E$ &$\{0,-3,-3,-2\}$ & Stable\\
\hline
$F$ & $\left\{-6,-\frac{1}{2},\frac{3 \left(-\lambda ^2-\sqrt{24 \lambda ^2-7 \lambda ^4}\right)}{4 \lambda ^2},\frac{3 \left(\sqrt{24 \lambda ^2-7 \lambda ^4}-\lambda ^2\right)}{4 \lambda ^2}\right\}$ & Stable for $-2 \sqrt{\frac{6}{7}}\leq \lambda <-\sqrt{3}\lor \sqrt{3}<\lambda \leq 2 \sqrt{\frac{6}{7}}$\\
\hline
$G$ & $\left\{-6,-\frac{1}{2},\frac{3 \left(-\lambda ^2-\sqrt{24 \lambda ^2-7 \lambda ^4}\right)}{4 \lambda ^2},\frac{3 \left(\sqrt{24 \lambda ^2-7 \lambda ^4}-\lambda ^2\right)}{4 \lambda ^2}\right\}$ & Stable for $-2 \sqrt{\frac{6}{7}}\leq \lambda <-\sqrt{3}\lor \sqrt{3}<\lambda \leq 2 \sqrt{\frac{6}{7}}$\\
\hline
$H$ & $\left\{-2 \lambda ^2,\frac{1}{2} \left(\lambda ^2-6\right),\frac{1}{2} \left(\lambda ^2-4\right),\lambda ^2-3\right\}$ & Stable for $-\sqrt{3}<\lambda <0\lor 0<\lambda <\sqrt{3}$\\
\hline
$I$ & $\left\{-2 \lambda ^2,\frac{1}{2} \left(\lambda ^2-6\right),\frac{1}{2} \left(\lambda ^2-4\right),\lambda ^2-3\right\}$ & Stable for $-\sqrt{3}<\lambda <0\lor 0<\lambda <\sqrt{3}$\\
\hline
$J$ &$\{-8,-1,1,0\}$ & Unstable \\
\hline
$K$ &$\{-8,-1,1,0\}$ & Unstable\\
[1ex] 
\hline 
\end{tabular}
\label{TABLE-XII}
\end{table}

The stability conditions for this case $\alpha=2$ are presented in Table~\ref{TABLE-XII}. Table observations conclude that critical points $H$ and $I$ show stability for parametric condition $-\sqrt{3}<\lambda <0\lor 0<\lambda <\sqrt{3}$ and can describe dark energy dominated era. The critical points $F$ and $G$ show stability at $-2 \sqrt{\frac{6}{7}}\leq \lambda <-\sqrt{3}\lor \sqrt{3}<\lambda \leq 2 \sqrt{\frac{6}{7}}$ and describe cold dark matter dominated era. The critical point $A$, $B$ and $C$ with deceleration parameter in positive range are unstable since eigenvalues are with both positive and negative signature. The critical points $D$ and $E$ represent de Sitter solution and show stable behaviour. The critical points $J$ and $K$ are representing radiation dominated era at $\lambda=2$ and are unstable in nature.\\

\begin{figure}[H]
    \centering
    \includegraphics[width=75mm]{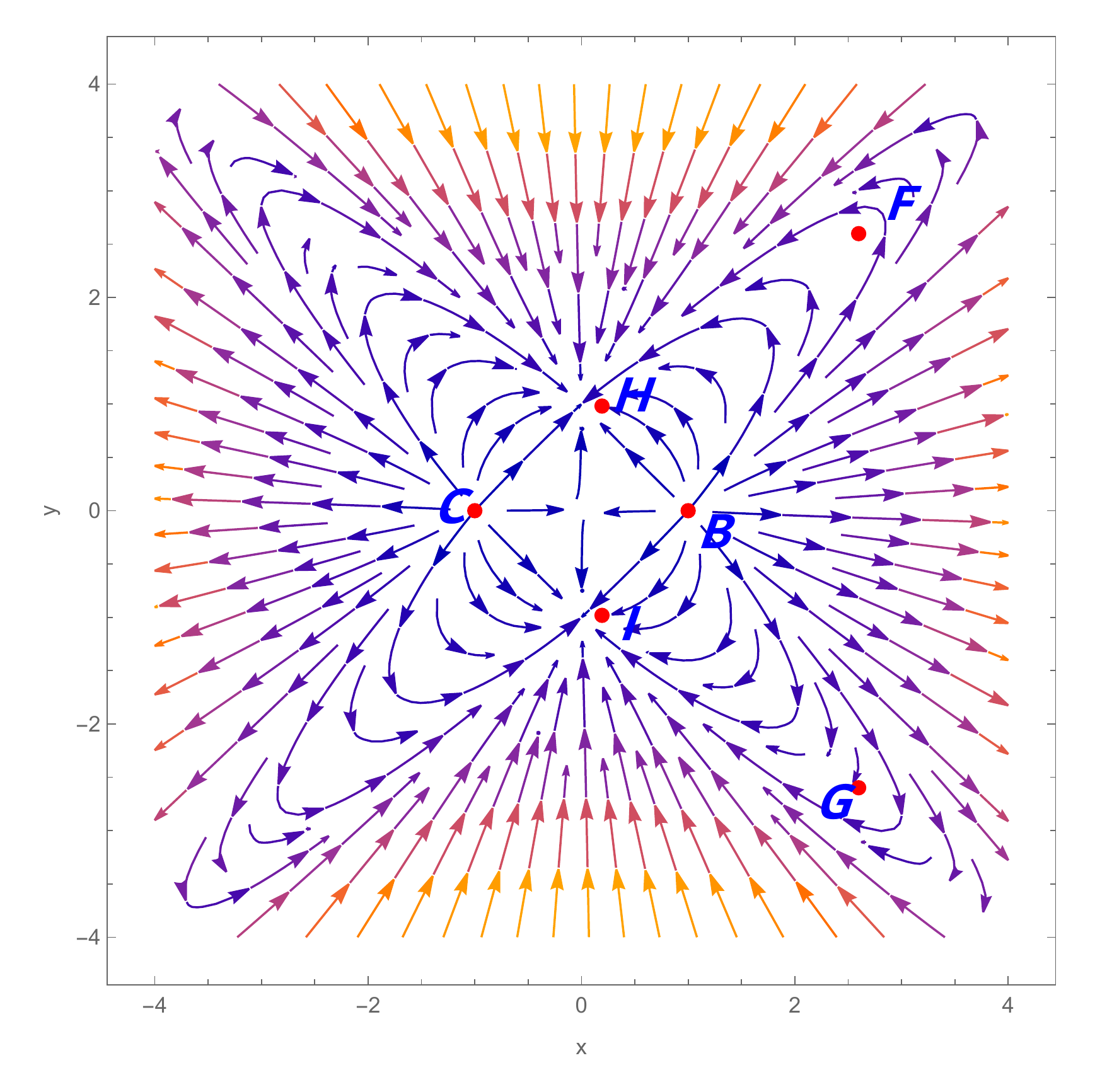}
    \includegraphics[width=75mm]{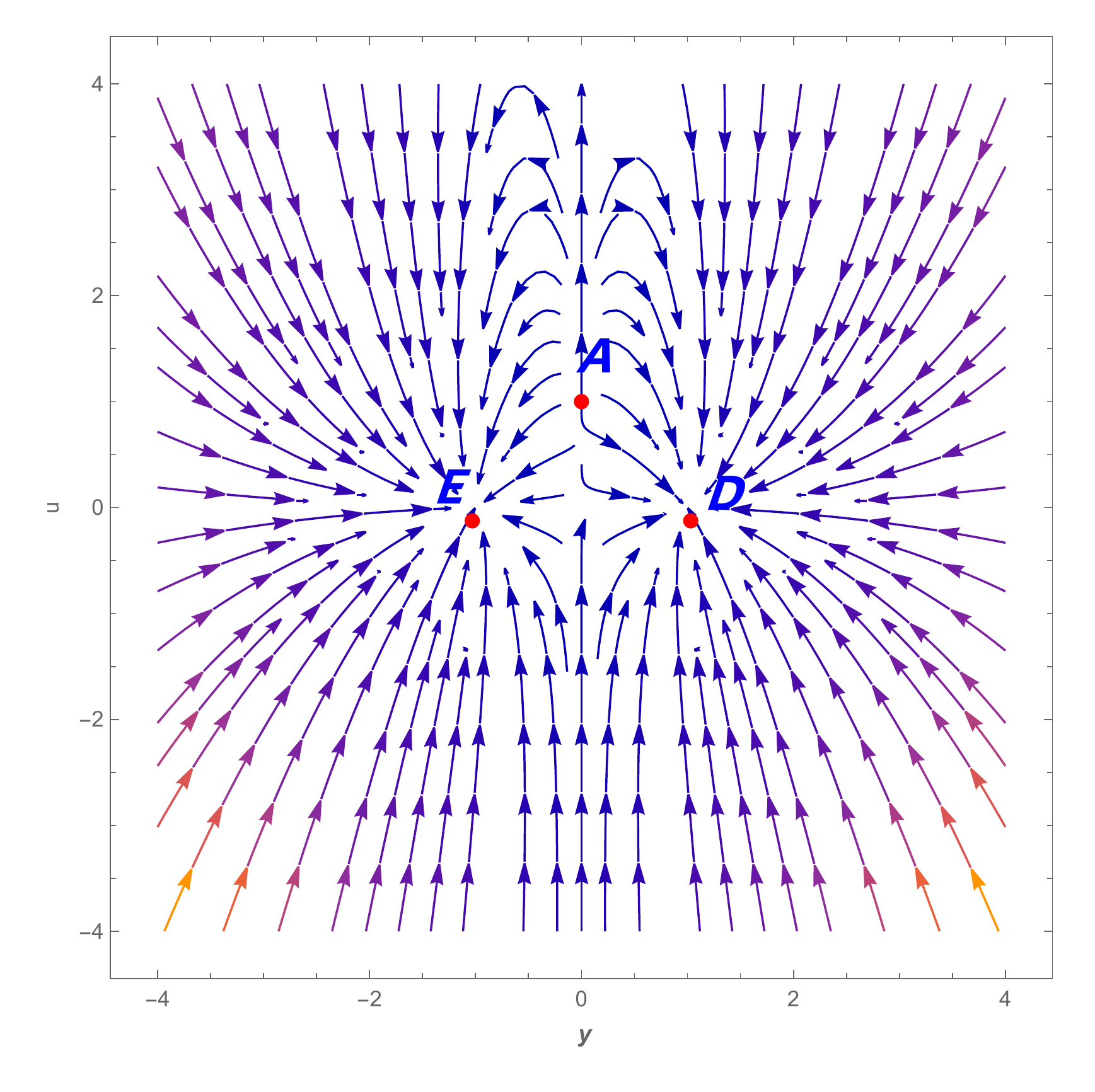}
    \includegraphics[width=75mm]{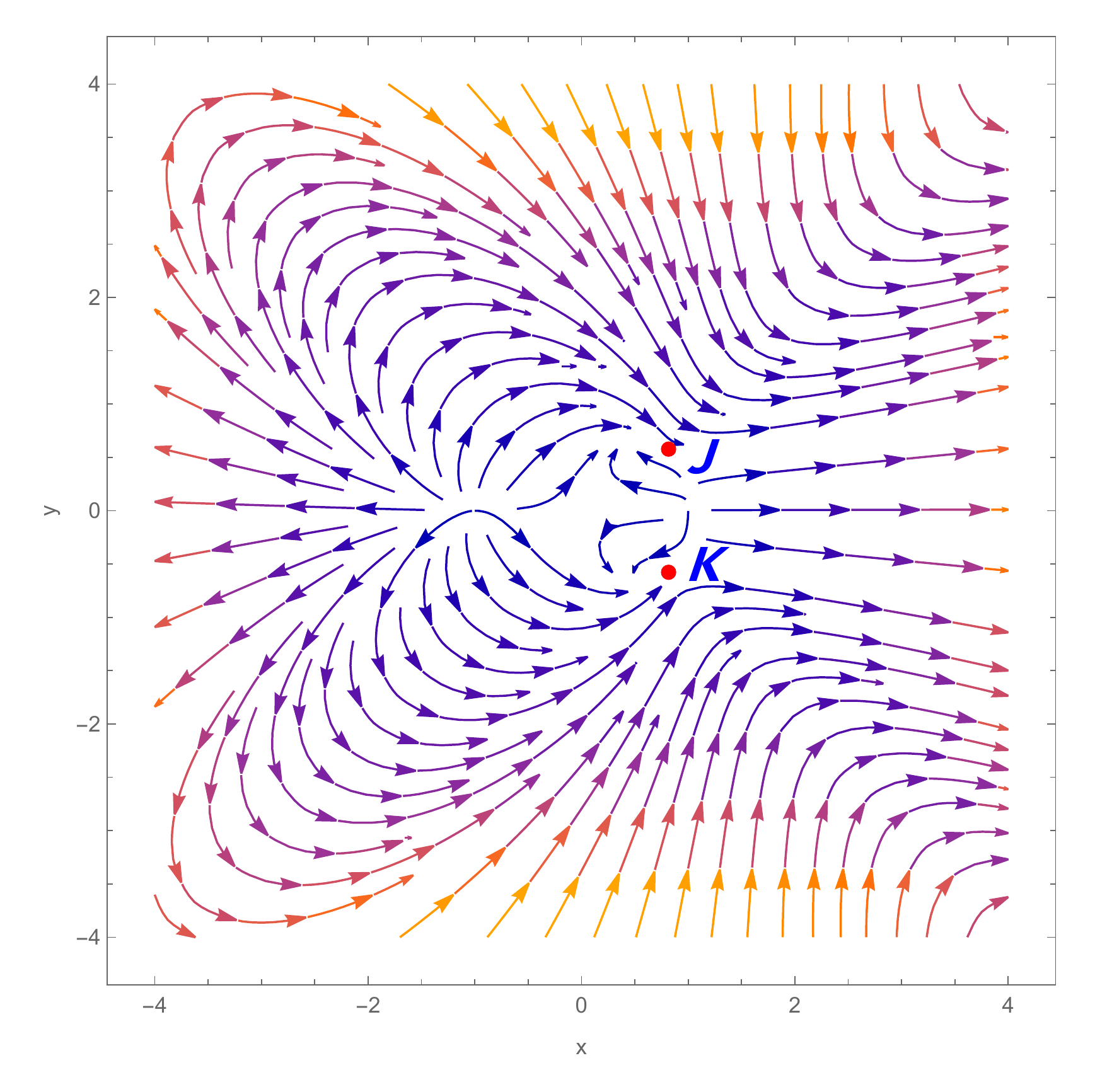}
    \caption{Phase portrait upper left is plotted for the parametric values $u=0$, $\rho=0$ and $\lambda=\sqrt{\frac{2}{9}}$ and for upper right side plot parametric values are  $x=0$, $\rho=0$, $\sigma=\frac{1}{4}$. The lower phase portrait is for the parametric values $u=0$, $\rho=0$, $\lambda=2$.} \label{Fig8}
\end{figure}

The phase space diagram for dynamical system Eqs.~(\ref{eq:model_2_case_2_dx_dN}--\ref{eq:model_2_case_2_dl_dN}) are plotted in Fig.~\ref{Fig8}. The critical points $D$ and $E$ showing attracting stable de Sitter solution and represent dark energy dominated era. While at critical point $A$ phase space trajectories are moving away hence unstable addressing saddle point behaviour. Similarly critical points $J$ and $K$ are showing unstable behaviour representing radiation dominated era. The upper right plot is presented for critical points $B$, $C$, $H$, $I$, $F$ and $G$. The critical points $B$ and $C$ behaves as unstable node with respect to the existence of positive  eigenvalues. The phase space trajectories are moving away from critical points $F$ and $G$, these critical points represent unstable behaviour and critical points $H$ and $I$ display attracting behaviour of phase space trajectories and show consistent with current observational studies.

\section{Conclusion}\label{sec:conclusion}

In this work we have explored the cosmological dynamics of dark energy through the prism of scalar-torsion gravity \cite{Paliathanasis:2021nqa,Leon:2022oyy} in the context of power-law couplings with the kinetic term. In particular, we have studied one nontrivial extension of the recently proposed teleparallel analogue of Horndeski gravity \cite{Bahamonde:2019shr}. This new frameworks offers a pathway to circumvent the severe constraints on the speed of gravitational wave constraint. This is possible due to the lower order nature of TG where the curvature associated with the Levi-Civita connection is exchanged with the torsion from the teleparallel connection. In the present case, we are interested in extending the analysis of dynamical systems by looking into how power-law-like couplings with the nonvanishing terms for background FLRW cosmologies. The teleparallel analogue of Horndeski theory offers a broad spectrum of possible extensions to regular Horndeski gravity and thus the space of possible functional forms are vast. This work goes some way to elucidating the behaviour that such functions adhere to.

For our FLRW background cosmology, we explore these models in the presence of both radiation and cold dark matter through the density parameters in Eq.~(\ref{eq:41}). The Friedmann equations in Eqs.~(\ref{eq:30},\ref{eq:31}) and Klein-Gordon equation in Eq.~(\ref{eq:kg_eq}) directly lead to a set of autonomous equations for each of the models under investigation. These are then used in each case to derive the critical points of the particular cosmologies from which we can expose the model behaviour using the dynamical analysis in the parameter phase space. This also opens the doorway to understanding the stability of the models in question.

In the first model for the action in Eq.~(\ref{eq:action_model_1}), we utilize the dynamical variables defined in Eq.~(\ref{eq:49}), which using the constraint in Eq.~(\ref{eq:50}) together with the equations of motion, are then used to derive the system of autonomous equations given in Eqs.~(\ref{eq:dx_dN_model_1}--\ref{eq:dl_dN_model_1}) which express the behaviour of the model in phase space. The critical points are then arrived at by imposing that each of these derivatives vanishes. These first order equations of motion of the dynamical variables are represented as derivatives with respect to $N = \ln a$ which shows the behaviour of the system in a more direct way. The result of this analysis is shown in Table~\ref{TABLE-I}. In this table we show the values of the dynamical variables at which these critical points occur together with the value of the deceleration and EoS parameters which already show an indication of the cosmological behaviour at those points in the evolution of the cosmological model. Each of these critical points is then further analyzed for their stability in Table~\ref{TABLE-II}. In some circumstances, stability occurs for a smaller change of parameter values as described in the last column of the table. For transparency we also show the corresponding Eigenvalues at each critical point. In Fig.~\ref{Fig1}, we show the phase portraits of this model for four specific examples of representative parameter values. In these plots the nature of the critical points is further exposed through their impact on the evolution contours.

We further probe the behaviour of this model in the specific cases of the kinetic term being linear and quadratic which represent the first cases that a Taylor expansion would open to. We do this in Sec.~\ref{sec:model_1_case_1} and Sec.~\ref{sec:model_1_case_2} respectively. We also show the phase portraits for these cases in Fig.~\ref{Fig2} and Figs.~\ref{Fig3}, \ref{Fig4} for the two respective cases. Finally, we show the nuanced critical points for both cases in Tables~\ref{TABLE-III}, \ref{TABLE-V} and their respective stability in Tables.~\ref{TABLE-IV}, \ref{TABLE-VI}.

The second model we explore the coupling between a power-law kinetic term and $I_2$ scalar written in Eq.~(\ref{eq:I_2_scalar}). This scalar represents the only nonvanishing term that is linear in its contractions with the torsion tensor. In this case, we take the action Eq.~(\ref{eq:action_model_2}) where TEGR and the canonical scalar field are complemented by this new coupling term together with the matter and radiation contributions. This leads directly to the Friedmann equations in Eq.~(\ref{eq:70}, \ref{eq:71}), and the Klein-Gordon equation in Eq.~(\ref{eq:72}). Now, by defining the dynamical variables in Eq.~(\ref{eq:model_2_dynamic_var}) we can explore the dynamical system that the background cosmology represents through the system of linear autonomous differential equations given in Eqs.~(\ref{eq:model_2_dx_dN}--\ref{eq:model_2_dl_dN}). By performing a similar critical point analysis as in first case, we find the critical points as listed in Table~\ref{TABLE-VII}. These are then further analyzed for their stability nature in Table~\ref{TABLE-VIII}. As in the first case, we find a rich structure of critical points for the various model parameter values, which we show through the phase portrait in Fig.~\ref{Fig5}. As in the first model, we consider the cases where the power index takes linear and quadratic forms in the model action. For the linear case, the dynamical system is represented by Eqs.~(\ref{eq:model_2_csae_1_dx_dN}--\ref{eq:model_2_csae_1_dl_dN}) which produce the critical points in Table~\ref{TABLE-IX} with natures shown in Table~\ref{TABLE-X}. This interesting scenario produces the phase portraits given in Figs.~(\ref{Fig6}--\ref{Fig7}). On the other hand, the quadratic case is represented through the dynamical system given in Eqs.~(\ref{eq:model_2_case_2_dx_dN}--\ref{eq:model_2_case_2_dl_dN}). The corresponding critical point analysis produces Table~\ref{TABLE-XI} which have stability conditions described in Table~\ref{TABLE-XII}. The final phase portraits for this case are then shown in Fig.~\ref{Fig8}.

The action presented Eq.~(\ref{action}) produces the most general second order theory that contains only one scalar field in TG. Naturally, this will produce a wealth of dynamics for which reason we explore two prominent models and expose their dynamical systems in the ensuing sections. While the models produce a vast array of critical points in Tables.~\ref{TABLE-I},\ref{TABLE-VII}, these are not all realized in each possible evolution of the individual models. Another aspect to highlight is the impact of boundary conditions on these potential evolution histories, that is, some critical points will not be accessible for certain boundary. This is further highlighted through the phase portrait examples where some critical points are common to all cosmological histories while other only appear for some iterations of the system. Thus, as explored in the interesting review in Ref.~\cite{Bahamonde:2017ize}, while some behaviours remain common to the model produced by one of the actions such as a de Sitter future critical point, others only partially appear in the broad range of possible cosmological evolution histories available.

If we glance at a study made in Ref.\cite{manuel2021cosmological}, we can easily compare the cosmological implications and stability conditions of critical points from Table-1 in Ref. \cite{manuel2021cosmological} and Table \ref{TABLE-I} for the Model-I general $\alpha$ case. From this comparison, we can note that we get four more critical points than in Ref.\cite{manuel2021cosmological}, this may be possible because of the model construction and background formalism in the teleparallel Horndeski theory. This comparison allows us to know more about minor differences; in our study we get four critical points which are able to describe the dark energy era that is critical points  H, I D, E. From these, the two critical points, D and E, are the stable de-Sitter solution (the study for the  Model-II, Table \ref{TABLE-VII} also showing similar comparison conclusion with sixteen critical points), but in Ref.\cite{manuel2021cosmological}, we will likely deal with four critical points explaining the dark energy era with only one de-Sitter solution, critical point "l." With this progress we also have the number of critical points describing the matter-dominated solution with $\omega=0$ is more than in Ref.\cite{manuel2021cosmological}. These novel critical points plays an important role to get more clarity in the description of the matter and the dark energy dominated era. Also, we can compare our analysis with the study of various dark energy models in the modified theory of gravity in Ref.\cite{Bahamonde:2017ize}.

\section*{Acknowledgements}
SAK acknowledges the financial support provided by University Grants Commission (UGC) through Junior Research Fellowship (UGC Ref. No.: 191620205335) to carry out the research work. BM acknowledges IUCAA, Pune, India for hospitality and support during an academic visit where a part of this work has been accomplished. JLS would like to acknowledge funding from Cosmology@MALTA which is supported by the University of Malta. The authors are thankful to the honorable referee for the comments and suggestions to improve the quality of the paper.

\bibliographystyle{utphys}
\bibliography{references}

\providecommand{\href}[2]{#2}\begingroup\raggedright\begin{thebibliography}{10}

\bibitem{SupernovaSearchTeam:1998fmf}
{\bf Supernova Search Team} Collaboration, A.~G. Riess {\em et al.},
  ``{Observational evidence from supernovae for an accelerating universe and a
  cosmological constant},'' \href{http://dx.doi.org/10.1086/300499}{{\em
  Astron. J.} {\bf 116} (1998)  1009--1038},
  \href{http://arxiv.org/abs/astro-ph/9805201}{{\tt arXiv:astro-ph/9805201}}.

\bibitem{SupernovaCosmologyProject:1998vns}
{\bf Supernova Cosmology Project} Collaboration, S.~Perlmutter {\em et al.},
  ``{Measurements of $\Omega$ and $\Lambda$ from 42 high redshift
  supernovae},'' \href{http://dx.doi.org/10.1086/307221}{{\em Astrophys. J.}
  {\bf 517} (1999)  565--586},
  \href{http://arxiv.org/abs/astro-ph/9812133}{{\tt arXiv:astro-ph/9812133}}.

\bibitem{Abdalla:2022yfr}
E.~Abdalla {\em et al.}, ``{Cosmology Intertwined: A Review of the Particle
  Physics, Astrophysics, and Cosmology Associated with the Cosmological
  Tensions and Anomalies},'' in {\em {2022 Snowmass Summer Study}}.
\newblock 3, 2022.
\newblock \href{http://arxiv.org/abs/2203.06142}{{\tt arXiv:2203.06142
  [astro-ph.CO]}}.

\bibitem{DiValentino:2021izs}
E.~Di~Valentino, O.~Mena, S.~Pan, L.~Visinelli, W.~Yang, A.~Melchiorri, D.~F.
  Mota, A.~G. Riess, and J.~Silk, ``{In the realm of the Hubble
  tension\textemdash{}a review of solutions},''
  \href{http://dx.doi.org/10.1088/1361-6382/ac086d}{{\em Class. Quant. Grav.}
  {\bf 38} (2021) no.~15, 153001}, \href{http://arxiv.org/abs/2103.01183}{{\tt
  arXiv:2103.01183 [astro-ph.CO]}}.

\bibitem{Brout:2022vxf}
D.~Brout {\em et al.}, ``{The Pantheon+ Analysis: Cosmological Constraints},''
  \href{http://arxiv.org/abs/2202.04077}{{\tt arXiv:2202.04077 [astro-ph.CO]}}.

\bibitem{Bernal:2016gxb}
J.~L. Bernal, L.~Verde, and A.~G. Riess, ``{The trouble with $H_0$},''
  \href{http://dx.doi.org/10.1088/1475-7516/2016/10/019}{{\em JCAP} {\bf 10}
  (2016)  019}, \href{http://arxiv.org/abs/1607.05617}{{\tt arXiv:1607.05617
  [astro-ph.CO]}}.

\bibitem{Benisty:2021cmq}
D.~Benisty and A.-C. Davis, ``{Dark energy interactions near the Galactic
  Center},'' \href{http://dx.doi.org/10.1103/PhysRevD.105.024052}{{\em Phys.
  Rev. D} {\bf 105} (2022) no.~2, 024052},
  \href{http://arxiv.org/abs/2108.06286}{{\tt arXiv:2108.06286 [astro-ph.CO]}}.

\bibitem{Riess:2021jrx}
A.~G. Riess {\em et al.}, ``{A Comprehensive Measurement of the Local Value of
  the Hubble Constant with 1 km/s/Mpc Uncertainty from the Hubble Space
  Telescope and the SH0ES Team},'' \href{http://arxiv.org/abs/2112.04510}{{\tt
  arXiv:2112.04510 [astro-ph.CO]}}.

\bibitem{Wong:2019kwg}
K.~C. Wong {\em et al.}, ``{H0LiCOW \textendash{} XIII. A 2.4 per cent
  measurement of H0 from lensed quasars: 5.3\ensuremath{\sigma} tension between
  early- and late-Universe probes},''
  \href{http://dx.doi.org/10.1093/mnras/stz3094}{{\em Mon. Not. Roy. Astron.
  Soc.} {\bf 498} (2020) no.~1, 1420--1439},
  \href{http://arxiv.org/abs/1907.04869}{{\tt arXiv:1907.04869 [astro-ph.CO]}}.

\bibitem{Planck:2018vyg}
{\bf Planck} Collaboration, N.~Aghanim {\em et al.}, ``{Planck 2018 results.
  VI. Cosmological parameters},''
  \href{http://dx.doi.org/10.1051/0004-6361/201833910}{{\em Astron. Astrophys.}
  {\bf 641} (2020)  A6}, \href{http://arxiv.org/abs/1807.06209}{{\tt
  arXiv:1807.06209 [astro-ph.CO]}}. [Erratum: Astron.Astrophys. 652, C4
  (2021)].

\bibitem{DES:2021wwk}
{\bf DES} Collaboration, T.~M.~C. Abbott {\em et al.}, ``{Dark Energy Survey
  Year 3 results: Cosmological constraints from galaxy clustering and weak
  lensing},'' \href{http://dx.doi.org/10.1103/PhysRevD.105.023520}{{\em Phys.
  Rev. D} {\bf 105} (2022) no.~2, 023520},
  \href{http://arxiv.org/abs/2105.13549}{{\tt arXiv:2105.13549 [astro-ph.CO]}}.

\bibitem{DiValentino:2020vhf}
E.~Di~Valentino {\em et al.}, ``{Snowmass2021 - Letter of interest cosmology
  intertwined I: Perspectives for the next decade},''
  \href{http://dx.doi.org/10.1016/j.astropartphys.2021.102606}{{\em Astropart.
  Phys.} {\bf 131} (2021)  102606}, \href{http://arxiv.org/abs/2008.11283}{{\tt
  arXiv:2008.11283 [astro-ph.CO]}}.

\bibitem{DiValentino:2020zio}
E.~Di~Valentino {\em et al.}, ``{Snowmass2021 - Letter of interest cosmology
  intertwined II: The hubble constant tension},''
  \href{http://dx.doi.org/10.1016/j.astropartphys.2021.102605}{{\em Astropart.
  Phys.} {\bf 131} (2021)  102605}, \href{http://arxiv.org/abs/2008.11284}{{\tt
  arXiv:2008.11284 [astro-ph.CO]}}.

\bibitem{DiValentino:2020vvd}
E.~Di~Valentino {\em et al.}, ``{Cosmology intertwined III: $f\sigma_8$ and
  $S_8$},'' \href{http://dx.doi.org/10.1016/j.astropartphys.2021.102604}{{\em
  Astropart. Phys.} {\bf 131} (2021)  102604},
  \href{http://arxiv.org/abs/2008.11285}{{\tt arXiv:2008.11285 [astro-ph.CO]}}.

\bibitem{Sotiriou:2008rp}
T.~P. Sotiriou and V.~Faraoni, ``{f(R) Theories Of Gravity},''
  \href{http://dx.doi.org/10.1103/RevModPhys.82.451}{{\em Rev. Mod. Phys.} {\bf
  82} (2010)  451--497}, \href{http://arxiv.org/abs/0805.1726}{{\tt
  arXiv:0805.1726 [gr-qc]}}.

\bibitem{Clifton:2011jh}
T.~Clifton, P.~G. Ferreira, A.~Padilla, and C.~Skordis, ``{Modified Gravity and
  Cosmology},'' \href{http://dx.doi.org/10.1016/j.physrep.2012.01.001}{{\em
  Phys. Rept.} {\bf 513} (2012)  1--189},
  \href{http://arxiv.org/abs/1106.2476}{{\tt arXiv:1106.2476 [astro-ph.CO]}}.

\bibitem{CANTATA:2021ktz}
{\bf CANTATA} Collaboration, E.~N. Saridakis {\em et al.}, ``{Modified Gravity
  and Cosmology: An Update by the CANTATA Network},''
  \href{http://arxiv.org/abs/2105.12582}{{\tt arXiv:2105.12582 [gr-qc]}}.

\bibitem{Nojiri:2017ncd}
S.~Nojiri, S.~D. Odintsov, and V.~K. Oikonomou, ``{Modified Gravity Theories on
  a Nutshell: Inflation, Bounce and Late-time Evolution},''
  \href{http://dx.doi.org/10.1016/j.physrep.2017.06.001}{{\em Phys. Rept.} {\bf
  692} (2017)  1--104}, \href{http://arxiv.org/abs/1705.11098}{{\tt
  arXiv:1705.11098 [gr-qc]}}.

\bibitem{Faraoni:2008mf}
V.~Faraoni, ``{f(R) gravity: Successes and challenges},'' in {\em {18th SIGRAV
  Conference}}.
\newblock 10, 2008.
\newblock \href{http://arxiv.org/abs/0810.2602}{{\tt arXiv:0810.2602 [gr-qc]}}.

\bibitem{Capozziello:2011et}
S.~Capozziello and M.~De~Laurentis, ``{Extended Theories of Gravity},''
  \href{http://dx.doi.org/10.1016/j.physrep.2011.09.003}{{\em Phys. Rept.} {\bf
  509} (2011)  167--321}, \href{http://arxiv.org/abs/1108.6266}{{\tt
  arXiv:1108.6266 [gr-qc]}}.

\bibitem{misner1973gravitation}
C.~Misner, K.~Thorne, and J.~Wheeler, {\em Gravitation}.
\newblock No.~pt. 3 in Gravitation. W. H. Freeman, 1973.
\newblock \url{https://books.google.com.mt/books?id=w4Gigq3tY1kC}.

\bibitem{Bahamonde:2021gfp}
S.~Bahamonde, K.~F. Dialektopoulos, C.~Escamilla-Rivera, G.~Farrugia, V.~Gakis,
  M.~Hendry, M.~Hohmann, J.~L. Said, J.~Mifsud, and E.~Di~Valentino,
  ``{Teleparallel Gravity: From Theory to Cosmology},''
  \href{http://arxiv.org/abs/2106.13793}{{\tt arXiv:2106.13793 [gr-qc]}}.

\bibitem{Aldrovandi:2013wha}
R.~Aldrovandi and J.~G. Pereira,
  \href{http://dx.doi.org/10.1007/978-94-007-5143-9}{{\em {Teleparallel
  Gravity}: {An Introduction}}}.
\newblock Springer, 2013.

\bibitem{Cai:2015emx}
Y.-F. Cai, S.~Capozziello, M.~De~Laurentis, and E.~N. Saridakis, ``{f(T)
  teleparallel gravity and cosmology},''
  \href{http://dx.doi.org/10.1088/0034-4885/79/10/106901}{{\em Rept. Prog.
  Phys.} {\bf 79} (2016) no.~10, 106901},
  \href{http://arxiv.org/abs/1511.07586}{{\tt arXiv:1511.07586 [gr-qc]}}.

\bibitem{Krssak:2018ywd}
M.~Krssak, R.~J. van~den Hoogen, J.~G. Pereira, C.~G. B\"ohmer, and A.~A.
  Coley, ``{Teleparallel theories of gravity: illuminating a fully invariant
  approach},'' \href{http://dx.doi.org/10.1088/1361-6382/ab2e1f}{{\em Class.
  Quant. Grav.} {\bf 36} (2019) no.~18, 183001},
  \href{http://arxiv.org/abs/1810.12932}{{\tt arXiv:1810.12932 [gr-qc]}}.

\bibitem{Weitzenbock1923}
R.~Weitzenb\"{o}ock, {\em `Invariantentheorie'}.
\newblock Noordhoff, Gronningen, 1923.

\bibitem{Lovelock:1971yv}
D.~Lovelock, ``{The Einstein tensor and its generalizations},''
  \href{http://dx.doi.org/10.1063/1.1665613}{{\em J. Math. Phys.} {\bf 12}
  (1971)  498--501}.

\bibitem{Gonzalez:2015sha}
P.~A. Gonzalez and Y.~Vasquez, ``{Teleparallel Equivalent of Lovelock
  Gravity},'' \href{http://dx.doi.org/10.1103/PhysRevD.92.124023}{{\em Phys.
  Rev. D} {\bf 92} (2015) no.~12, 124023},
  \href{http://arxiv.org/abs/1508.01174}{{\tt arXiv:1508.01174 [hep-th]}}.

\bibitem{Bahamonde:2019shr}
S.~Bahamonde, K.~F. Dialektopoulos, and J.~Levi~Said, ``{Can Horndeski Theory
  be recast using Teleparallel Gravity?},''
  \href{http://dx.doi.org/10.1103/PhysRevD.100.064018}{{\em Phys. Rev. D} {\bf
  100} (2019) no.~6, 064018}, \href{http://arxiv.org/abs/1904.10791}{{\tt
  arXiv:1904.10791 [gr-qc]}}.

\bibitem{Ferraro:2006jd}
R.~Ferraro and F.~Fiorini, ``{Modified teleparallel gravity: Inflation without
  inflaton},'' \href{http://dx.doi.org/10.1103/PhysRevD.75.084031}{{\em Phys.
  Rev. D} {\bf 75} (2007)  084031},
  \href{http://arxiv.org/abs/gr-qc/0610067}{{\tt arXiv:gr-qc/0610067}}.

\bibitem{Ferraro:2008ey}
R.~Ferraro and F.~Fiorini, ``{On Born-Infeld Gravity in Weitzenbock
  spacetime},'' \href{http://dx.doi.org/10.1103/PhysRevD.78.124019}{{\em Phys.
  Rev. D} {\bf 78} (2008)  124019}, \href{http://arxiv.org/abs/0812.1981}{{\tt
  arXiv:0812.1981 [gr-qc]}}.

\bibitem{Bengochea:2008gz}
G.~R. Bengochea and R.~Ferraro, ``{Dark torsion as the cosmic speed-up},''
  \href{http://dx.doi.org/10.1103/PhysRevD.79.124019}{{\em Phys. Rev. D} {\bf
  79} (2009)  124019}, \href{http://arxiv.org/abs/0812.1205}{{\tt
  arXiv:0812.1205 [astro-ph]}}.

\bibitem{Linder:2010py}
E.~V. Linder, ``{Einstein's Other Gravity and the Acceleration of the
  Universe},'' \href{http://dx.doi.org/10.1103/PhysRevD.81.127301}{{\em Phys.
  Rev. D} {\bf 81} (2010)  127301}, \href{http://arxiv.org/abs/1005.3039}{{\tt
  arXiv:1005.3039 [astro-ph.CO]}}. [Erratum: Phys.Rev.D 82, 109902 (2010)].

\bibitem{Chen:2010va}
S.-H. Chen, J.~B. Dent, S.~Dutta, and E.~N. Saridakis, ``{Cosmological
  perturbations in f(T) gravity},''
  \href{http://dx.doi.org/10.1103/PhysRevD.83.023508}{{\em Phys. Rev. D} {\bf
  83} (2011)  023508}, \href{http://arxiv.org/abs/1008.1250}{{\tt
  arXiv:1008.1250 [astro-ph.CO]}}.

\bibitem{Bahamonde:2019zea}
S.~Bahamonde, K.~Flathmann, and C.~Pfeifer, ``{Photon sphere and perihelion
  shift in weak $f(T)$ gravity},''
  \href{http://dx.doi.org/10.1103/PhysRevD.100.084064}{{\em Phys. Rev. D} {\bf
  100} (2019) no.~8, 084064}, \href{http://arxiv.org/abs/1907.10858}{{\tt
  arXiv:1907.10858 [gr-qc]}}.

\bibitem{Paliathanasis:2021nqa}
A.~Paliathanasis, ``{Dynamics in Interacting Scalar-Torsion Cosmology},''
  \href{http://dx.doi.org/10.3390/universe7070244}{{\em Universe} {\bf 7}
  (2021) no.~7, 244}, \href{http://arxiv.org/abs/2107.05880}{{\tt
  arXiv:2107.05880 [gr-qc]}}.

\bibitem{Leon:2022oyy}
G.~Leon, A.~Paliathanasis, E.~N. Saridakis, and S.~Basilakos, ``{Unified dark
  sectors in scalar-torsion theories of gravity},''
  \href{http://arxiv.org/abs/2203.14866}{{\tt arXiv:2203.14866 [gr-qc]}}.

\bibitem{Duchaniya:2022rqu}
L.~K. Duchaniya, S.~V. Lohakare, B.~Mishra, and S.~K. Tripathy, ``{Dynamical
  stability analysis of accelerating f(T) gravity models},''
  \href{http://dx.doi.org/10.1140/epjc/s10052-022-10406-w}{{\em Eur. Phys. J.
  C} {\bf 82} (2022) no.~5, 448}, \href{http://arxiv.org/abs/2202.08150}{{\tt
  arXiv:2202.08150 [gr-qc]}}.

\bibitem{Farrugia:2016qqe}
G.~Farrugia and J.~Levi~Said, ``{Stability of the flat FLRW metric in $f(T)$
  gravity},'' \href{http://dx.doi.org/10.1103/PhysRevD.94.124054}{{\em Phys.
  Rev. D} {\bf 94} (2016) no.~12, 124054},
  \href{http://arxiv.org/abs/1701.00134}{{\tt arXiv:1701.00134 [gr-qc]}}.

\bibitem{Finch:2018gkh}
A.~Finch and J.~L. Said, ``{Galactic Rotation Dynamics in f(T) gravity},''
  \href{http://dx.doi.org/10.1140/epjc/s10052-018-6028-1}{{\em Eur. Phys. J. C}
  {\bf 78} (2018) no.~7, 560}, \href{http://arxiv.org/abs/1806.09677}{{\tt
  arXiv:1806.09677 [astro-ph.GA]}}.

\bibitem{Farrugia:2016xcw}
G.~Farrugia, J.~Levi~Said, and M.~L. Ruggiero, ``{Solar System tests in f(T)
  gravity},'' \href{http://dx.doi.org/10.1103/PhysRevD.93.104034}{{\em Phys.
  Rev. D} {\bf 93} (2016) no.~10, 104034},
  \href{http://arxiv.org/abs/1605.07614}{{\tt arXiv:1605.07614 [gr-qc]}}.

\bibitem{Iorio:2012cm}
L.~Iorio and E.~N. Saridakis, ``{Solar system constraints on f(T) gravity},''
  \href{http://dx.doi.org/10.1111/j.1365-2966.2012.21995.x}{{\em Mon. Not. Roy.
  Astron. Soc.} {\bf 427} (2012)  1555},
  \href{http://arxiv.org/abs/1203.5781}{{\tt arXiv:1203.5781 [gr-qc]}}.

\bibitem{Deng:2018ncg}
X.-M. Deng, ``{Probing f(T) gravity with gravitational time advancement},''
  \href{http://dx.doi.org/10.1088/1361-6382/aad391}{{\em Class. Quant. Grav.}
  {\bf 35} (2018) no.~17, 175013}.

\bibitem{Copeland:2006wr}
E.~J. Copeland, M.~Sami, and S.~Tsujikawa, ``{Dynamics of dark energy},''
  \href{http://dx.doi.org/10.1142/S021827180600942X}{{\em Int. J. Mod. Phys. D}
  {\bf 15} (2006)  1753--1936}, \href{http://arxiv.org/abs/hep-th/0603057}{{\tt
  arXiv:hep-th/0603057}}.

\bibitem{Bamba:2012cp}
K.~Bamba, S.~Capozziello, S.~Nojiri, and S.~D. Odintsov, ``{Dark energy
  cosmology: the equivalent description via different theoretical models and
  cosmography tests},'' \href{http://dx.doi.org/10.1007/s10509-012-1181-8}{{\em
  Astrophys. Space Sci.} {\bf 342} (2012)  155--228},
  \href{http://arxiv.org/abs/1205.3421}{{\tt arXiv:1205.3421 [gr-qc]}}.

\bibitem{Horndeski:1974wa}
G.~W. Horndeski, ``{Second-order scalar-tensor field equations in a
  four-dimensional space},'' \href{http://dx.doi.org/10.1007/BF01807638}{{\em
  Int. J. Theor. Phys.} {\bf 10} (1974)  363--384}.

\bibitem{Kobayashi:2019hrl}
T.~Kobayashi, ``{Horndeski theory and beyond: a review},''
  \href{http://dx.doi.org/10.1088/1361-6633/ab2429}{{\em Rept. Prog. Phys.}
  {\bf 82} (2019) no.~8, 086901}, \href{http://arxiv.org/abs/1901.07183}{{\tt
  arXiv:1901.07183 [gr-qc]}}.

\bibitem{Deffayet:2009wt}
C.~Deffayet, G.~Esposito-Farese, and A.~Vikman, ``{Covariant Galileon},''
  \href{http://dx.doi.org/10.1103/PhysRevD.79.084003}{{\em Phys. Rev. D} {\bf
  79} (2009)  084003}, \href{http://arxiv.org/abs/0901.1314}{{\tt
  arXiv:0901.1314 [hep-th]}}.

\bibitem{Kobayashi:2011nu}
T.~Kobayashi, M.~Yamaguchi, and J.~Yokoyama, ``{Generalized G-inflation:
  Inflation with the most general second-order field equations},''
  \href{http://dx.doi.org/10.1143/PTP.126.511}{{\em Prog. Theor. Phys.} {\bf
  126} (2011)  511--529}, \href{http://arxiv.org/abs/1105.5723}{{\tt
  arXiv:1105.5723 [hep-th]}}.

\bibitem{Bahamonde:2019ipm}
S.~Bahamonde, K.~F. Dialektopoulos, V.~Gakis, and J.~Levi~Said, ``{Reviving
  Horndeski theory using teleparallel gravity after GW170817},''
  \href{http://dx.doi.org/10.1103/PhysRevD.101.084060}{{\em Phys. Rev. D} {\bf
  101} (2020) no.~8, 084060}, \href{http://arxiv.org/abs/1907.10057}{{\tt
  arXiv:1907.10057 [gr-qc]}}.

\bibitem{Bahamonde:2020cfv}
S.~Bahamonde, K.~F. Dialektopoulos, M.~Hohmann, and J.~Levi~Said,
  ``{Post-Newtonian limit of Teleparallel Horndeski gravity},''
  \href{http://dx.doi.org/10.1088/1361-6382/abc441}{{\em Class. Quant. Grav.}
  {\bf 38} (2020) no.~2, 025006}, \href{http://arxiv.org/abs/2003.11554}{{\tt
  arXiv:2003.11554 [gr-qc]}}.

\bibitem{Bahamonde:2021dqn}
S.~Bahamonde, M.~Caruana, K.~F. Dialektopoulos, V.~Gakis, M.~Hohmann,
  J.~Levi~Said, E.~N. Saridakis, and J.~Sultana, ``{Gravitational-wave
  propagation and polarizations in the teleparallel analog of Horndeski
  gravity},'' \href{http://dx.doi.org/10.1103/PhysRevD.104.084082}{{\em Phys.
  Rev. D} {\bf 104} (2021) no.~8, 084082},
  \href{http://arxiv.org/abs/2105.13243}{{\tt arXiv:2105.13243 [gr-qc]}}.

\bibitem{Bernardo:2021bsg}
R.~C. Bernardo, J.~L. Said, M.~Caruana, and S.~Appleby, ``{Well-tempered
  Minkowski solutions in teleparallel Horndeski theory},''
  \href{http://dx.doi.org/10.1088/1361-6382/ac36e4}{{\em Class. Quant. Grav.}
  {\bf 39} (2022) no.~1, 015013}, \href{http://arxiv.org/abs/2108.02500}{{\tt
  arXiv:2108.02500 [gr-qc]}}.

\bibitem{Bernardo:2021izq}
R.~C. Bernardo, J.~L. Said, M.~Caruana, and S.~Appleby, ``{Well-tempered
  teleparallel Horndeski cosmology: a teleparallel variation to the
  cosmological constant problem},''
  \href{http://dx.doi.org/10.1088/1475-7516/2021/10/078}{{\em JCAP} {\bf 10}
  (2021)  078}, \href{http://arxiv.org/abs/2107.08762}{{\tt arXiv:2107.08762
  [gr-qc]}}.

\bibitem{Dialektopoulos:2021ryi}
K.~F. Dialektopoulos, J.~L. Said, and Z.~Oikonomopoulou, ``{Classification of
  teleparallel Horndeski cosmology via Noether symmetries},''
  \href{http://dx.doi.org/10.1140/epjc/s10052-022-10201-7}{{\em Eur. Phys. J.
  C} {\bf 82} (2022) no.~3, 259}, \href{http://arxiv.org/abs/2112.15045}{{\tt
  arXiv:2112.15045 [gr-qc]}}.

\bibitem{Bahamonde:2017ize}
S.~Bahamonde, C.~G. B\"ohmer, S.~Carloni, E.~J. Copeland, W.~Fang, and
  N.~Tamanini, ``{Dynamical systems applied to cosmology: dark energy and
  modified gravity},''
  \href{http://dx.doi.org/10.1016/j.physrep.2018.09.001}{{\em Phys. Rept.} {\bf
  775-777} (2018)  1--122}, \href{http://arxiv.org/abs/1712.03107}{{\tt
  arXiv:1712.03107 [gr-qc]}}.

\bibitem{Gonzalez-Espinoza:2020jss}
M.~Gonzalez-Espinoza and G.~Otalora, ``{Cosmological dynamics of dark energy in
  scalar-torsion $f(T,\phi )$ gravity},''
  \href{http://dx.doi.org/10.1140/epjc/s10052-021-09270-x}{{\em Eur. Phys. J.
  C} {\bf 81} (2021) no.~5, 480}, \href{http://arxiv.org/abs/2011.08377}{{\tt
  arXiv:2011.08377 [gr-qc]}}.

\bibitem{Hayashi:1979qx}
K.~Hayashi and T.~Shirafuji, ``{New General Relativity},''
  \href{http://dx.doi.org/10.1103/PhysRevD.19.3524}{{\em Phys. Rev. D} {\bf 19}
  (1979)  3524--3553}. [Addendum: Phys.Rev.D 24, 3312--3314 (1982)].

\bibitem{Chandrasekhar:1984siy}
S.~Chandrasekhar, ``{The Mathematical Theory of Black Holes},''
  \href{http://dx.doi.org/10.1007/978-94-009-6469-3_2}{{\em Fundam. Theor.
  Phys.} {\bf 9} (1984)  5--26}.

\bibitem{Krssak:2015oua}
M.~Kr\v{s}\v{s}\'ak and E.~N. Saridakis, ``{The covariant formulation of f(T)
  gravity},'' \href{http://dx.doi.org/10.1088/0264-9381/33/11/115009}{{\em
  Class. Quant. Grav.} {\bf 33} (2016) no.~11, 115009},
  \href{http://arxiv.org/abs/1510.08432}{{\tt arXiv:1510.08432 [gr-qc]}}.

\bibitem{Bahamonde:2015zma}
S.~Bahamonde, C.~G. B\"ohmer, and M.~Wright, ``{Modified teleparallel theories
  of gravity},'' \href{http://dx.doi.org/10.1103/PhysRevD.92.104042}{{\em Phys.
  Rev. D} {\bf 92} (2015) no.~10, 104042},
  \href{http://arxiv.org/abs/1508.05120}{{\tt arXiv:1508.05120 [gr-qc]}}.

\bibitem{DeFelice:2010aj}
A.~De~Felice and S.~Tsujikawa, ``{f(R) theories},''
  \href{http://dx.doi.org/10.12942/lrr-2010-3}{{\em Living Rev. Rel.} {\bf 13}
  (2010)  3}, \href{http://arxiv.org/abs/1002.4928}{{\tt arXiv:1002.4928
  [gr-qc]}}.

\bibitem{RezaeiAkbarieh:2018ijw}
A.~Rezaei~Akbarieh and Y.~Izadi, ``{Tachyon Inflation in Teleparallel
  Gravity},'' \href{http://dx.doi.org/10.1140/epjc/s10052-019-6819-z}{{\em Eur.
  Phys. J. C} {\bf 79} (2019) no.~4, 366},
  \href{http://arxiv.org/abs/1812.06649}{{\tt arXiv:1812.06649 [gr-qc]}}.

\bibitem{Bahamonde:2017wwk}
S.~Bahamonde, C.~G. B\"ohmer, and M.~Kr\v{s}\v{s}\'ak, ``{New classes of
  modified teleparallel gravity models},''
  \href{http://dx.doi.org/10.1016/j.physletb.2017.10.026}{{\em Phys. Lett. B}
  {\bf 775} (2017)  37--43}, \href{http://arxiv.org/abs/1706.04920}{{\tt
  arXiv:1706.04920 [gr-qc]}}.

\bibitem{Gonzalez:2019tky}
P.~A. Gonz\'alez, S.~Reyes, and Y.~V\'asquez, ``{Teleparallel Equivalent of
  Lovelock Gravity, Generalizations and Cosmological Applications},''
  \href{http://dx.doi.org/10.1088/1475-7516/2019/07/040}{{\em JCAP} {\bf 07}
  (2019)  040}, \href{http://arxiv.org/abs/1905.07633}{{\tt arXiv:1905.07633
  [gr-qc]}}.

\bibitem{TheLIGOScientific:2017qsa}
{\bf LIGO Scientific, Virgo} Collaboration, B.~P. Abbott {\em et al.},
  ``{GW170817: Observation of Gravitational Waves from a Binary Neutron Star
  Inspiral},'' \href{http://dx.doi.org/10.1103/PhysRevLett.119.161101}{{\em
  Phys. Rev. Lett.} {\bf 119} (2017) no.~16, 161101},
\href{http://arxiv.org/abs/1710.05832}{{\tt arXiv:1710.05832 [gr-qc]}}.

\bibitem{AAKoley1999}
A.~A. Coley, ``Dynamical systems in cosmology,''.
  \url{https://arxiv.org/abs/gr-qc/9910074}.

\bibitem{manuel2021cosmological}
G.-E. Manuel and O.~Giovanni, ``Cosmological dynamics of dark energy in
  scalar-torsion f (t, $\phi$) gravity,'' {\em Eur. Phys. J. C} {\bf 81} (2021)
   480.

\end{thebibliography}\endgroup


\providecommand{\href}[2]{#2}\begingroup\raggedright\endgroup

\end{document}